\documentclass[%
aps,
prx,
longbibliography,
reprint,
twocolumn,
nobalancelastpage,
superscriptaddress,
amsmath,amssymb,
floatfix
]{revtex4-2}

\usepackage[utf8]{inputenc} 
\usepackage[T1]{fontenc}
\usepackage[looser]{newtxtext}
\usepackage{microtype}
\usepackage[defaultlines=2,all]{nowidow}
\setcitestyle{numbers,square,citesep={,\kern-.20em}}

\usepackage{graphicx}
\usepackage{dcolumn}
\usepackage{booktabs}
\usepackage[dvipsnames]{xcolor}
\usepackage[colorlinks,urlcolor=Blue,linkcolor=Blue,citecolor=Blue]{hyperref}
\usepackage{orcidlink}
\usepackage[nameinlink,capitalise]{cleveref}
\crefname{equation}{Eq.\!}{Eqs.\!}

\usepackage{gensymb}
\usepackage{textcomp}
\usepackage{CJK}
\usepackage{bm}
\usepackage{braket}

\newcommand*{\figref}[2][]{%
  \hyperref[{#2}]{%
    Fig.\,\ref*{#2}%
    \ifx\\#1\\%
    \else
      (#1)%
    \fi
  }%
}

\newcommand*{\figureref}[2][]{%
  \hyperref[{#2}]{%
    Figure~\ref*{#2}%
    \ifx\\#1\\%
    \else
      (#1)%
    \fi
  }%
}
\mathchardef\mhyphen="2D
\newcommand{\op}[1]{\hat{#1}}
\renewcommand{\v}[1]{\ensuremath{\bm{\mathrm{#1}}}}
\newcommand{\SrClock}{\ensuremath{{^1\mathrm{S}_0}\!-\!{^3\mathrm{P}_0}}}%
\newcommand{\SrRed}{\ensuremath{{^1\mathrm{S}_0}\!-\!{^3\mathrm{P}_1}}}%
\newcommand{\SrBlue}{\ensuremath{{^1\mathrm{S}_0}\!-\!{^1\mathrm{P}_1}}}%
\newcommand{\ThreePhoton}{\ensuremath{{^1\mathrm{S}_0}\! \rightarrow\! {^3\mathrm{P}_1} \!\rightarrow\! {^3\mathrm{S}_1} \!\rightarrow\! {^3\mathrm{P}_0}}}%
\newcommand{\oneSzero}{\ensuremath{{^1\mathrm{S}_0}}}%
\newcommand{\threePzero}{\ensuremath{{^3\mathrm{P}_0}}}%
\newcommand{\threePone}{\ensuremath{{^3\mathrm{P}_1}}}%
\newcommand{\threePoneZero}{\ensuremath{{^3\mathrm{P}_1^{(0)}}}}%
\newcommand{\threePtwo}{\ensuremath{{^3\mathrm{P}_2}}}%
\newcommand{\threePJ}{\ensuremath{{^3\mathrm{P}_J}}}%
\newcommand{\threeSone}{\ensuremath{{^3\mathrm{S}_1}}}%
\newcommand{\onePone}{\ensuremath{{^1\mathrm{P}_1}}}%
\newcommand{\onePoneZero}{\ensuremath{{^1\mathrm{P}_1^{(0)}}}}%

\newcommand{\ketLS}[4]{\ket{#1,#2}_{\ell}\ket{#3,#4}_{s}}
\newcommand{\braLS}[4]{{}_{\ell}\!\bra{#1,#2}{}_{s}\!\bra{#3,#4}}

\newcommand{\braL}[2]{{}_{\ell}\!\bra{#1,#2}}

\newcommand{\ketJ}[3]{\ket{#2,#3}_{#1}}
\newcommand{\braJ}[3]{{}_{#1}\!\bra{#2,#3}}

\usepackage{mleftright}
\usepackage{multirow}

\begin{document}

\title{Collinear Three-Photon Excitation of a Strongly Forbidden Optical Clock Transition}

\begin{CJK*}{UTF8}{gbsn}

\author{Samuel P.\ Carman\,\orcidlink{0000-0003-3624-558X}}
 \thanks{These authors contributed equally to this work.}
\affiliation{Department of Physics, Stanford University, Stanford, California 94305, USA}

\author{Jan Rudolph\,\orcidlink{0000-0002-9083-162X}}
 \thanks{These authors contributed equally to this work.}
\affiliation{Department of Physics, Stanford University, Stanford, California 94305, USA}
\affiliation{Fermi National Accelerator Laboratory, Batavia, Illinois 60510, USA}

\author{\mbox{Benjamin E.\ Garber\,\orcidlink{0000-0002-1000-6874}}}
 \thanks{These authors contributed equally to this work.}
\affiliation{Department of Physics, Stanford University, Stanford, California 94305, USA}

\author{Michael J.\ Van de Graaff\,\orcidlink{0000-0002-3626-2251}}
\affiliation{Department of Physics, Stanford University, Stanford, California 94305, USA}

\author{Hunter Swan\,\orcidlink{0000-0002-1840-288X}}
\affiliation{Department of Physics, Stanford University, Stanford, California 94305, USA}

\author{Yijun Jiang (姜一君)\,\orcidlink{0000-0002-1170-7736}}
\affiliation{Department of Applied Physics, Stanford University, Stanford, California 94305, USA}

\author{Megan Nantel\,\orcidlink{0000-0002-9684-3314}}
\affiliation{Department of Applied Physics, Stanford University, Stanford, California 94305, USA}

\author{Mahiro Abe\,\orcidlink{0000-0003-4670-0347}}
\affiliation{Department of Physics, Stanford University, Stanford, California 94305, USA}

\author{Rachel L.\ Barcklay\,\orcidlink{0009-0003-0287-4633}}
\affiliation{Department of Electrical Engineering, Stanford University, Stanford, California 94305, USA}

\author{Jason M.\ Hogan\,\orcidlink{0000-0003-1218-2692}}
\email[]{hogan@stanford.edu}
\affiliation{Department of Physics, Stanford University, Stanford, California 94305, USA}

\begin{abstract}
The \SrClock{} clock transition in strontium serves as the foundation for the world's best atomic clocks and for gravitational wave detector concepts in clock atom interferometry.
This transition is weakly allowed in the fermionic isotope $^{87}$Sr but strongly forbidden in bosonic isotopes.
Here, we demonstrate coherent excitation of the clock transition in bosonic ${}^{88}$Sr using a novel collinear three-photon process in a weak magnetic field.
We observe Rabi oscillations with frequencies of up to $50~\text{kHz}$ using $\text{W}/\text{cm}^{2}$ laser intensities and Gauss-level magnetic field amplitudes.
The absence of nuclear spin in bosonic isotopes offers decreased sensitivity to magnetic fields and optical lattice light shifts, enabling atomic clocks with reduced systematic errors.
The collinear propagation of the laser fields permits the interrogation of spatially separated atomic ensembles with common laser pulses, a key requirement for dark matter searches and gravitational wave detection with next-generation quantum sensors.
\end{abstract}

\date{\today}

\maketitle
\end{CJK*}

Narrow optical transitions to long-lived atomic states are essential for many applications in metrology~\cite{Katori2011,Kozlov2018,Grotti2018}, precision timekeeping~\cite{Bloom2014, Ludlow2015, Marti2018}, and tests of fundamental physics~\cite{Blatt2008, Godun2014, Takamoto2020, Bothwell2022}.
The most accurate optical atomic clocks use \SrClock{} singlet-to-triplet transitions in alkaline-earth-like neutral atoms and ions (e.g.\ Sr, Yb, Mg, Al$^+$)~\cite{Bothwell2019, McGrew2018, Kulosa2015, Brewer2019}.
These ultranarrow transitions are weakly allowed in fermions because of the hyperfine interaction, but are strongly forbidden in bosons, where a large external magnetic field is required to induce a comparable coupling strength~\cite{Taichenachev2006, Barber2006, Akatsuka2008,Poli2014}.
Bosonic isotopes of neutral atoms promise considerable advantages, such as their lack of nuclear spin, simplified laser cooling and state preparation, as well as the scalar polarizability of their clock states~\cite{Santra2005, Ludlow2015}.
To access these desirable properties, coherent multiphoton processes in bosons have been proposed~\cite{Hong2005, Santra2005, Vitaly2007, Alden2014}.
Of particular interest is the three-photon excitation \ThreePhoton{}~\cite{Hong2005, Barker2016}, using a set of laser frequencies readily available for laser cooling, repumping, and imaging.

Ostensibly, this coherent three-photon excitation cannot be driven with collinear laser beams due to angular momentum selection rules~\cite{Vitaly2007, Grynberg1980}.
Instead, at least one of the beams must be sent from a different direction.
This is easily accommodated in most clock applications, where the beams can even be arranged to eliminate the net momentum transfer to the atoms~\cite{Barker2016, Panelli2024}.
In contrast, collinear laser beams are advantageous in applications where several atomic ensembles are ideally addressed with identical laser pulses, such as multiqubit entanglement in optical tweezer arrays~\cite{Madjarov2020, Schine2022, Eckner2023, Cao2024}, differential operation of atomic clocks~\cite{Hinkley2013, Kolkowitz2016, Schioppo2016, Oelker2019, BACON2021, Zheng2022}, and gradiometer configurations in clock atom interferometry~\cite{Hu2017, Abe2021}.
In gradiometers, common laser pulses are particularly important for laser frequency noise suppression, which requires that all spectral content must be delivered from the same direction~\cite{Graham2013}. 
Like clocks, clock atom interferometry proposals generally rely on naturally occurring narrow-line transitions in fermions~\cite{Abe2021}. 
Because of the weak coupling, interferometry pulses on these transitions are slow and the interferometer sensitivity is ultimately limited by the number of pulses that can be applied.
Unlocking bosonic isotopes for long-baseline clock atom interferometry via multiphoton processes could lead to a substantial increase in sensitivity by engineering an effective transition with both strong coupling and long excited state lifetime.

Here, we demonstrate a coherent, collinear three-photon process \ThreePhoton{} in bosonic $^{88}$Sr mediated by a weak magnetic bias field.
The constant field lifts the degeneracy of the intermediate \threePone{} sublevels that otherwise leads to destructive interference of the excitation paths.
We show why this transition cannot be driven with collinear laser beams without a magnetic field and illustrate the transformation of the effective angular wave function of \threePone{} under the associated external torque.
We drive this three-photon transition using a single trichromatic laser pulse, delivered via a polarization-maintaining optical fiber.
We observe Rabi oscillations between the clock states \oneSzero{} and \threePzero{} with frequencies of up to $50~\text{kHz}$, substantially surpassing what could be achieved with similar laser intensities on the naturally occurring single-photon transition in fermionic $^{87}$Sr and the magnetic-field-induced transition in $^{88}$Sr.

With this new technique, we demonstrate the first multi-photon clock atom interferometer using a Mach-Zehnder pulse sequence.
Since all spectral content for the three-photon transition is copropagating, this interferometer scheme is compatible with long-baseline clock gradiometers like \mbox{MAGIS-100}~\cite{Abe2021}.
Bosonic isotopes can now be employed in such experiments, which extends the utility of next-generation quantum sensors.

Alkaline-earth-like atoms such as strontium have two valence electrons and their atomic states are split into a singlet and a triplet manifold [see \figref[a]{Fig:LevelDiagrams}].
While strong electric dipole transitions are possible within each manifold, spin-flipping electric dipole transitions between singlet and triplet states are generally forbidden.
A notable exception is the \SrRed{} transition, which is weakly allowed through spin-orbit coupling.
The three-photon process described here combines this weak intercombination line and two strong electric dipole transitions into one coherent excitation.

\looseness=-1
The coupling strength of this multiphoton transition can be expressed as the product of the individual single-photon couplings $\Omega_i$ divided by the intermediate laser frequency detunings $\Delta_i$ [\figref[b]{Fig:LevelDiagrams}].
When using linearly polarized light fields, the effective three-photon coupling can be written as
\begin{equation}\label{eq:ThreePhotonCoupling}
    \Omega_\text{eff} = \frac{\Omega_1 \Omega_2 \Omega_3}{2\Delta_2} \, \epsilon_{jkl}  \, D_{jn} \, e^{(1)}_{n\phantom{l}} e^{(2)}_{k\phantom{l}} e^{(3)}_{l\phantom{l}},
\end{equation}
where $\epsilon_{jkl}$ is the Levi-Civita symbol, and $e_n^{(i)}$ is the $n^{\text{th}}$ Cartesian component of the $i^{\,\text{th}}$ laser's polarization vector, with an implicit sum over repeated indices (see \cref{Sec:EffCoupling}).
The entries of the matrix $D_{jn}$ depend on the inverse detunings $1/\Delta_{1,m}$ to the three magnetic sublevels $m$ of \threePone{}.
When these sublevels are degenerate, $D_{jn}=\frac{1}{\Delta_1}\delta_{jn}$ is diagonal and \Cref{eq:ThreePhotonCoupling} reduces to
\begin{equation}\label{eq:TripleProduct}
\Omega_{\text{eff}} = \frac{\Omega_1\Omega_2\Omega_3}{2\Delta_1\Delta_2} \;\; \v{e}^{(1)} \cdot 
 \left(\v{e}^{(2)}\times\v{e}^{(3)}\right).
\end{equation}
The three-photon coupling therefore scales with the volume spanned by the electric field polarization vectors, which vanishes for collinearly propagating optical fields.

Applying an external magnetic field $\v{B}$ along $\v{\hat{z}}$ lifts the degeneracy of the \threePone{} sublevels via a Zeeman shift \mbox{$\delta \omega_B = g_J \, \mu_\text{B} \, B/\hbar$}, where $g_J$ is the Land{\'e} $g$ factor and $\mu_\text{B}$ is the Bohr magneton [see \figref[b]{Fig:LevelDiagrams}].
Choosing a set of polarizations $\v{e}^{(1)} = \v{e}^{(2)} = \v{\hat{x}}$ and $\v{e}^{(3)} = \v{\hat{z}}$ with collinear propagation direction $\v{\hat{y}}$, the coupling becomes
\begin{equation}\label{eq:DestInterference}
\Omega_{\text{eff}} = \frac{\Omega_1 \Omega_2 \Omega_3}{4\Delta_2} \left(\frac{1}{\Delta_1 - \delta\omega_B} - \frac{1}{\Delta_1 + \delta\omega_B} \right).
\end{equation}
The minus sign between the terms conveys the destructive interference of the two excitation paths (via $m=-1$ and $m=+1$) in the absence of a magnetic field.

\begin{figure}[t]
    \centering
    \vspace*{12pt}\includegraphics[width=0.76\columnwidth]{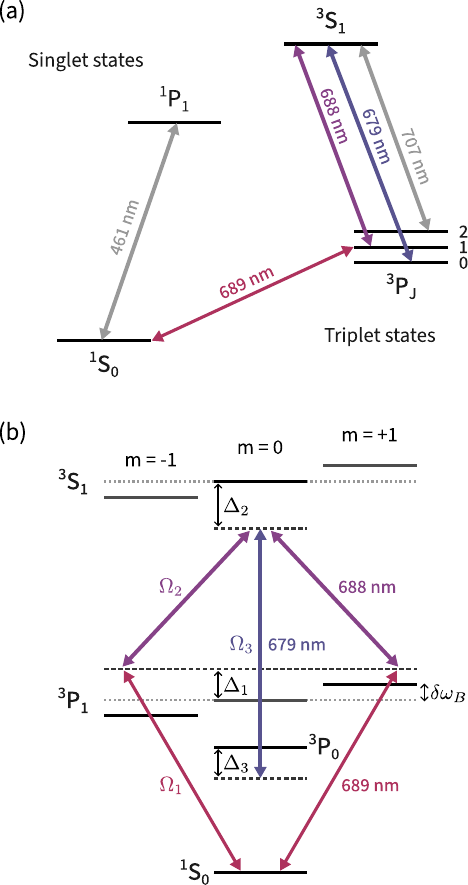}
    \caption{(a) Singlet and triplet states of $^{88}$Sr and the relevant transition wavelength connecting the atomic states.
    We describe a three-photon transition between \oneSzero{} and \threePzero{} using laser light at 689~nm (magenta), 688~nm (purple), and 679~nm (blue).\linebreak[4]
    (b) Energy levels involved in the three-photon transition with Zeeman sublevels $m$ in the presence of a magnetic field, causing a relative shift $\delta\omega_B$. The polarizations of the optical fields are linear, with 689~nm\,($\Omega_1$) and 688~nm\,($\Omega_2$) normal, and 679~nm\,($\Omega_3$) parallel to the quantization axis.
    Cumulative frequency detunings of the lasers from the respective $m=0$ states are denoted by $\Delta_1$, $\Delta_2$, and $\Delta_3$.}
\label{Fig:LevelDiagrams}
\end{figure}

\begin{figure*}[t]
    \centering
    \includegraphics[width=0.75\textwidth]{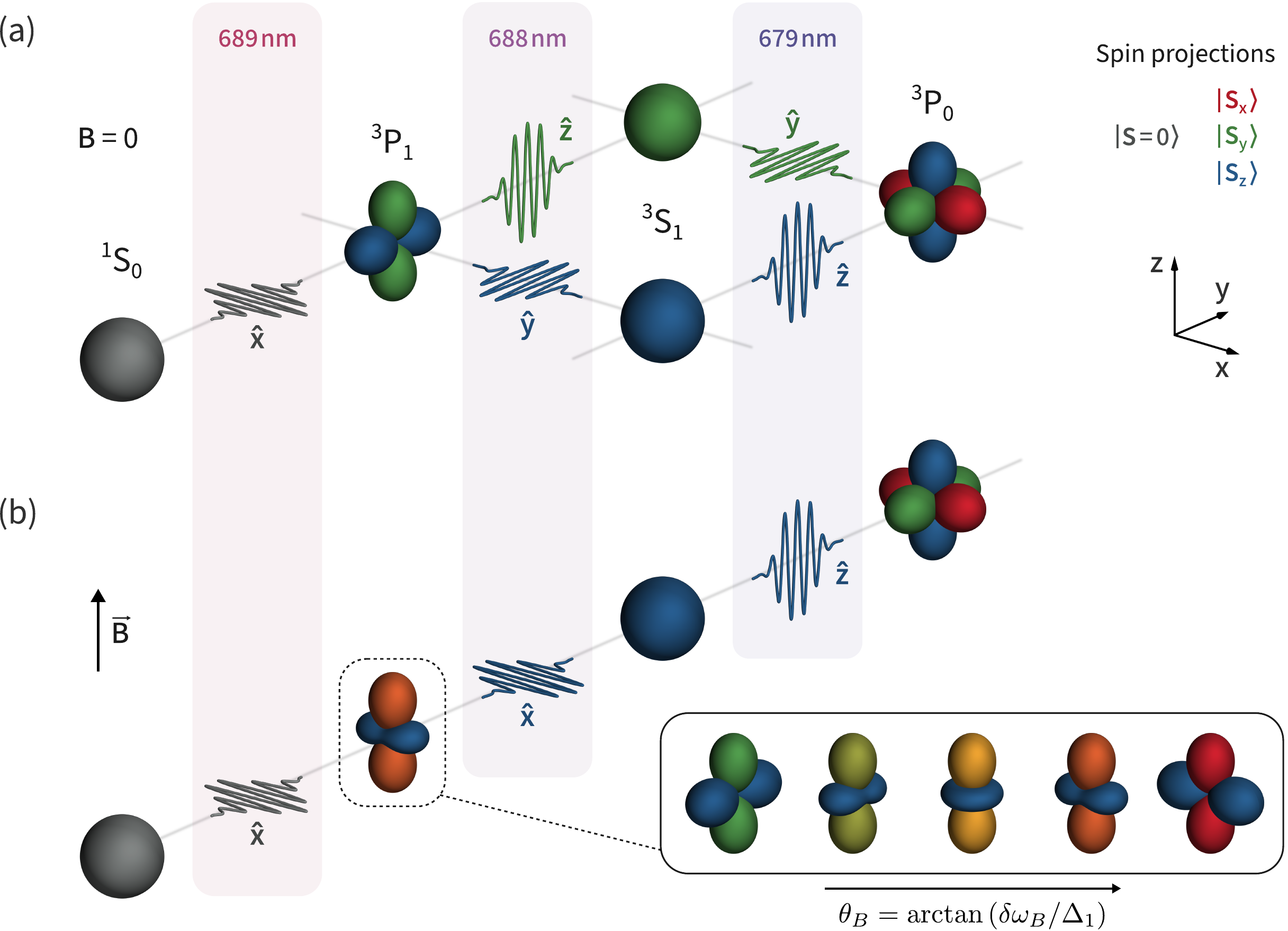}
    \caption{Engineering a collinear three-photon transition.
    Cartesian basis representation of the orbital angular momentum (shape) and spin (color) of the atomic wave function during the three-photon transition \ThreePhoton{}.
    While we depict the three-photon process as a sequence of interactions, the coherent excitation is nonsequential and all fields are applied simultaneously.
    The light fields linking the states are shown with the required polarizations, and spin-preserving transitions ($688$ and $679~\text{nm}$) are colored according to the spin state $\ket{S_x}$\,(red), $\ket{S_y}$\,(green), and $\ket{S_z}$\,(blue).
    (a) Contrary to the dipole-allowed transitions, the light at $689~\text{nm}$ causes an excitation orthogonal to its polarization along $\v{\hat{x}}$ and changes the spin state.
    The other two transitions are spin- (color) preserving and the light polarizations need to be aligned with the respective $p$ orbitals.
    To reach the appropriate spin projection of \threePzero{} via either the $\v{\hat{x}}\,\mhyphen\,\v{\hat{y}}\,\mhyphen\,\v{\hat{z}}$ (blue) or $\v{\hat{x}}\,\mhyphen\,\v{\hat{z}}\,\mhyphen\,\v{\hat{y}}$ (green) combination, at least one of the polarization vectors must have a noncoplanar component with respect to the other two.
    (b) Applying a magnetic field along $\v{\hat{z}}$ deforms the effective angular wave function of \threePone{} and permits driving the three-photon transition with collinear light via $\v{\hat{x}}\,\mhyphen\,\v{\hat{x}}\,\mhyphen\,\v{\hat{z}}$ (blue). Inset: the effective wave function for various angles $\theta_B$ from 0{\textdegree} to 90{\textdegree} (maximal projection along \v{\hat{x}}). Note that the collinear combination $\v{\hat{x}}\,\mhyphen\,\v{\hat{z}}\,\mhyphen\,\v{\hat{x}}$ (red) would now also be permitted (not shown).}
    \label{Fig:TransitionPaths}
\end{figure*}

The optical field requirements without external magnetic field can be illustrated using the angular part of the atomic wave function, which consists of an entangled superposition of orbital angular momentum $\v{L}$ and spin angular momentum $\v{S}$ [see \figref[a]{Fig:TransitionPaths}].
The light field at $689~\text{nm}$ along $\v{\hat{x}}$ drives a weakly allowed singlet-to-triplet transition to \threePone{} possible only through spin-orbit coupling.
The triplet component of the resulting state has the form $\left(\ket{L_y}\ket{S_z}-\ket{L_z}\ket{S_y}\right)$, where $\ket{L_n}$ and $\ket{S_n}$ are the Cartesian basis states of the orbital angular momentum and spin operators (see \cref{Sec:CartBasis}).
Here, the symmetry axes of these $p$ orbitals $\ket{L_n}$ are orthogonal to the polarization of the exciting laser.
In contrast, the other two light fields at $688$ and $679~\text{nm}$ drive spin-preserving, dipole-allowed transitions where the polarization of the light has to match the $p$ orbital symmetry axis of the associated spin projection $\ket{S_n}$ (color).
Since the final state of the three-photon process \threePzero{} has the form $\left(\ket{L_x}\ket{S_x}+\ket{L_y}\ket{S_y}+\ket{L_z}\ket{S_z}\right)$, two mutually orthogonal polarization components are required to reach the appropriate spin state (either via the green path or the blue path).
Thus, \figref[a]{Fig:TransitionPaths} is a visual representation of the scalar triple product in \Cref{eq:TripleProduct}, illustrating the need for noncopropagating light fields~\footnote{Though we only show the case where the first photon is $\v{\hat{x}}$ polarized, the requirement for noncoplanar polarizations is independent of the choice of the first polarization for any $\v{e}^{(1)}\perp\v{B}$.}.

Applying a magnetic field along $\v{\hat{z}}$ alters the angular part of the effective \threePone{} wave function
\begin{align}
\begin{split}
    \Ket{\threePone{}, \theta_B} =  & \tfrac{1}{\sqrt{2}}\Big(\!\sin\theta_B\Ket{L_x}+i\cos\theta_B\Ket{L_y}\!\Big)\Ket{S_z} \\
    -& \tfrac{1}{\sqrt{2}}\Ket{L_z}\Big(\!\sin\theta_B\Ket{S_x}+i\cos\theta_B\Ket{S_y}\!\Big),
    \label{Eq:PsiTheta}
\end{split}
\end{align}
where we define the angle $\theta_B \equiv \arctan{\beta}$, with dimensionless ratio $\beta\equiv\delta\omega_B/\Delta_1$.
With increasing $\theta_B$ the state acquires both a spin and an orbital angular momentum component along $\v{\hat{x}}$.
The three-photon transition can now be driven with collinear laser beams, e.g.\ using $\v{\hat{x}}\,\mhyphen\,\v{\hat{x}}\,\mhyphen\,\v{\hat{z}}$ polarizations \mbox{[see \figref[b]{Fig:TransitionPaths}]}.

\begin{figure*}[t]
    \centering
    \includegraphics[width=0.9\textwidth]{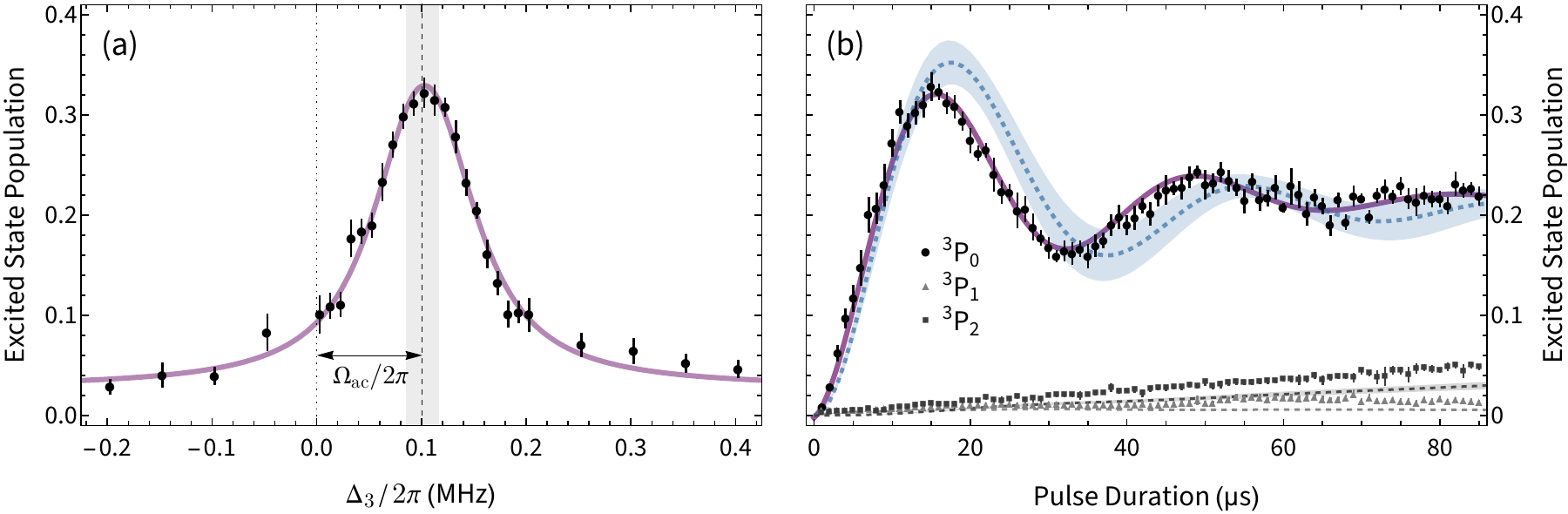}
    \caption{Coherent three-photon excitation of the \SrClock{} clock transition. 
    (a) Line scan of the three-photon transition showing the fractional excited state population versus the cumulative laser detuning $\Delta_3$, using $\Delta_{1}/2\pi = 9.95(1)~\text{MHz}$, $\Delta_2/2\pi = -2.54~\text{GHz}$, and $\delta\omega_{B}/2\pi = 21.13(1)~\text{MHz}$.
    Each data point is an average of five measurements and the error bars represent the standard error of the mean.
    The location of $\Delta_3=0$ is a prediction based on the measured absolute frequencies of the lasers and the literature values for the strontium transition energies.
    The dashed line and shaded region indicate the expected resonance location at $\Delta_\text{eff} = 0$ and its uncertainty, with a calculated ac-Stark shift of $\Omega_\text{ac}/2\pi = -101(16)~\text{kHz}$ from \Cref{Eq:ACStark}. This agrees well with the Voigt profile fit to the data (solid curve), from which we extract a center frequency of $103(1)~\text{kHz}$.
    (b) Rabi oscillation at the measured three-photon resonance frequency, showing the fractional excited state populations versus the pulse duration.
    Circles, triangles, and squares indicate the population in the states \threePzero{}, \threePone{}, and \threePtwo{}, respectively.
    Each data point is an average of five measurements and the error bars represent the standard error of the mean.
    The dashed curves (blue, dark gray, light gray) correspond to the time traces predicted
    by the density matrix simulation averaged over the measured broadening, and the bands
    are the $\pm1\sigma$ predictions based on the uncertainties in the simulation parameters.
    The solid curve (purple) is a fit to an exponentially damped sinusoid used to extract the observed Rabi frequency of $30.2(2)~\text{kHz}$.
    }
    \label{Fig:LineRabi}
\end{figure*}

We demonstrate this coherent three-photon excitation using a thermal ensemble of bosonic $^{88}$Sr atoms.
In a constant magnetic field aligned with $\v{\hat{z}}$, a single laser pulse containing all three wavelengths is applied to the atoms along the $\v{\hat{y}}$ direction.
The laser light at $689$ and $688~\text{nm}$ is linearly polarized along $\v{\hat{x}}$, while the $679~\text{nm}$ component is linearly polarized along $\v{\hat{z}}$.
To avoid populating any of the intermediate states, the lasers are each detuned from their respective single-photon resonance by several hundred natural linewidths using an optical frequency comb as a reference.
To estimate the three-photon pulse fidelity, we measure the populations in the ground state \oneSzero{} and in all metastable exited states \threePJ{} using a state-selective imaging scheme (see \cref{Sec:Imaging}).

The resonance frequency for the three-photon transition is given by the effective detuning $\Delta_{\text{eff}} \equiv \Delta_3 + \Omega_{\text{ac}}$, with cumulative laser detuning $\Delta_3$ and ac-Stark shift (see \cref{Sec:Detunings}):
\begin{equation}\label{Eq:ACStark}
    \Omega_{\text{ac}} = - \frac{\Omega_1^2}{2}\frac{\Delta_{1}}{\Delta_{1}^2-\delta\omega_B^2} + \frac{\Omega_3^2}{4\Delta_2}.
\end{equation}
We vary the laser detuning $\Delta_3$ around the calculated resonance frequency and find the three-photon transition as expected.
\figureref[a]{Fig:LineRabi} shows such a line scan with peak excited state transfer of $32\%$ and a line shape that fits to a Voigt profile with HWHM of $57(6)\,\text{kHz}$. 
At the measured three-photon resonance, we scan the pulse duration and observe Rabi oscillations in the \threePzero{} population with a fitted frequency of $30.2(2)\,\text{kHz}$ [see \figref[b]{Fig:LineRabi}].
This is in good agreement with the predicted three-photon Rabi frequency of $28(1)\,\text{kHz}$ from \Cref{eq:DestInterference}, using laser intensities $(I_1, I_2, I_3) = (0.16, 1.4, 0.91)~\text{W}/\text{cm}^{2}$ and a magnetic field amplitude of $10.1~\text{G}$ (see  \cref{Sec:Methods}).
The low state transfer and noticeable damping of the Rabi oscillation stem from the temperature of the atoms, laser frequency noise, and intensity inhomogeneity across the cloud, since we intentionally focus the laser beam to increase the peak intensity.
Much of this damping can be avoided simply by using a larger beam size (see \cref{Sec:DensityMatrix}).
The populations in the other metastable excited states \threePone{} and \threePtwo{} are small, consistent with minimal off-resonant single-photon excitation.

We compare the observed three-photon dynamics to a density matrix simulation that includes all dipole-allowed couplings and decay channels between the 13 magnetic sublevels of the \oneSzero{}, \threeSone{}, and \threePJ{} states. 
For each state population, we compute a weighted average that accounts for Doppler shifts, laser frequency noise, and intensity inhomogeneity.
In \figureref[b]{Fig:LineRabi}, we show the predicted populations as a function of time, using the measured cloud size and temperature, laser beam size, and cumulative laser frequency noise.
This parameter-free model qualitatively agrees with the experimentally observed dynamics, which feature a slightly larger damping rate of the Rabi oscillation as well as larger residual populations in the other triplet states.
We attribute these deviations in part to the simplified laser noise model which assumes a Gaussian distribution for the atom-laser detuning.
Other factors include inhomogeneity of the imaging beam intensity and residual crosstalk between the image measurement ports (see \cref{Sec:Imaging}).
The majority of the damping ($97\%$) is explained by the temperature of the atoms, the laser beam size, and laser noise.
Off-resonant scattering inherent to the three-photon process accounts for only a small portion of the damping ($3\%$).
We summarize the individual contributions to the damping rate of the modeled three-photon Rabi oscillation in \Cref{Tab:Damping}.
Our density matrix simulation shows that this loss can be reduced further with an appropriate choice of laser intensities and detunings to support a state transfer of over $99\%$ while maintaining a Rabi frequency of at least $10~\text{kHz}$ (see \cref{Sec:DensityMatrix}).

We further characterize the three-photon process at various magnetic fields and laser detunings by measuring the resonant Rabi frequencies and comparing their magnitudes.
It is convenient to parametrize the coupling via the ratio of the detunings $\beta = \delta\omega_B/\Delta_1$ such that \Cref{eq:DestInterference} can be written as
\begin{equation}
\Omega_\text{eff} = \Omega_0 \; \frac{\beta}{1 - \beta^2},
\label{eq:meta1}
\end{equation}
where $\Omega_0 \equiv \Omega_1\Omega_2\Omega_3/(2\Delta_1\Delta_2)$ is the maximum Rabi coupling in \Cref{eq:TripleProduct}.
In \figureref[a]{Fig:Metaplots}, we show the normalized three-photon coupling $\Omega_\text{eff}/\Omega_0$ over a wide range of detuning ratios $\beta$.
We find excellent agreement with the expected coupling strength for the effective two-level system based on the inferred single-photon couplings (see \cref{Sec:Methods}).
For ratios close to $\beta=1$, direct excitation to \threePone{} is no longer negligible and the observed Rabi frequencies are slightly lower than predicted.
While the strong coupling near this divergence is desirable, sufficient detuning from the single-photon resonance is required to maintain efficient three-photon excitation.
We achieve Rabi frequencies as high as $50~\text{kHz}$ while maintaining a detuning of over $500$ natural linewidths.

\begin{figure*}[t]
    \centering
    \includegraphics[width=0.9\textwidth]{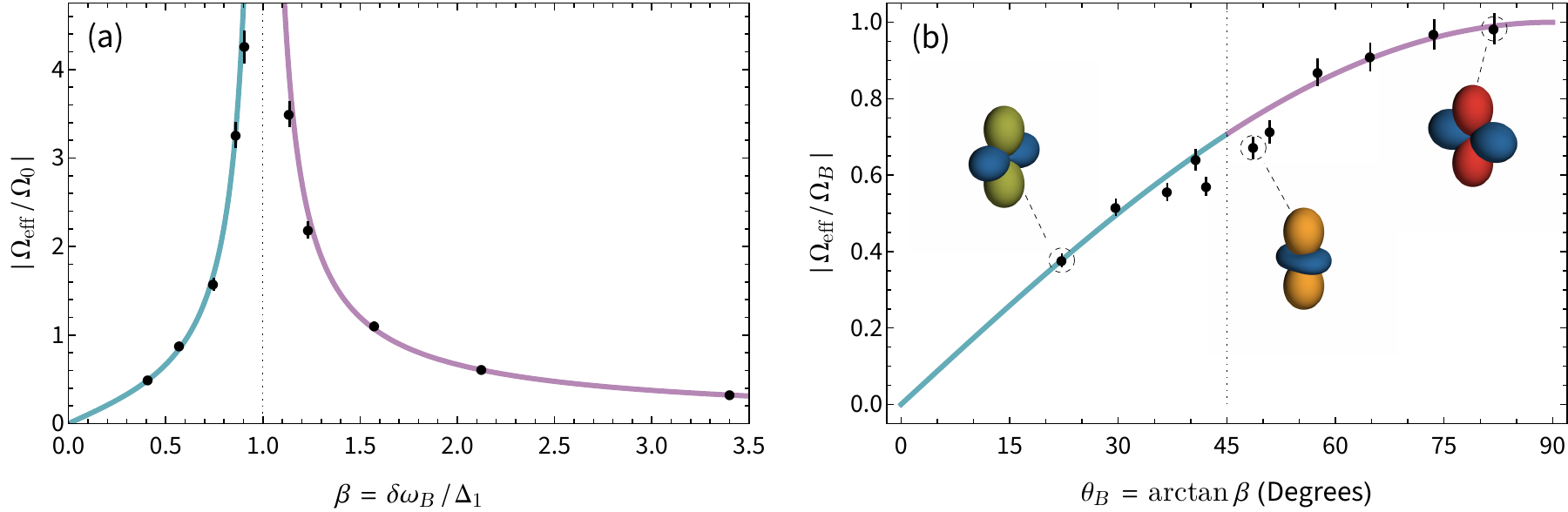}
    \caption{Dimensionless three-photon coupling and projection angle.
    (a) Measured normalized Rabi frequencies $|\Omega_{\text{eff}}/\Omega_0|$ as a function of $\beta$, where $\Omega_0$ is the predicted maximum coupling strength at $\beta=0$.
    Error bars indicate one standard deviation of the combined uncertainty in the measurement of $\Omega_{\text{eff}}$ and the theoretical prediction of $\Omega_0$.
    The solid curves correspond to the expected scaling from \Cref{eq:meta1}.
    The parameter space is divided into two distinct regions, $\beta<1$~(teal) and $\beta>1$~(purple).
    The coupling strength diverges at $\beta=1$~(dotted line), which corresponds to the single-photon resonance to the $\Ket{\threePone{}, m=+1}$ state.
    For this data, the laser detuning $\Delta_1$ is always positive, but an identical divergence around the $m=-1$ state can be found for negative detunings.
    (b) The same data in a different normalization $|\Omega_\text{eff}/\Omega_B|$, where $\Omega_B$ is the maximum coupling at a given magnetic field, versus the projection angle $\theta_B$. 
    This parametrization emphasizes the wave-plate-like rotation of the atom's effective dipole moment relative to the polarization of the light, as depicted in \figref[b]{Fig:TransitionPaths}. 
    The solid curve corresponds to the sinusoid in \Cref{eq:meta2}, with the same colored regions as in (a).
    Since there is no divergence in this parametrization, the slight deviation from the model around $\theta_B = 45^{\degree}$ (dotted line) is more apparent.
    The callouts show the shape of the effective wave function $\ket{\threePone{}, \theta_B}$ for a selection of data points.}
    \label{Fig:Metaplots}
\end{figure*}

We expect the three-photon coupling to depend on the shape of the effective orbital $\ket{\threePone{}, \theta_B}$ and the projection of the light onto it [see \figref[b]{Fig:TransitionPaths}].
We define the maximum Rabi coupling at a given magnetic field \mbox{$\Omega_B \equiv \Omega_0 \, \sqrt{1 + \beta^2}/(1 - \beta^2)$}, such that the three-photon coupling becomes
\begin{equation}
    \Omega_\text{eff} = \Omega_B \,\sin{\theta_B}, \quad \text{for} \;  -\tfrac{\pi}{2} \leq \theta_B \leq \tfrac{\pi}{2}.
\label{eq:meta2}
\end{equation}
In \figureref[b]{Fig:Metaplots}, we plot the previous dataset in the alternative parametrization $\Omega_\text{eff}/\Omega_B$ versus the projection angle $\theta_B$.
This normalization avoids the divergence at $\beta=1$ and shows the sinusoidal variation of the projection, which asymptotes to 1 for large values of $\beta$.

At a given $\Omega_B$, the coupling depends on the projection of the atom's dipole moment onto the light polarization vector.
Tuning $\theta_B$ has the same effect on the coupling as using a polarization for the second light field $\v{e}^{(2)}$ that is impossible to attain with collinear light.
For a general elliptical polarization $\v{e}^{(2)} = \sin{\theta}\; \v{\hat{x}} + i \cos{\theta}\; \v{\hat{y}}$ with ellipticity angle $\theta$ in the \mbox{$xy$-plane}, the three-photon coupling becomes $\Omega_{\text{eff}} = \Omega_B\cos(\theta_B-\theta)$.
In this work we use $\theta=\pi/2$, which is the only polarization choice compatible with collinear propagation along $\v{\hat{y}}$.
Thus, the external torque from the magnetic field tunes the effective dipole moment of the atom in much the same way as a wave plate rotates the polarization of light (see \cref{Sec:Metaplots}).

\begin{figure}[t]
    \centering\vspace*{-5pt}
    \includegraphics[width=0.9\columnwidth]{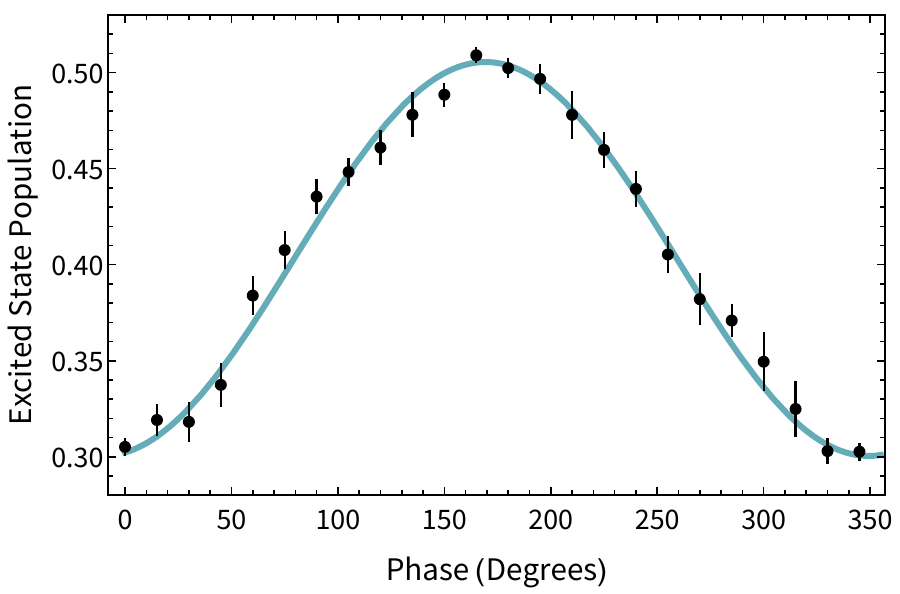}\vspace*{-4pt}
    \caption{A three-photon clock atom interferometer with bosonic $^{88}$Sr. Normalized \threePzero{} excited state population after a Mach-Zehnder pulse sequence versus the laser phase of the first pulse.
    The pulses are separated by an interrogation time $T = 200~\micro\text{s}$.
    Every data point is the mean of ten measurements and error bars indicate the standard error of the mean.
    The solid line is a sinusoidal fit with a visibility (peak-to-peak amplitude) of $20.5(3)\%$.
    The three-photon pulses are applied with $\Delta_{1}/2\pi = 28.0(1)\,\text{MHz}$ and $\delta\omega_{B}/2\pi = 21.15(1)\,\text{MHz}$.}
\label{Fig:MZ}
\end{figure}

Next, we demonstrate a proof-of-principle clock atom interferometer using the collinear three-photon transition [\figref{Fig:MZ}].
We apply a symmetric Mach-Zehnder pulse sequence~\cite{Kasevich1991} consisting of a beam splitter pulse ($\pi/2$), followed by a mirror pulse ($\pi$), and a recombination pulse ($\pi/2$).
These pulses are separated by an interrogation time $T = 200~\micro\text{s}$.
The visibility of the interferometer signal is maintained for interrogation times much longer than the intermediate state lifetimes, confirming coherent three-photon excitation of the long-lived clock state.
This visibility can be improved by using a colder atomic ensemble as well as a larger beam size to reduce intensity inhomogeneity across the cloud.

This is the first demonstration of a \SrClock{} clock atom interferometer in $^{88}$Sr without the use of a large magnetic bias field~\cite{Hu2017}.
Compared to previous work, we observe a 30-fold increase in Rabi frequency using only 17\% of the laser intensity and 3\% of the magnetic field strength~\cite{Hu2019}.
Because of the low magnetic field requirement, our method is applicable in long-baseline atom interferometers, where efficient atom-light interaction is desired anywhere along the baseline~\cite{Hartwig2015,Kovachy2015}.
Additionally, the small magnetic field amplitude sets a \threePzero{} lifetime of over $10^6\,\text{s}$, significantly longer than the natural lifetime of $118~\text{s}$ in $^{87}$Sr~\cite{Muniz2021}.
Long coherence times are essential for probing gravity~\cite{Xu2019} and detecting gravitational waves with long baselines~\cite{Graham2013, Kolkowitz2016}.
We also observe more than ten times the Rabi frequency than what could be achieved on the naturally occurring $^{87}$Sr clock transition with the same total intensity.
High Rabi frequencies ensure broad velocity acceptance~\cite{Wilkason2022} and allow for more pulses in a given free-fall time, which is beneficial for any large momentum transfer (LMT) atom interferometer~\cite{Kovachy2015, Rudolph2020}, regardless of scale.
LMT atom interferometry between the \oneSzero{} and \threePzero{} clock states could now have utility in small-scale devices due to the speed advantage of three-photon atom optics, which relaxes the atom temperature requirements compared to Bragg transitions~\cite{Szigeti2012}. 
Moreover, a modified three-photon transition with the $679~\text{nm}$ beam counterpropagating with respect to the other two beams would yield a net momentum transfer of $3\,\hbar k$ per pulse, allowing for a larger sensitivity enhancement than two-photon atom optics for the same number of pulses.
As such, LMT-enhanced three-photon clock atom interferometers can be employed in similar applications as compact Bragg and Raman atom interferometers, including gravimetry, gravity gradiometry, inertial navigation, and geodesy~\cite{Cronin2009}.

Next-generation long-baseline clock atom interferometer experiments, including MAGIS-100~\cite{Abe2021} and AION~\cite{Badurina2020}, rely on LMT atom optics to significantly enhance their sensitivity.
For these applications, interferometry pulses with transfer fidelities above $99\%$ are required~\cite{Wilkason2022,Beguin2023}.
Density matrix simulations suggest that our method can support such pulse efficiencies with $\text{W}/\text{cm}^2$-scale laser intensities and magnetic fields below $10~\text{G}$, yielding Rabi frequencies of at least $10~\text{kHz}$ (see \cref{Sec:Fidelity}).
This pulse fidelity could be reached by using a larger beam waist (e.g., in the millimeter range) to reduce intensity inhomogeneity across the cloud, combined with either a lower-temperature cloud, a velocity-selective initial pulse \cite{Kovachy2015}, or a composite pulse sequence to reduce Doppler broadening \cite{Dunning2014}.
Moreover, collinear three-photon transitions enable the use of bosonic isotopes that feature higher natural abundance, simpler level structure, and reduced magnetic field sensitivity.
All of these attributes are favorable for reaching the challenging sensitivity targets for gravitational wave detection and dark matter searches with clock atom interferometers.

Driving the \SrClock{} clock transition in bosonic atoms also has many advantages in atomic clocks, including insensitivity to the polarization of trapping light, and lack of a first-order Zeeman shift~\cite{Baillard2007, Akatsuka2008, Ludlow2015}.
Existing $^{88}$Sr clocks are limited by shifts from the second-order Zeeman effect and from the probe light~\cite{Akatsuka2010, Poli2014, Origlia2018, Norcia2019}.
The method described here employs a magnetic field strength that is substantially smaller per unit Rabi frequency, reducing the Zeeman shift for a given coupling strength.
In addition, the overall probe light shift can be eliminated to first order with an appropriate choice of laser detunings (see \cref{Sec:acStark}).
Thus, the three-photon transition demonstrated here appears promising for improving the accuracy of bosonic clocks.

Many of these advantages are also favorable for applications in quantum information science~\cite{Stock2008}, where the bosonic \threePzero{}~\cite{Young2020, Schine2022, Pagano2022, Cao2024} and \threePtwo{}~\cite{Okuno2022, Trautmann2023} states show promise for qubit storage and manipulation by leveraging their long coherence times and environmental insensitivity.
While not demonstrated in this work, a similar three-photon excitation $\oneSzero{}\!\rightarrow\!\threePone{}\!\rightarrow\!\threeSone{}\!\rightarrow\!\threePtwo{}$ can be achieved by substituting $707~\text{nm}$ for $679~\text{nm}$ light~\cite{Barker2016}, which may have advantages over magnetic quadrupole excitation to \threePtwo{}~\cite{Pucher2024, Klusener2024}.

The method demonstrated here is broadly applicable in other systems that use ultranarrow singlet-to-triplet transitions, especially when high Rabi coupling is beneficial and bosonic isotopes are preferred.
In particular, precision measurements based on the differential interrogation of distant atoms benefit from the application of common, copropagating laser pulses.

\textit{Note added.} -- Upon completion of this manuscript, we became aware of related work on non-collinear three-photon transitions in Bose-Einstein condensates of $^{84}$Sr~\cite{He2024}.

\section*{Acknowledgements}
We wish to thank Mark Kasevich and Shaun Burd for helpful discussions.
We thank Leo Hollberg for letting us use his frequency reference.
B.G.\ acknowledges support from the Office of Naval Research through the NDSEG fellowship.
This work was supported by the Gordon and Betty Moore Foundation Grant No.\ GBMF7945, the NSF QLCI Award No.\ OMA-2016244, and partially supported by the U.S.\ Department of Energy, Office of Science, National Quantum Information Science Research Centers, Superconducting Quantum Materials and Systems Center (SQMS) under Contract No.\ DE-AC02-07CH11359.

\appendix
\section{METHODS}\label{Sec:Methods}

\subsection{Experimental setup}
We prepare a thermal ensemble of $10^6$ ${}^{88}$Sr atoms at a temperature of $2~\micro\text{K}$, using a two-stage magneto-optical trap (MOT).
A more detailed description of the setup and experimental sequence for the atomic cloud preparation can be found in Ref.~\cite{Rudolph2020}.
After the atoms are released from the final MOT, we turn off the magnetic quadrupole field and apply a constant bias field with an amplitude between $4$ and $19~\text{G}$.
Then, a three-photon laser pulse of variable duration and detuning is applied to the atoms.
The atomic cloud has an rms radius of $130(5)~\micro\text{m}$ at the time of the pulse.
The final atomic state is characterized using a state-selective imaging scheme described below.

The three-photon pulses consist of light from three external cavity diode lasers (ECDLs) at $689$, $688$, and $679~\text{nm}$.
Each ECDL is locked to an optical frequency comb (Menlo Systems FC1500-ULN) that is stabilized to an optical cavity (Menlo Systems 1550 ORS).
From the sum of the residual error signals of these locks, we infer the rms frequency noise of $\Delta_3/2\pi$ to be $11~\text{kHz}$ within the three-photon pulse bandwidth.
The repetition rate of the comb is calibrated using Rabi spectroscopy on the \SrRed{} transition.
The relative polarizations of the lasers are set by combining the beams on a polarizing beam splitter.
The combined output is sent through an acousto-optic modulator and the diffracted order is coupled into a polarization-maintaining optical fiber. 
The final three-photon beam at the location of the atoms has a $1/e^2$ radial waist of $392.5(3)~\micro\text{m}$.
The average optical power in the desired linear polarizations are $0.79$, $6.9$, and $2.2~\text{mW}$ at $689$, $688$, and $679~\text{nm}$, respectively.

The $679~\text{nm}$ laser detuning was held fixed for all experiments at $-2536~\text{MHz}$ with respect to the $\threePzero{}\!-\!\threeSone{}$ resonance, while the $689~\text{nm}$ detuning $\Delta_1/2\pi$ was varied between $4$ and $43~\text{MHz}$.
In each case, the $688~\text{nm}$ detuning was adjusted to achieve three-photon resonance ($\Delta_{\text{eff}}=0$) for the given value of $\Delta_1$, with a range of $-2556\pm22~\text{MHz}$ with respect to the $\threePone{}\!-\!\threeSone{}$ resonance.

\begin{figure}[t]
    \centering
    \includegraphics[width=0.9\columnwidth]{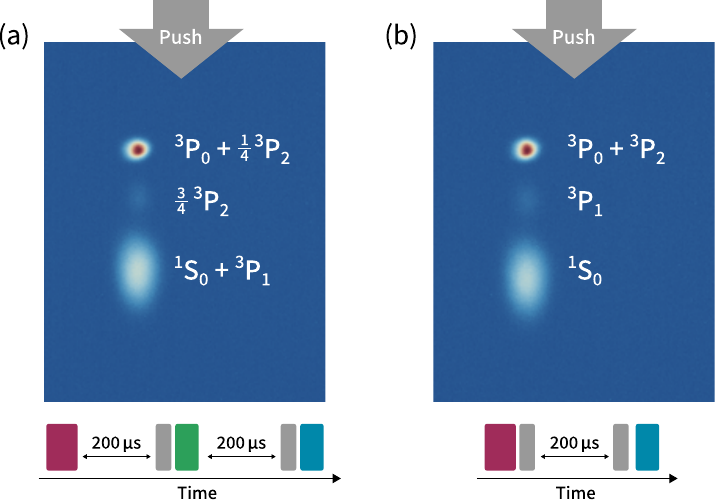}
    \caption{State-selective imaging schemes. Each scheme produces three spatially separated clouds through the successive application of push pulses, as indicated in the pulse sequences below. Here, red indicates a three-photon pulse and gray a push pulse, while green ($707~\text{nm}$) and blue ($679$ + $707~\text{nm}$) are repump pulses.\linebreak[4] (a) The first imaging scheme is designed to determine the \threePzero{} and \threePtwo{} populations, while the \oneSzero{} and \threePone{} populations cannot be measured independently.
    (b) The second scheme measures the \threePone{} and \oneSzero{} populations.}
\label{Fig:Imaging}
\end{figure}

\subsection{State-selective imaging}\label{Sec:Imaging}
To characterize the atomic state after a three-photon pulse, we spatially separate the atoms into three clouds using a sequence of push and repump pulses (see \figref{Fig:Imaging}).
Push pulses consist of $461~\text{nm}$ light resonant with the \SrBlue{} transition, which imparts momentum to atoms in the ground state \oneSzero{}. 
The first push pulse occurs $200~\micro\text{s}$ after the end of the three-photon pulse.
This time is chosen to give atoms in the \threePone{} state sufficient time to decay to the ground state.
To optically pump atoms out of \threePtwo{}, we apply a repump pulse resonant with the $\threePtwo{}\!-\!\threeSone{}$ transition at $707~\text{nm}$.
Because of the $3\!:\!1$ branching ratio of this process, most of the atoms return to the ground state, while the rest are shelved in \threePzero{}.
After an additional $200~\micro\text{s}$, a second push pulse imparts momentum to the fraction of ground state atoms that were pumped out of \threePtwo{} and adds additional momentum to the first pushed cloud.
Finally, repump light at both $707~\text{nm}$ and $679~\text{nm}$ optically pumps all remaining atoms back into the ground state. 
We wait $4.5~\text{ms}$ for the three clouds to spatially separate and then count the number of atoms in each cloud using fluorescence imaging on the $461~\text{nm}$ transition.
Because of the $3\!:\!1$ branching ratio, the true \threePtwo{} population is $4/3$ of the raw atom number in the singly-pushed cloud.
Conversely, the true \threePzero{} population is obtained by subtracting $1/3$ of the atoms in the singly-pushed cloud from the nominal atom number in the unpushed cloud.

The \threePone{} population is obtained using a modified version of the above imaging sequence (see \figref[b]{Fig:Imaging}).
A push pulse immediately following the three-photon pulse separates unexcited ground state atoms from those anywhere in the \threePJ{} manifold.
After a wait time of $200~\micro\text{s}$, a second push pulse imparts momentum to ground state atoms that have decayed from \threePone{}.
Finally, a repump pulse at both $707$ and $679~\text{nm}$ optically pumps the combined populations of \threePzero{} and \threePtwo{} back into the ground state.
The atoms are then imaged after a $4.5~\text{ms}$ wait time as in the other scheme.

We count the populations using appropriate bins centered around each cloud pictured in \figref{Fig:Imaging}.
While the clouds are well separated, residual populations in neighboring bins can occur.
Additionally, intensity inhomogeneity of the imaging beams causes a nonuniform scattering rate, leading to reduced counts, particularly for atoms in the \oneSzero{} bin.
Small correction factors (approximately $10\%$) are applied to the totals to minimize the covariance between the total atom number and the Rabi oscillations.

\subsection{Systematic errors}
An important caveat to the \threePone{} imaging sequence illustrated in \figref[b]{Fig:Imaging} is that it undercounts the loss due to unwanted excitation, since atoms in this state can decay back into the ground state during the three-photon pulse.
An estimate for the total loss due to off-resonant scattering from \threePone{} is obtained from the number of scattering events $N$ over the pulse duration $t_{\text{p}}$, given by
\begin{equation*}
    N=\frac{1}{\tau}\int_0^{t_{\text{p}}}P_{e}(t)\, dt \,\approx\, \frac{t_{\text{p}}}{\tau} \left<P_e\right>.
\end{equation*}
Here, $\tau$ is the lifetime and $P_e(t)$ is the state population in \threePone{}, with time-averaged fraction $\left<P_e\right>$.
For the data in \figref[b]{Fig:LineRabi}, we estimate $\left<P_e\right>=1.3(1)\%$ and $N=5.1(5)\%$ for the total loss from this channel over the full $85~\micro\text{s}$.

We measure a laser polarization impurity of $2\%$, which can lead to parasitic excitation to the $\Ket{\threePone{},m=0}$ state if the $689~\text{nm}$ detuning is near resonance ($\Delta_1 = 0$).
We maintain a detuning of over $500$ natural linewidths to avoid competing single-photon processes. 

We observe transient magnetic field oscillations when turning on the bias magnetic field.
We wait $4.5~\text{ms}$ for the field to settle before applying three-photon pulses.
The magnetic field amplitudes $B$ are obtained using Rabi spectroscopy of the \threePone{} Zeeman splitting, which is given by $\delta\omega_B = g_J\mu_\text{B} B/\hbar$, where $g_J=3/2$ is the Land\'{e} $g$ factor and $\mu_\text{B}$ is the Bohr magneton.
We measure the Zeeman splittings at the start and $100~\micro\text{s}$ after the three-photon pulse and observe a time variation of less than $1\%$.

\subsection{Calculating the three-photon coupling}

\begin{table*}[t]
\begingroup
\renewcommand{\arraystretch}{1.2}
\setlength{\tabcolsep}{4pt}
\centering
\small
\caption{Parameters for the relevant atomic transitions in $^{88}\text{Sr}$ used to calculate the expected single-photon Rabi frequencies $\Omega_{i}$.}\vspace{10pt}
\begin{tabular}{l c c c c c c} 
 \toprule
 Transition $\ket{j}\leftrightarrow \ket{k}$ & Wavelength & $\omega_{i}/2\pi~\text{(THz)}$ & $\eta_{jk}$ & $\tau_{k}~\text{(s)}$ & $P_{i}~\text{(mW)}$ & $\Omega_i/2\pi~\text{(MHz)}$ \\ 
 \midrule
 $\ket{\oneSzero{}, m=0}\hspace{7pt}\leftrightarrow\;\ket{\threePone{}, m=\pm1}$ & $689~\text{nm}$ & 434.829\,121\,311(010)~\cite{Ferrari2003} & 1 & $21.28(3)\times10^{-6}$~\cite{Nicholson2015} & 0.79 & 1.239(1)\\
 $\ket{\threePone{},m=\pm1}\leftrightarrow\;\ket{\threeSone{}, m=0}$ & $688~\text{nm}$ & 435.731\,697\,200(500)~\cite{Courtillot2005} & 1/6 & \multirow{2}{*}{$15.0(8)\times10^{-9}$~\cite{Jonsson1984}} & 6.9 & 56.0(4)\\
 $\ket{\threePzero{}, m=0}\hspace{7pt}\leftrightarrow\;\ket{\threeSone{},m=0}$ & $679~\text{nm}$ & 441.332\,751\,300(700)~\cite{Courtillot2005} & 1/9 &  & 2.2 & 35.8(3)\\
 \bottomrule
\end{tabular}
\label{Tab:RabiFrequencyParams}
\endgroup
\end{table*}

The expected three-photon Rabi frequencies are calculated using \Cref{eq:DestInterference} and the three single-photon couplings
\begin{equation}
    \Omega_i = \sqrt{\frac{6\pi c^2 \,\Gamma_{jk} \, I_i}{\hbar\, \omega_i^3}}, \quad\text{with} \;\; I_i=\frac{P_{i}}{\tfrac{\pi}{2}\,w_0^2}.
\end{equation}
Here, $I_i$ are the peak intensities of the Gaussian beams with $1/e^2$ radial waist $w_0$, and $P_{i}$ is the optical power in the spherical basis component that connects the specific $m$ states [\figref[b]{Fig:LevelDiagrams}].
The decay rate $\Gamma_{jk}$ between states $\ket{k}$ and $\ket{j}$ is related to the lifetime of the excited state $\tau_{k}$ by 
\begin{align}\label{Eq:DecayRates}
    \Gamma_{jk}&=\eta_{jk}\frac{1}{\tau_{k}},
\end{align}
where $\eta_{jk}$ are the branching ratios, which can be calculated using Wigner 3-$j$ and 6-$j$ symbols~\cite{Metcalf1999}.
\Cref{Tab:RabiFrequencyParams} summarizes the parameters used to calculate the single-photon couplings, including the experimentally determined transition frequencies and lifetimes.

To calibrate the laser intensities, we drive resonant Rabi oscillations between $\Ket{\oneSzero{}}$ and $\Ket{\threePone{}, m=0}$ with a measured optical power at $689~\text{nm}$.
The peak Rabi frequency from this measurement serves as an \textit{in situ} measurement of the effective beam size, with a $1/e^2$ radial waist of $w_0 = 392.5(3)~\micro\text{m}$.
This effective waist is then approximately common to all three wavelengths, since they are delivered via the same optical fiber and collimation optics.

The detuning of each laser from its respective atomic resonance is inferred using the frequency comb, which we calibrate with the observed $689~\text{nm}$ resonance frequency in conjunction with its best experimental estimate~\cite{Ferrari2003}.
The uncertainties in $\Delta_1$ and $\Delta_2$ used to calculate $\Omega_{\text{eff}}$, $\Omega_0$, and $\Omega_B$ reflect the combined uncertainties from our frequency comb measurements and the uncertainties in the literature transition frequencies~\cite{Ferrari2003, Courtillot2005}. 
Our uncertainty in the cumulative laser detuning $\Delta_3$ is lower than the reported uncertainties for the $688$ and $679~\text{nm}$ transition frequencies. We take advantage of the fact that their difference is equal to the energy splitting between the \threePone{} and \threePzero{} states, which is known more precisely~\cite{Ferrari2003, Morzynski2015}.
Our uncertainty in the effective three-photon detuning $\Delta_{\text{eff}}$, which corresponds to the width of the shaded region in \figref[a]{Fig:LineRabi}, reflects the combined uncertainties of $\Delta_3$ and $\Omega_\text{ac}$.

\subsection{The ac-Stark shift}\label{Sec:acStark}
The three-photon transition experiences a net ac-Stark shift given by Eqs.\,(\ref{Eq:OmegaAC0}) and (\ref{Eq:OmegaAC3}). To lowest order, the shift is
\begin{align}
    \Omega_{\text{ac}}&=-\Omega^{\text{ac}}_{0}+\Omega^{\text{ac}}_{3} \nonumber\\
    &= -\frac{|\Omega_1|^2}{2}\frac{\Delta_1}{\Delta_1^2-\delta\omega_B^2}+\frac{|\Omega_3|^2}{4\Delta_2}.\label{Eq:OmegaAC}
\end{align}
The higher-order terms in Eqs.\,(\ref{Eq:OmegaAC0}) and (\ref{Eq:OmegaAC3}) are small compared to the lowest-order shifts and constitute a percent-level correction which we neglect here. In terms of the individual laser detunings $\delta_1$ and $\delta_2$ defined below [\Cref{eq:delta1def,eq:delta2def,eq:delta3def}], \Cref{Eq:OmegaAC} is
\begin{equation}
    \Omega_{\text{ac}}=-\frac{|\Omega_1|^2}{2}\frac{\delta_1}{\delta_1^2-\delta\omega_B^2}+\frac{|\Omega_3|^2}{4(\delta_1+\delta_2)}.\label{Eq:acStark}
\end{equation}
Assuming fixed laser intensities, this shift vanishes for an appropriate choice of the 688~nm laser detuning, labeled $\delta_2^{\ast}$:
\begin{equation}
    \delta_2^{\ast}=\delta_1\left[\frac{\Omega_3^2}{2\Omega_1^2}\left(1-\beta^2\right)-1\right],\label{Eq:StarkShiftFreeDetuning}
\end{equation}
where, as usual, $\beta=\delta\omega_B/\delta_1$. Since the $679~\text{nm}$ transition is 3 orders of magnitude stronger than the $689~\text{nm}$ transition, the ratio $\Omega_3^2/\Omega_1^2$ yields a large $\delta_2^{\ast}$ for laser powers typically available in most experiments. Consequently, the ac-Stark shift-free condition can be fulfilled while maintaining a large detuning from \threeSone{}. 

\section{DENSITY MATRIX SIMULATIONS}\label{Sec:DensityMatrix}

\subsection{Ensemble average}
We simulate the three-photon dynamics by numerically solving the Lindblad master equation for all 13 magnetic sublevels of \oneSzero{}, \threeSone{}, and \threePJ{}, including all possible dipole-allowed transitions between the simulated states and excited state decay via spontaneous emission.
We include the effects of intensity inhomogeneity and other broadening mechanisms by averaging the density matrix over variations in the single-photon Rabi frequencies and atom-laser detunings.

We denote $r$ as the atom's radial position with respect to the peak intensity of the beam and $\Delta$ as the detuning between the laser frequency and the atom's resonance frequency at rest.  We take the normalized probability distributions in both coordinates to be Gaussian,
\begin{align*}
        n(r)&=\frac{1}{2\pi \sigma_r^2}\,e^{-r^2/2\sigma_r^2},\\
        \mu(\Delta)&=\frac{1}{\sqrt{2\pi}\sigma_\Delta}\,e^{-\Delta^2/2\sigma_\Delta^2},
\end{align*}
where $\sigma_r$ and $\sigma_\Delta$ are the widths of the respective distributions.
We label the density matrix of an atom at coordinates $(r,\Delta)$ as $\rho(t;r,\Delta)$.
The weighted average density matrix is then given by
\begin{equation*}
    \left<\rho(t) \right>=\int_{-\infty}^{\infty}\int_0^{\infty}\rho(t;r,\Delta)\,n(r)\, \mu(\Delta)\,2\pi r\, dr \, d\Delta.
\end{equation*}

\begin{table}[t]
\centering
\small\vspace{-12pt}
    \caption{Damping rates of the three-photon Rabi oscillation in the averaged density matrix model [see \figref[b]{Fig:LineRabi}]. Damping mechanisms include off-resonant scattering (OS) from intermediate states in the three-photon process, as well as other technical contributions from laser frequency noise (LN), Doppler shifts (DS), and intensity inhomogeneity (IN).}\label{Tab:Damping}
    \begin{tabular}{lcc}
        \toprule
        Effects included & Damping rate~$(\text{ms}^{-1})$ & \; Curve in \cref{Fig:dampedOscillations}\\
        \midrule
        OS (three-photon) & $\phantom{1}1.2\pm0.1$ & Purple\\\addlinespace[0.1em]
        OS + LN & $\phantom{1}4.3\pm0.5$ & Orange \\
        OS + LN + DS & $12.6\pm0.3$ & Green\\
        OS + LN + DS + IN & $36\pm2$ & Blue\\
        \bottomrule
     \end{tabular}
\end{table}

\begin{figure}[t]
    \centering
        \includegraphics[width=0.48\textwidth]{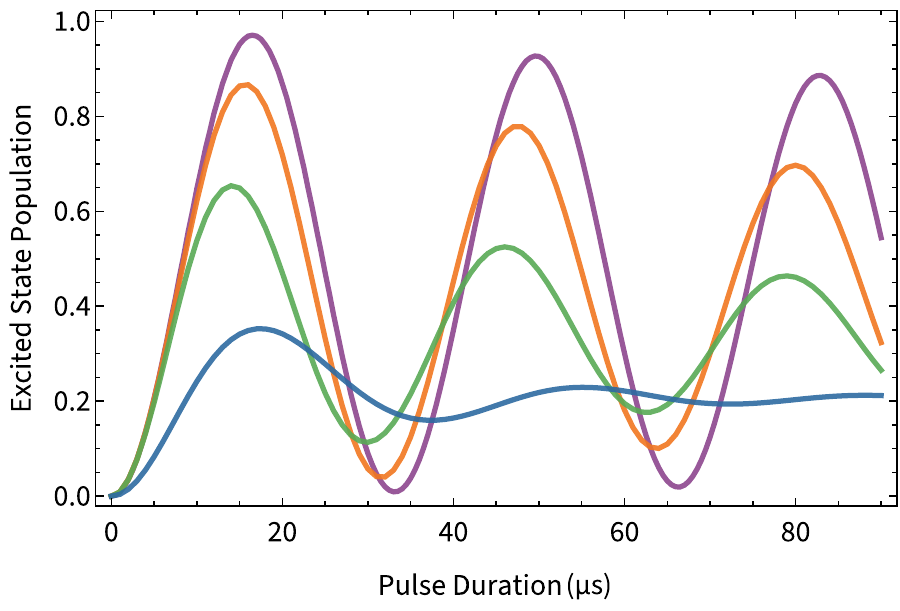}
    \caption{Simulated Rabi oscillations in the presence of the damping mechanisms listed in \Cref{Tab:Damping}. The oscillation with all effects included (blue) corresponds to the mean density matrix prediction in \figref[b]{Fig:LineRabi}. The peak transfer efficiencies are considerably higher without intensity inhomogeneity (green), and without intensity inhomogeneity or Doppler shifts (orange).
    The purple curve only includes damping from off-resonant scattering inherent in the three-photon process and features a peak transfer efficiency of $97\%$.}
    \label{Fig:dampedOscillations}
\end{figure}

We further assume that the lasers' intensity profiles are Gaussian, the single-photon Rabi frequencies vary in $r$ according to
\begin{align*}
        \Omega_i(r)&=\Omega_{i} \; e^{-r^2/w_0^2},
\end{align*}
where $\Omega_{i}$ is the peak single-photon Rabi frequency, and $w_0$ is the laser beam waist.
For an atom with velocity $v$ along the light propagation direction, the three lasers have detunings given by 
\begin{align*}
    \delta_i &=\delta_{i,0}+k_i\, v+\delta\omega_i,
\end{align*}
where $\delta_{i,0}$ is the nominal single-photon detuning for the $i^{\,\text{th}}$ laser, $k_i\, v$ is the Doppler shift for laser wave number $k_i$, and $\delta\omega_i$ is a possible offset caused by laser frequency noise. 
For simplicity, we assume that the laser frequency is constant but that $\delta\omega_i$ can vary shot to shot. 
The primary effect of laser noise and the Doppler shift is to detune the three-photon resonance, resulting in $\Delta_3=(\delta_{679,0}-\delta_{688,0}-\delta_{689,0}) -\Delta$, where $\Delta\equiv (k_{689}+k_{688}-k_{679})v + (\delta\omega_{689}+\delta\omega_{688}-\delta\omega_{679})$ is the cumulative Doppler shift and laser noise.
Furthermore, if we ignore independent effects of the different $\delta_i$ (e.g., single-beam ac-Stark shifts), then we can model the cumulative broadening of the three-photon transition by adding $\Delta$ to any one of the three laser frequencies.
In the simulation, we add the Doppler shift and noise to the 688~\text{nm} laser detuning, so that $\delta_{688} =\delta_{688,0}+\Delta$ and we set $\delta_i =\delta_{i,0}$ for the other lasers.

In \Cref{Fig:dampedOscillations}, we show the impact of each of the broadening effects included in the model used in \figref[b]{Fig:LineRabi}.
To quantify the contribution of each effect to the observed damping, 
in \Cref{Tab:Damping} we report the simulated damping rates for different combinations of broadening effects.
In the absence of laser frequency noise, temperature-dependent Doppler shifts, and intensity inhomogeneities present in our experiment, the maximum state transfer fidelity of $97\%$ is limited by the off-resonant scattering loss set by the laser parameters and magnetic field strength.
The fidelity can therefore be improved substantially by appropriate choice of laser parameters in an optimized experiment.

\subsection{Minimizing off-resonant scattering losses}\label{Sec:Fidelity}
In the ideal case where inhomogeneities are small, the transfer efficiency of a three-photon pulse is ultimately limited by off-resonant scattering from \threePone~and \threeSone.
Since population in these states is primarily due to direct single-photon excitation, this loss is proportional to the $689~\text{and}~679~\text{nm}$ laser intensities.  To first order, the $688~\text{nm}$ laser intensity does not contribute to scattering, and it only begins to contribute at second order via two-photon Raman scattering.
This observation can be exploited to reduce off-resonant scattering loss at a given three-photon coupling strength by appropriate allocation of intensities (i.e., using higher intensity for $688~\text{nm}$).

The instantaneous scattering rates for the two single-photon processes can be approximately modeled as
\begin{align}
\begin{split}\label{Eq:Rscat}
    R(\threePone{}) &=\frac{1}{\tau_1}P_1(t)P_0(t),\\
    R(\threeSone{}) &=\frac{1}{\tau_2}P_2(t)P_3(t),
\end{split}
\end{align}
where $P_i(t)$ are the populations and $\tau_k$ the lifetimes of states $i=\{0,1,2,3\}=\{\oneSzero{},\threePone{},\threeSone{},\threePzero{}\}$.
Here the initial and final states of the three-photon process, $P_0(t)$ and $P_3(t)$, act as weighting factors that set the fraction of atoms that can participate in the single-photon processes driving $P_1(t)$ and $P_2(t)$ at any given time.
When the scattering rate is small, the steady-state populations of $\threePone{}$ and $\threeSone{}$ may be estimated from the two-level optical Bloch equations:
\begin{align}\label{Eq:SteadyStatePops}
    P_{i}=\frac{(\Omega_i/\Gamma_i)^2}{1+2(\Omega_i/\Gamma_i)^2+(2\delta_i/\Gamma_i)^2}\approx\left(\frac{\Omega_i}{2\delta_i}\right)^2 \quad \text{for}~i=1,2,
\end{align}
where in this limit the single-photon detunings $\delta_i$ are much larger than $\Omega_i$ and $\Gamma_{jk}$.
Assuming a resonant three-photon excitation with minimal damping, 
\begin{align}\label{Eq:ClockStatePops}
\begin{split}
    P_3(t)&=\sin^2\frac{\Omega_\text{eff}\,t}{2},\\
    P_0(t)&=1-P_3(t).
\end{split}
\end{align}
These can be inserted into \Cref{Eq:Rscat} and integrated up to $t=t_\pi\equiv\pi/\Omega_\text{eff}$ to give the estimated infidelity of a $\pi$ pulse
\begin{align}\label{Eq:Nscat}
    N_\text{scat}\approx\frac{\pi}{8\,\Omega_\text{eff}}\left[\frac{1}{\tau_2}\frac{\Omega_3^2}{\delta_3^2}+\frac{1}{\tau_1}\sum_{m=\pm1}\frac{\Omega_1^2}{(\delta_{1}+m\,\delta\omega_B)^2}\right]. 
\end{align}
The assumption that the single-photon dynamics reach steady state within $t_\pi$ is well justified for the $\threePzero\!-\!\threeSone$ transition, since the lifetime of the upper state $\tau_2\ll t_\pi$ and since $\delta_3$ is several orders of magnitude larger than both $\Omega_3$ and $\Gamma_3$. 
This steady-state approximation is less accurate for the narrow $\oneSzero\!-\!\threePone$ transition.
As a result, \Cref{Eq:Nscat} overestimates the loss due to scattering from \threePone, but is nevertheless useful for calculating initial guesses for laser parameters which can be further optimized to yield high-fidelity Rabi oscillations in the density matrix simulation.

\begin{figure}[t]
    \centering
        \includegraphics[width=0.48\textwidth]{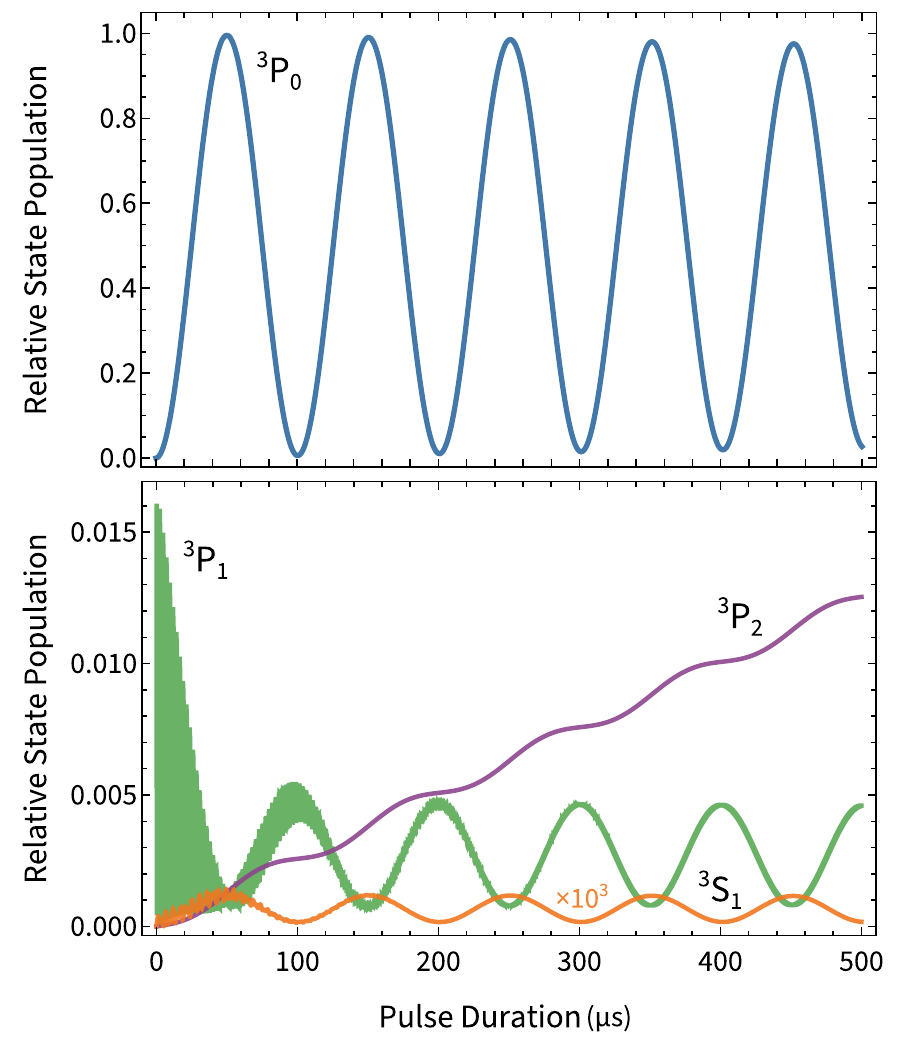}
    \caption{Predicted populations in the relevant excited states from optimized density matrix simulations that minimize off-resonant scattering (parameters from third row in \Cref{Tab:HighFidelityParams}).  (a) A three-photon Rabi oscillation with a frequency of $10~\text{kHz}$ and peak transfer to the target state \threePzero{} (blue) of $99.5\%$. (b) Residual populations in \threePone{} (green), \threePtwo{} (purple), and \threeSone{} (orange). The small transient population in \threeSone{} is scaled up ($\times 10^3$) for visibility.}
    \label{Fig:HighFidelitySim}
\end{figure}

\Cref{Tab:HighFidelityParams} details a few examples of parameters found for three-photon Rabi oscillations with reduced off-resonant scattering optimized using \Cref{Eq:Nscat}.  All examples maintain a high Rabi frequency ($\Omega_{\text{eff}}=2\pi\times10~\text{kHz}$) and use modest laser intensities ($1-10~\text{W}/\text{cm}^2$) and magnetic field strengths ($<5~\text{G}$).
\Cref{Fig:HighFidelitySim} shows the result of a full density matrix simulation using one of these parameter sets, validating the loss model.
These examples suggest that it is possible to achieve high-fidelity three-photon transitions between the \oneSzero{} and \threePzero{} clock states while maintaining high Rabi frequency. By comparison, to reach the same coupling strength for the single-photon transition in $^{87}$Sr would require about 10 times the total laser intensity assumed here.

Additionally, it is informative to compare the technical requirements for the three-photon approach to the magnetic-field-induced single-photon transition in bosons at $698~\text{nm}$~\cite{Taichenachev2006}.  To achieve the same coupling strength at the same magnetic field level ($5~\text{G}$) would require a $698~\text{nm}$ laser intensity of more than $10^7~\text{W}/\text{cm}^2$.  Alternatively, using the same total intensity assumed in \Cref{Tab:HighFidelityParams}, reaching a Rabi frequency of $10~\text{kHz}$ would require magnetic field strengths from $5,000$ to $11,000~\text{G}$ ($0.5$ to $1.1~\text{T}$).

\begin{table*}[t]
\begingroup
\setlength{\tabcolsep}{5pt}
\centering
\small
    \caption{Example parameters for three-photon Rabi oscillations with reduced off-resonant scattering.
    All parameter sets yield a Rabi frequency of $\Omega_{\text{eff}}=2\pi\times10~\text{kHz}$.
    By comparison, to achieve the same coupling strength in the single-photon \SrClock{} transition in $^{87}$Sr, a peak intensity of approximately $70~\text{W}/\text{cm}^2$ would be required.
    The magnitude of the laser detunings to the $m=\pm1$ magnetic sublevels of \threePone{} are given by $|\Delta_1\pm\delta\omega_B|$,  shown in the fifth column.
    For these simulations, we take $\Delta_1=0$.
    }\vspace{8pt}
    \begin{tabular}{c c c c c c c c c}
         \toprule
         $I_{689}$ & $I_{688}$ & $I_{679}$ & $I_{\text{Total}}$ & $|\Delta_1\pm\delta\omega_B|/2\pi$ & $\Delta_{2}/2\pi$ & $B$ & Peak transfer \\
         $(\text{W}/\text{cm}^2)$ & $(\text{W}/\text{cm}^2)$ & $(\text{W}/\text{cm}^2)$ & $(\text{W}/\text{cm}^2)$ & (MHz) & (GHz) & (G) & $(\%)$ \\
         \midrule
         0.13 & 1.75 & 0.09 & 1.97 & 6.3 & $-3.10$ & 3.0 & 99.0 \\
         0.13 & 3.57 & 0.27 & 3.97 & 7.4 & $-6.50$ & 3.5 & 99.2 \\
         0.13 & 8.98 & 0.45 & 9.55 & 8.4 & $-11.5$ & 4.0 & 99.5 \\
         \bottomrule
     \end{tabular}
    \label{Tab:HighFidelityParams}
\endgroup
\end{table*}

\section{FREQUENCY COMB CALIBRATION}

\subsection{Cumulative detuning \texorpdfstring{$\bm{\Delta_3}$}{Delta3}}\label{Sec:Detunings}

Using the definitions of the cumulative detunings \Crefrange{Eq:Cumulative1}{Eq:Cumulative3} with individual laser detunings $\delta_1 = \delta_{689}$, $\delta_2 = \delta_{688}$, and $\delta_3=\delta_{679}$, the total cumulative detuning $\Delta_3$ is
\begin{align}
\begin{split}
    \Delta_3
    &=\delta_{679}-(\delta_{689}+\delta_{688}) \label{eq:Delta3definition}
\end{split}
\end{align}
The uncertainties in the single-photon detunings are limited by the literature uncertainties in the respective transitions \cite{Courtillot2005, Ferrari2003}. However, we can accurately calculate $\Delta_3$ by taking advantage of the fact that the difference in the $679$ and $688~\text{nm}$ transition frequencies is equal to the difference in the $689$ and $698~\text{nm}$ transition frequencies, i.e.\ $f_{679}-f_{688}=f_{689}-f_{698}$. The latter two transitions are known to higher accuracy than the former two, which provides us with a significantly better estimate of the overall detuning. In terms of the frequency comb repetition rate $f_{\text{RR}}$ and tooth integers $n_i$, the difference in these detunings is
\begin{align}
\begin{split}
    \frac{\delta_{679}-\delta_{688}}{2\pi}
    &= \left(\delta f_{679}-\delta f_{688}\right)-\left(f_{689}-f_{698}\right)\nonumber\\ &+(n_{679}-n_{688})f_{\text{RR}}
\end{split}
\end{align}
Here, $\delta f_i$ is the beat note between the $i^{\text{th}}$ laser and the nearest comb tooth $n_i$, and $f_i$ is the (literature) resonance frequency of the $i^{\text{th}}$ transition. We then have
\begin{align}\label{Eq:Delta3Comb}
    \frac{\Delta_3}{2\pi} &=\left(\delta f_{679}-\delta f_{688}\right)-\frac{\delta_{689}}{2\pi}-\left(f_{689}-f_{698}\right) \nonumber\\ &+ (n_{679}-n_{688})f_{\text{RR}}
\end{align}
This expression for $\Delta_3$ is independent of the uncertainty in the $679$ and $688~\text{nm}$ transition frequencies. 
Note that we have assumed here that the $679$ and $688~\text{nm}$ lasers are are offset from resonance by the same number of comb teeth, as is the case in our experiment.

\subsection{Clock implementation}\label{Sec:Clock}

Here we briefly outline how the spectroscopic information obtained from observing the three-photon resonance could be combined with an optical frequency comb to realize a clock.  In particular, we show conceptually how to stabilize a frequency comb to the atomic resonance.  This process differs from a conventional optical atomic clock because there are now three different laser frequencies instead of one, and none of them are on resonance with the desired $\oneSzero{}\rightarrow\threePzero{}$ clock transition.  Consider the cumulative three-photon detuning defined previously:
\begin{align}
    \Delta_3 &= \delta_{3} - (\delta_{1} + \delta_{2}) \nonumber \\
    &= \big(\omega_3 - (\omega^a_{\ket{2,0}} - \omega^a_{\ket{3}})\big) \nonumber \\ & \quad - \Big(\omega_1 - (\omega^a_{\ket{1,0}} - \omega^a_{\ket{0}}) + \omega_2 - (\omega^a_{\ket{2,0}} - \omega^a_{\ket{1,0}}) \Big) \nonumber\\
    &=(\omega^a_{\ket{3}} - \omega^a_{\ket{0}}) - ((\omega_1 + \omega_2) - \omega_3)
\end{align}
where we used the definitions of the detunings from \Cref{eq:delta1def,eq:delta2def,eq:delta3def}.  In terms of the Sr transitions used in this work, the three laser frequencies are $\omega_1=\omega_{689}$, $\omega_2=\omega_{688}$, and $\omega_3=\omega_{679}$, and so we have
\begin{align}
    \Delta_3 &= \omega^a_{698} - \big((\omega_{689} + \omega_{688}) - \omega_{679}\big) \label{eq:Delta3clock}
\end{align}
where $\omega^a_{698} \equiv \omega^a_{\ket{3}} - \omega^a_{\ket{0}}$ is the atomic energy splitting of the $\oneSzero{}\rightarrow\threePzero{}$ clock transition at $698~\text{nm}$.  Thus, $\Delta_3$ measures the difference between the energies of the three lasers and the desired clock transition that we wish to lock them to.

Performing spectroscopy on the three-photon transition results in a measurement of the effective three-photon detuning $\Delta_\text{eff} = \Delta_3 -\Omega_0^\text{ac} + \Omega_3^\text{ac}$.  Assuming that the ac-Stark shift terms are appropriately stabilized, this measurement of $\Delta_\text{eff}$ then provides information that can be used to stabilize $\Delta_3$ by applying feedback to any of the three lasers.  As a concrete example, we could apply feedback to the 688~nm laser frequency $\omega_{688}$ to lock $\Delta_3 = 0$.  This would stabilize the linear combination of laser frequencies $\omega_{689} + \omega_{688} - \omega_{679}$, while leaving the other two linearly independent laser degrees of freedom unstabilized.

After stabilizing the lasers to the atom in this way, we would next stabilize the comb to the lasers in order to realize the clock.  
Consider the following beat note measurements between the three lasers and the nearest respective comb teeth:
\begin{align}
    \delta \omega_{689} &= \omega_{689} - 2\pi (n_{689} f_\text{RR} + f_\text{ceo})\\
    \delta \omega_{688} &= \omega_{688} - 2\pi (n_{688} f_\text{RR} + f_\text{ceo})\\
    \delta \omega_{679} &= \omega_{679} - 2\pi (n_{679} f_\text{RR} + f_\text{ceo})
\end{align}
where $n_{689}$, $n_{688}$, and $n_{679}$ are the integer comb tooth indices, $f_\text{RR}$ is the comb repetition rate, and $f_\text{ceo}$ is the carrier-envelope offset frequency.  By measuring each of these beat note frequencies, we construct the effective comb detuning $\Delta_\text{comb}$ defined as
\begin{align}
    \Delta_\text{comb} &\equiv \delta \omega_{689} + \delta \omega_{688} - \delta \omega_{679}
\end{align}
Note that $\Delta_\text{comb}$ should be thought of as a measured quantity that derives from the three beat note measurements.  Thus, the measured value of $\Delta_\text{comb}$ can be used to provide feedback to stabilize the comb.  Using the expressions above, we have
\begin{align}
    \Delta_\text{comb} &= \omega_{689} + \omega_{688} - \omega_{679} \nonumber\\& - 2\pi \Big((n_{689} + n_{688} - n_{679}) f_\text{RR} + f_\text{ceo}\Big)
\end{align}
Finally, using \Cref{eq:Delta3clock} we have
\begin{align}
    \Delta_\text{comb} &= \omega^a_{698} - 2\pi \Big((n_{689} + n_{688} - n_{679}) f_\text{RR} + f_\text{ceo}\Big) - \Delta_3 \nonumber
\end{align}
Assuming $\Delta_3 = 0$ because of the laser lock to the atom as described above, we see that $\Delta_\text{comb}$ depends only on the comb parameters $f_\text{RR}$ and $f_\text{ceo}$ and the known integers $n_{689}$, $n_{688}$, and $n_{679}$.  Furthermore, $(n_{689} + n_{688} - n_{679}) f_\text{RR} + f_\text{ceo}$ is indeed the frequency of the nearest comb tooth to the 698~nm clock transition, and so $\Delta_\text{comb}$ measures the detuning of this comb tooth from the atomic energy $\omega^a_{698}$.  Thus, the comb could be stabilized by applying feedback to $f_\text{RR}$ in order to lock $\Delta_\text{comb} = 0$.

\section{CALCULATION OF THE THREE-PHOTON COUPLING}\label{Sec:Derivation}

\subsection{Three-photon Rabi oscillations}
Here we show that the dynamics of the three-photon transition reduces to an effective coupling between the two clock states $\oneSzero{}$ and $\threePzero{}$.
We include all eight relevant atomic energy levels as well as the three laser fields.
We then adiabatically eliminate the intermediate states, yielding an effective two-level system that undergoes Rabi oscillations.

The Hamiltonian for the multilevel atom is
\begin{align}
    \op{H} &= \op{H}_a + \op{V}
\end{align}
where $\op{V} = - \v{\op{\mu}}\cdot\v{E}(t)$ is the electric dipole coupling with electric dipole operator $\v{\op{\mu}}$, and $\op{H}_a$ is the atomic Hamiltonian that describes the internal degrees of freedom of the atom.  The applied light electric field is
\begin{align}
    \v{E}(t) &= \sum_{\alpha=1}^3 \text{Re}{ \left[\v{\widetilde{E}}_\alpha e^{i \omega_\alpha t} \right] } = \frac{1}{2}\sum_{\alpha=1}^3 \left( \v{\widetilde{E}}_\alpha e^{i \omega_\alpha t} + \v{\widetilde{E}}^*_\alpha e^{-i \omega_\alpha t}\right) \nonumber
\end{align}
where the $\alpha$ sum is over the laser frequencies $\omega_\alpha$ used to drive the three-photon transition.  The components of the field amplitudes $\v{\widetilde{E}}_\alpha$ are, in general, complex to allow for arbitrary elliptical polarization.   The atomic Hamiltonian has solutions of the form $\ket{\Phi_i(t)} = \ket{i} e^{-i \omega^a_i t}$ which satisfy the Schr{\"o}dinger equation $i\hbar\partial_t \ket{\Phi_i(t)} = \op{H}_a\ket{\Phi_i(t)}$, where $\ket{i}$ are the energy eigenstates of the atom with associated energy levels $\hbar \omega^a_i$.

Switching to the interaction picture, we write the state as $\ket{\psi(t)} = \sum_i c_i(t) \ket{\Phi_i(t)}$
with time-dependent amplitudes $c_i(t)$.  Substituting this into the full Schr{\"o}dinger equation $i\hbar\,\partial_t \ket{\psi(t)} = \op{H}\ket{\psi(t)}$ then yields
\begin{align}
    \dot{c}_i &= \frac{1}{i\hbar} \sum_j c_j e^{i \omega_{ij}^a t} \bra{i} \big( \!-\!\v{\op{\mu}}\cdot  \v{E}(t) \big)  \ket{j} \nonumber \\
    \dot{c}_i &= \frac{1}{2i}\sum_{j,\alpha}\,  c_j \Big( \Omega_{ij}^{(\alpha)}  e^{i(\omega_\alpha + \omega_{ij}^a) t} + \Omega_{ij}^{(\alpha*)} e^{-i(\omega_\alpha - \omega_{ij}^a) t} \Big) \nonumber
\end{align}
where $\Omega_{ij}^{(\alpha)} \equiv \tfrac{1}{\hbar}\bra{i} (-\v{\op{\mu}}\cdot\v{\widetilde{E}}_\alpha) \ket{j}$ is the Rabi coupling between levels $\ket{i}$ and $\ket{j}$ of field $\v{\widetilde{E}}_\alpha$ and $\omega_{ij}^a \equiv \omega_{i}^a - \omega_{j}^a$ is their energy difference.  Similarly, $\Omega_{ij}^{(\alpha*)} \equiv \tfrac{1}{\hbar}\bra{i} (-\v{\op{\mu}}\cdot\v{\widetilde{E}}^*_\alpha) \ket{j}$ and we have $\left(\Omega_{ij}^{(\alpha)}\right)^* = \Omega_{ji}^{(\alpha*)}$.

To model the three-photon dynamics, we consider the following states of the Sr atom: $\ket{0}\equiv\ket{\oneSzero{}}$, $\ket{1,m_1}\equiv\ket{\threePone{},m_1}$, $\ket{2,m_2}\equiv\ket{\threeSone{},m_2}$, $\ket{3}\equiv\ket{\threePzero{}}$.
Here, $m_1$ and $m_2$ label the magnetic sublevels for states $\ket{1}$ and $\ket{2}$, respectively, and both range from $-1$ to $+1$.  This results in a set of eight coupled equations,
\begin{align}
    \dot{c}_0 &= \frac{1}{2i}\sum_{j,\alpha}\, c_j \Big( \Omega_{0j}^{(\alpha)} \, e^{i(\omega_\alpha + \omega_{0j}^a) t} + \Omega_{0j}^{(\alpha*)} \, e^{-i(\omega_\alpha - \omega_{0j}^a) t} \Big) \nonumber  \\
    \dot{c}_{1,m_1} &= \frac{1}{2i}\sum_{j,\alpha}\, c_j \Big( \Omega_{1m_1 j}^{(\alpha)} \, e^{i(\omega_\alpha + \omega_{1m_1 j}^a) t} \nonumber \\[-5pt] & \qquad \qquad \qquad + \Omega_{1m_1 j}^{(\alpha*)} \, e^{-i(\omega_\alpha - \omega_{1m_1 j}^a) t} \Big) \nonumber \\
    \dot{c}_{2,m_2} &= \frac{1}{2i}\sum_{j,\alpha}\, c_j \Big( \Omega_{2m_2 j}^{(\alpha)} \, e^{i(\omega_\alpha + \omega_{2m_2 j}^a) t} \nonumber \\[-5pt] & \qquad \qquad \qquad + \Omega_{2m_2 j}^{(\alpha*)} \, e^{-i(\omega_\alpha - \omega_{2m_2 j}^a) t} \Big) \nonumber \\
    \dot{c}_3 &= \frac{1}{2i}\sum_{j,\alpha}\, c_j \Big( \Omega_{3j}^{(\alpha)} \, e^{i(\omega_\alpha + \omega_{3j}^a) t} + \Omega_{3j}^{(\alpha*)} \, e^{-i(\omega_\alpha - \omega_{3j}^a) t}\Big) \nonumber
\end{align}
where we adopt the following subscript convention for the couplings and energy differences involving the $\ket{1,m_1}$ and $\ket{2,m_2}$ states: $\Omega_{1m_1 j}^{(\alpha)} = \tfrac{1}{\hbar}\bra{1,m_1} (-\v{\op{\mu}}\cdot\v{\widetilde{E}}_\alpha) \ket{j}$, $\Omega_{2m_2 j}^{(\alpha)} = \tfrac{1}{\hbar}\bra{2,m_2} (-\v{\op{\mu}}\cdot\v{\widetilde{E}}_\alpha) \ket{j}$, $\omega_{1m_1 j}^a = \omega_{\ket{1,m_1}}^a - \omega_{j}$, and $\omega_{2m_2 j}^a = \omega_{\ket{2,m_2}}^a - \omega_{j}$.
To simplify, we take advantage of the fact that each laser frequency interacts strongly only  with a limited number of states.  We consider only the following coupling terms:
\begin{align}
\begin{split}
    \Omega_{0m_1}^{(1)} &= \frac{1}{\hbar}\bra{0}(-\v{\mu}\cdot\v{\widetilde{E}}_1) \ket{1,m_1}\\
    \Omega_{1m_{1}2m_{2}}^{(2)} &= \frac{1}{\hbar}\bra{1,m_1}(-\v{\mu}\cdot\v{\widetilde{E}}_2) \ket{2,m_2}\\
    \Omega_{3m_2}^{(3)} &= \frac{1}{\hbar}\bra{3}(-\v{\mu}\cdot\v{\widetilde{E}}_3) \ket{2,m_2}   
\end{split} \nonumber
\end{align}
and we take all other $\Omega_{ij}^\alpha$ to be zero.  With this restriction, we simplify subscripts $1m_1\rightarrow m_1$ and $2m_2\rightarrow m_2$ whenever there is no ambiguity. After making the rotating wave approximation we find
\begin{align}
\begin{split}
    \dot{c}_0 &= \frac{1}{2i}\sum_{m_1}\, \Omega_{0m_1}^{(1)}  c_{1,m_1} e^{i(\omega_1 - \omega_{m_{1}0}^a) t}  \\
    \dot{c}_{1,m_1} &= \frac{1}{2i} \Omega_{m_{1}0}^{(1*)} c_0 e^{-i(\omega_1 - \omega_{m_{1}0}^a) t} \\ &+ \frac{1}{2i}\sum_{m_2}\, \Omega_{1m_{1}2m_2}^{(2)} c_{2,m_2} e^{i(\omega_2 - \omega_{2m_{2}1m_{1}}^a) t}   \\
    \dot{c}_{2,m_2} &= \frac{1}{2i} \Omega_{m_{2}3}^{(3*)} c_{3} e^{-i(\omega_3 - \omega_{m_{2}3}^a) t} \\ &+ \frac{1}{2i}\sum_{m_1}\, \Omega_{2m_{2}1m_{1}}^{(2*)} c_{1,m_1} e^{-i(\omega_2 - \omega_{2m_{2}1m_{1}}^a) t}   \\
    \dot{c}_3 &= \frac{1}{2i}\sum_{m_2}\, \Omega_{3m_{2}}^{(3)}  c_{2,m_2} e^{i(\omega_3 - \omega_{m_{2}3}^a) t}  
\end{split} \nonumber
\end{align}
where $\omega_{2m_{2}1m_{1}}^a = \omega^a_{\ket{2,m_2}} - \omega^a_{\ket{1,m_1}} > 0$.  We now define the detuning of each laser from the respective $m_1=0$ and $m_2=0$ states:
\begin{align}
    \delta_{1} &\equiv \omega_1 - (\omega^a_{\ket{1,0}} - \omega^a_{\ket{0}}) = \omega_1 - \omega_{00}^a \label{eq:delta1def}\\
    \delta_{2} &\equiv \omega_2 - (\omega^a_{\ket{2,0}} - \omega^a_{\ket{1,0}}) = \omega_2 - \omega_{2010}^a \label{eq:delta2def}\\
    \delta_3 &\equiv \omega_3 - (\omega^a_{\ket{2,0}} - \omega^a_{\ket{3}}) = \omega_3 - \omega_{03}^a\label{eq:delta3def}
\end{align}
In the experiment, the detuning $\delta_2$ must always be much greater than the Zeeman splittings of the $\ket{1,m_1}$ and $\ket{2,m_2}$ levels to avoid off-resonant scattering losses from $\ket{2,m_2}$.  As a result, we can neglect these Zeeman splittings to a good approximation and assume $(\omega_2 - \omega_{2m_{2}1m_{1}}^a) \approx \delta_2$.  Likewise, $\delta_3$ is also much larger than the Zeeman splitting of $\ket{2,m_2}$ so we have $(\omega_3 - \omega_{m_{2}3}^a)\approx \delta_3$.  In contrast, the Zeeman shift of $\ket{1,m_1}$ is comparable to $\delta_1$ and cannot be neglected:
\begin{align}
    \Delta_{1,m_1} &\equiv \omega_1 - \omega_{m_{1}0}^a = \delta_1 - m_1 \delta\omega_B
\end{align}
where $\delta\omega_B$ is the Zeeman splitting between $\ket{1,1}$ and $\ket{1,0}$,
\begin{align}
\begin{split}
    \dot{c}_0 &= \frac{1}{2i}\sum_{m_1}\, \Omega_{0m_1}^{(1)}  c_{1,m_1} e^{i \Delta_{1,m_1} t}  \\
    \dot{c}_{1,m_1} &= \frac{1}{2i} \Omega_{m_{1}0}^{(1*)} c_0 e^{-i \Delta_{1,m_1} t} + \frac{1}{2i}e^{i \delta_{2} t}\sum_{m_2}\, \Omega_{1m_{1}2m_2}^{(2)} c_{2,m_2}   \\
    \dot{c}_{2,m_2} &= \frac{1}{2i} \Omega_{m_{2}3}^{(3*)} c_{3} e^{-i \delta_3 t} + \frac{1}{2i} e^{-i \delta_{2} t}\sum_{m_1}\, \Omega_{2m_{2}1m_{1}}^{(2*)} c_{1,m_1}  \\
    \dot{c}_3 &= \frac{1}{2i}\sum_{m_2}\, \Omega_{3m_{2}}^{(3)}  c_{2,m_2} e^{i \delta_3 t}
\end{split} \nonumber
\end{align}
For convenience, we introduce the cumulative detunings from each of the three $m=0$ states,
\begin{align}
    \Delta_1 &\equiv \delta_1\label{Eq:Cumulative1}  \\
    \Delta_2 &\equiv \delta_1+\delta_2\label{Eq:Cumulative2}   \\
    \Delta_3 &\equiv \delta_3-(\delta_1+\delta_2)\label{Eq:Cumulative3}
\end{align}
Next, we eliminate the explicit time dependence by performing the following unitary transformation: 
\begin{align}
\begin{split}
    \widetilde{c}_0(t)&=c_0(t)\\
    \widetilde{c}_{1,m_1}(t)&=c_{1,m_1}(t) e^{i\Delta_{1,m_1} t}\\
    \widetilde{c}_{2,m_2}(t)&=c_{2,m_2}(t) e^{i \Delta_2 t}\\ 
    \widetilde{c}_3(t)&=c_3(t) e^{-i \Delta_3 t}
\end{split}
\end{align}
which results in
\begin{align}
\begin{split}
    \dot{\widetilde{c}}_0 &= \frac{1}{2i}\sum_{m_1}\, \Omega_{0m_1}^{(1)}  \widetilde{c}_{1,m_1} \\
    \dot{\widetilde{c}}_{1,m_1} &=i \Delta_{1,m_1}\widetilde{c}_{1,m_1} + \frac{1}{2i} \Omega_{m_{1}0}^{(1*)} \,\widetilde{c}_0  + \frac{1}{2i} \sum_{m_2}\, \Omega_{1m_{1}2m_2}^{(2)} \widetilde{c}_{2,m_2}   \\
    \dot{\widetilde{c}}_{2,m_2} &= i \Delta_2\,\widetilde{c}_{2,m_2} + \frac{1}{2i} \Omega_{m_{2}3}^{(3*)}\, \widetilde{c}_{3} + \frac{1}{2i} \sum_{m_1}\, \Omega_{2m_{2}1m_{1}}^{(2*)} \widetilde{c}_{1,m_1}   \\
    \dot{\widetilde{c}}_3 &= -i\Delta_3 \,\widetilde{c}_3 + \frac{1}{2i}\sum_{m_2}\, \Omega_{3m_{2}}^{(3)}  \widetilde{c}_{2,m_2}   
\end{split} \nonumber
\end{align}

Assuming that direct transitions to the intermediate states are sufficiently far detuned, we can now adiabatically eliminate the intermediate states since $|\dot{\widetilde{c}}_{1,m_1}|\ll|\Delta_{1,m_1}\widetilde{c}_{1,m_1}|$ and $|\dot{\widetilde{c}}_{2,m_2}|\ll|\Delta_{2}\,\widetilde{c}_{2,m_2}|$.  Taking $\dot{\widetilde{c}}_{1,m_1}\approx 0$ and $\dot{\widetilde{c}}_{2,m_2}\approx 0$ and solving gives
\begin{align}
    \widetilde{c}_{1,m_1} &\approx \frac{\Omega_{m_{1}0}^{(1*)}}{2\Delta_{1,m_1}}  \widetilde{c}_0  + \frac{1}{2\Delta_{1,m_1}} \sum_{m_2}\, \Omega_{1m_{1}2m_2}^{(2)} \widetilde{c}_{2,m_2} \label{eq:AdiabaticEliminationC2}  \\
    \widetilde{c}_{2,m_2} &\approx \frac{\Omega_{m_{2}3}^{(3*)}}{2\Delta_2}  \widetilde{c}_{3} + \frac{1}{2\Delta_2} \sum_{m_1}\, \Omega_{2m_{2}1m_{1}}^{(2*)} \widetilde{c}_{1,m_1} \label{eq:AdiabaticEliminationC3}
\end{align}
These equations are still coupled.  To solve for $\widetilde{c}_{1,m_1}$ and $\widetilde{c}_{2,m_1}$ explicitly, we first substitute each equation into the other to get a set of independent equations for $\ket{1,m_1}$ and $\ket{2,m_2}$:
\begin{align}
\begin{split}
    \widetilde{c}_{1,m_1} \approx & \frac{\Omega_{m_{1}0}^{(1*)}}{2\Delta_{1,m_1}}  \widetilde{c}_0  +   \sum_{m_2}\frac{\Omega_{1m_{1}2m_2}^{(2)}\Omega_{m_{2}3}^{(3*)}}{4\Delta_{1,m_1}\Delta_2}  \widetilde{c}_{3} \\ &+  \sum_{m_1'}\sum_{m_2}\, \frac{\Omega_{1m_{1}2m_2}^{(2)}  \Omega_{2m_{2}1m_{1}'}^{(2*)}}{4\Delta_{1,m_1}\Delta_2}  \widetilde{c}_{1,m_1'}   \\
    \widetilde{c}_{2,m_2} \approx & \frac{\Omega_{m_{2}3}^{(3*)}}{2\Delta_2}  \widetilde{c}_{3} +  \sum_{m_1}\frac{\Omega_{2m_{2}1m_{1}}^{(2*)}\Omega_{m_{1}0}^{(1*)}}{4\Delta_{1,m_1}\Delta_2}  \widetilde{c}_0  \\ &+ \sum_{m_2'}\sum_{m_1}\,\frac{\Omega_{2m_{2}1m_{1}}^{(2*)}\Omega_{1m_{1}2m_2'}^{(2)}}{4\Delta_{1,m_1}\Delta_2} \widetilde{c}_{2,m_2'}
\end{split}
\end{align}
A direct solution for $\widetilde{c}_{1,m_1}$ and $\widetilde{c}_{2,m_2}$ from these equations would be unwieldy, and we require only an approximate solution.  To make a controlled approximation, it is helpful to substitute in \Crefrange{eq:AdiabaticEliminationC2}{eq:AdiabaticEliminationC3} one more time, yielding
\begin{align}
    \widetilde{c}_{1,m_1} \approx & \frac{\Omega_{m_{1}0}^{(1*)}}{2\Delta_{1,m_1}}  \widetilde{c}_0  +   \sum_{m_2}\frac{\Omega_{1m_{1}2m_2}^{(2)}\Omega_{m_{2}3}^{(3*)}}{4\Delta_{1,m_1}\Delta_2}  \widetilde{c}_{3} \nonumber \\ & + \sum_{m_1',m_2}\! \frac{\Omega_{1m_{1}2m_2}^{(2)}  \Omega_{2m_{2}1m_{1}'}^{(2*)} \Omega_{m_{1}'0}^{(1*)}}{8\Delta_{1,m_1}\Delta_{1,m_1'}\Delta_2} \, \widetilde{c}_0 \nonumber \\
    & +\sum_{m_1',m_2',m_2}\!\! \frac{\Omega_{1m_{1}2m_2}^{(2)}  \Omega_{2m_{2}1m_{1}'}^{(2*)} \Omega_{1m_{1}'2m_2'}^{(2)}}{8\Delta_{1,m_1}\Delta_{1,m_1'}\Delta_2}   \widetilde{c}_{2,m_2'}  \nonumber \\
    \widetilde{c}_{2,m_2} \approx & \frac{\Omega_{m_{2}3}^{(3*)}}{2\Delta_2}  \widetilde{c}_{3} +  \sum_{m_1}\frac{\Omega_{2m_{2}1m_{1}}^{(2*)}\Omega_{m_{1}0}^{(1*)}}{4\Delta_{1,m_1}\Delta_2}  \widetilde{c}_0 \nonumber \\ & +  \sum_{m_2',m_1}\,\frac{\Omega_{2m_{2}1m_{1}}^{(2*)} \Omega_{1m_{1}2m_2'}^{(2)} \Omega_{m_{2}'3}^{(3*)}}{8\Delta_{1,m_1}\Delta_2^2} \, \widetilde{c}_{3} \nonumber \\
    & + \sum_{m_1',m_2',m_1}\,\frac{\Omega_{2m_{2}1m_{1}}^{(2*)}\Omega_{1m_{1}2m_2'}^{(2)} \Omega_{2m_{2}'1m_{1}'}^{(2*)} }{8\Delta_{1,m_1}\Delta_2^2} \widetilde{c}_{1,m_1'} \nonumber
\end{align}
We could continue to repeat this substitution process to generate additional terms. However, we see that this is indeed a decaying series, since $\frac{\Omega}{\delta} \ll 1$ for any coupling $\Omega$ and detuning $\delta$, and this small ratio appears with a higher power in each additional term.  As a result, we may safely drop the last term in the above expressions.
\begin{align}
    \widetilde{c}_{1,m_1} &\approx \frac{\Omega_{m_{1}0}^{(1*)}}{2\Delta_{1,m_1}}  \widetilde{c}_0  +   \sum_{m_2}\frac{\Omega_{1m_{1}2m_2}^{(2)}\Omega_{m_{2}3}^{(3*)}}{4\Delta_{1,m_1}\Delta_2}  \widetilde{c}_{3} \nonumber \\ & \quad + \sum_{m_1',m_2}\! \frac{\Omega_{1m_{1}2m_2}^{(2)}  \Omega_{2m_{2}1m_{1}'}^{(2*)} \Omega_{m_{1}'0}^{(1*)}}{8\Delta_{1,m_1}\Delta_{1,m_1'}\Delta_2} \, \widetilde{c}_0  \\
    \widetilde{c}_{2,m_2} &\approx \frac{\Omega_{m_{2}3}^{(3*)}}{2\Delta_2}  \widetilde{c}_{3} +  \sum_{m_1}\frac{\Omega_{2m_{2}1m_{1}}^{(2*)}\Omega_{m_{1}0}^{(1*)}}{4\Delta_{1,m_1}\Delta_2}  \widetilde{c}_0 \nonumber \\ & \quad +  \sum_{m_2',m_1}\,\frac{\Omega_{2m_{2}1m_{1}}^{(2*)} \Omega_{1m_{1}2m_2'}^{(2)} \Omega_{m_{2}'3}^{(3*)}}{8\Delta_{1,m_1}\Delta_2^2} \, \widetilde{c}_{3} 
\end{align}
We are ultimately interested in studying the effective two-level dynamics between levels $\ket{0}$ and $\ket{3}$ once the intermediate states have been adiabatically eliminated:
\begin{align}
    \dot{\widetilde{c}}_0 &= \frac{1}{2i}\sum_{m_1}\, \Omega_{0m_1}^{(1)}  \widetilde{c}_{1,m_1}  \\
    \dot{\widetilde{c}}_3 &= -i\Delta_3\,\widetilde{c}_3 + \frac{1}{2i}\sum_{m_2}\, \Omega_{3m_{2}}^{(3)}  \widetilde{c}_{2,m_2}   
\end{align}
Substituting the approximate solutions for $\widetilde{c}_{1,m_1}$ and $\widetilde{c}_{2,m_2}$,
\begin{align}
    \dot{\widetilde{c}}_0 &= -i\sum_{m_1}\, \frac{\Omega_{0m_1}^{(1)}\Omega_{m_{1}0}^{(1*)}}{4\Delta_{1,m_1}}  \widetilde{c}_0  \nonumber \\ & \quad  +   \frac{1}{2i}\sum_{m_1,m_2} \frac{\Omega_{0m_1}^{(1)}\Omega_{1m_{1}2m_2}^{(2)}\Omega_{m_{2}3}^{(3*)}}{4\Delta_{1,m_1}\Delta_2}  \widetilde{c}_{3} \nonumber \\
    & \quad - i \sum_{m_1,m_1',m_2}\! \frac{\Omega_{0m_1}^{(1)} \Omega_{1m_{1}2m_2}^{(2)}\Omega_{2m_{2}1m_{1}'}^{(2*)}\Omega_{m_{1}'0}^{(1*)}  }{16\Delta_{1,m_1}\Delta_{1,m_1'}\Delta_2} \, \widetilde{c}_0  \\
    \dot{\widetilde{c}}_3 &= -i\Delta_3\,\widetilde{c}_3 - i\sum_{m_2}\, \frac{\Omega_{3m_{2}}^{(3)}\Omega_{m_{2}3}^{(3*)}}{4\Delta_2}  \widetilde{c}_{3}  \nonumber \\ & \quad +  \frac{1}{2i}\sum_{m_1,m_2}\frac{\Omega_{3m_{2}}^{(3)}\Omega_{2m_{2}1m_{1}}^{(2*)}\Omega_{m_{1}0}^{(1*)}}{4\Delta_{1,m_1}\Delta_2}  \widetilde{c}_0 \nonumber \\
   & \quad - i\sum_{m_2,m_2',m_1}\!\!\frac{\Omega_{3m_{2}}^{(3)}\Omega_{2m_{2}1m_{1}}^{(2*)} \Omega_{1m_{1}2m_2'}^{(2)} \Omega_{m_{2}'3}^{(3*)}}{16\Delta_{1,m_1}\Delta_2^2} \, \widetilde{c}_{3} 
\end{align}\vspace{-2pt}
We now define the following:
\begin{align}
        \Omega_\text{eff} &\equiv \sum_{m_1,m_2}\frac{\Omega_{3m_{2}}^{(3)}\Omega_{2m_{2}1m_{1}}^{(2*)}\Omega_{m_{1}0}^{(1*)}}{4\Delta_{1,m_1}\Delta_2} 
\end{align}
\vspace{-10pt}
\begin{align}
    \Omega_0^\text{ac} &\equiv \sum_{m_1}\, \frac{|\Omega_{0m_1}^{(1)}|^2}{4\Delta_{1,m_1}}  + \!\!\!\!  \sum_{m_1,m_1',m_2}\!\!\!\!\!\frac{\Omega_{0m_1}^{(1)} \Omega_{1m_{1}2m_2}^{(2)}\Omega_{2m_{2}1m_{1}'}^{(2*)}\Omega_{m_{1}'0}^{(1*)} }{16\Delta_{1,m_1}\Delta_{1,m_1'}\Delta_2}\label{Eq:OmegaAC0}\\
    \Omega_3^\text{ac} &\equiv \sum_{m_2}\, \frac{|\Omega_{3m_{2}}^{(3)}|^2}{4\Delta_2} + \!\!\!\!\sum_{m_2,m_2',m_1}\!\!\!\!\!\frac{\Omega_{3m_{2}}^{(3)}\Omega_{2m_{2}1m_{1}}^{(2*)}\Omega_{1m_{1}2m_2'}^{(2)} \Omega_{m_{2}'3}^{(3*)}}{16\Delta_{1,m_1}\Delta_2^2}\label{Eq:OmegaAC3}
\end{align}
where $\Omega_\text{eff}$ is the effective three-photon Rabi frequency, and $\Omega_0^\text{ac}$ and $\Omega_3^\text{ac}$ are ac-Stark shifts of the states $\ket{0}$ and $\ket{3}$, respectively.  The effective two-level dynamics have the form
\begin{align}
    \dot{\widetilde{c}}_0 &= -i \, \Omega_0^\text{ac} \, \widetilde{c}_0  +   \frac{\Omega_\text{eff}^*}{2i}  \widetilde{c}_{3}  \\
    \dot{\widetilde{c}}_3 &= -i\left(\Delta_3+ \Omega_3^\text{ac}\right) \widetilde{c}_3  +  \frac{\Omega_\text{eff}}{2i} \,  \widetilde{c}_0 
\end{align}
It is helpful to perform another unitary transformation 
\begin{align}
    \widetilde{c}_0 &= \overline{c}_0 e^{-\frac{i}{2} (\Delta_3 + \Omega_3^\text{ac}+\Omega_0^\text{ac}) t}\\
    \widetilde{c}_3 &= \overline{c}_3 e^{-\frac{i}{2} (\Delta_3 + \Omega_3^\text{ac}+\Omega_0^\text{ac}) t}
\end{align}
which results in a symmetric set of equations
\begin{align}
    \partial_t\begin{pmatrix}
        \overline{c}_0 \\
        \overline{c}_3
    \end{pmatrix}
    &=
    \frac{1}{2i}\begin{pmatrix}
        -\Delta_\text{eff} & \Omega_\text{eff}^* \\
        \Omega_\text{eff} & \Delta_\text{eff} 
    \end{pmatrix}
    \begin{pmatrix}
        \overline{c}_0 \\
        \overline{c}_3
    \end{pmatrix}
\end{align}
where $\Delta_\text{eff} \equiv \Delta_3 + \Omega_{\text{ac}}$ is the three-photon detuning and $\Omega_{\text{ac}}\equiv-\Omega^{\text{ac}}_{0}+\Omega^{\text{ac}}_{3}$ is the net ac-Stark shift. This system behaves as an effective two-level atom that will Rabi oscillate between states $\ket{0}$ and $\ket{3}$ with resonant Rabi frequency $\Omega_\text{eff}$ and detuning $\Delta_\text{eff}$.

In terms of the dipole matrix elements, the effective three-photon Rabi coupling has the form
\begin{widetext}
\begin{align}
    \Omega_\text{eff} &= \sum_{m_1,m_2}\frac{\bra{3}\v{\mu}\cdot\v{\widetilde{E}}_3 \ket{2,m_2} \bra{2,m_2}\v{\mu}\cdot\v{\widetilde{E}}_2^* \ket{1,m_1}\bra{1,m_1}\v{\mu}\cdot\v{\widetilde{E}}_1^* \ket{0}}{4\Delta_{1,m_1}\Delta_2\hbar^3} 
\end{align}
or equivalently, in terms of the Sr energy levels
\begin{equation}
    \Omega_\text{eff} = \sum_{m_1,m_2}\frac{\bra{\threePzero{}}\v{\op{\mu}}\cdot\v{\widetilde{E}}_3 \ket{\threeSone{},m_2} \bra{\threeSone{},m_2}\v{\op{\mu}}\cdot\v{\widetilde{E}}_2^* \ket{\threePone{},m_1}\bra{\threePone{},m_1}\v{\op{\mu}}\cdot\v{\widetilde{E}}_1^* \ket{\oneSzero{}}}{4\Delta_{1,m_1}\Delta_2\hbar^3}\label{Eq:ThreePhotonCouplingSpherical}
\end{equation}
\end{widetext}

\subsection{Cartesian basis for J=1}\label{Sec:CartBasis}

When computing the three-photon coupling for linearly polarized light, it is convenient to express the angular momentum states using the Cartesian basis.  On the other hand, the energy level shifts from the magnetic field are most naturally described in the usual spherical basis. To help evaluate the dipole matrix elements for the intermediate states in \Cref{Eq:ThreePhotonCouplingSpherical}, we consider states with angular momentum $J=1$ and their representation in both the spherical and Cartesian bases.  Here, $\v{J}$ is a placeholder for any angular momentum, so the following applies to spin $\v{S}$ and orbital $\v{L}$, as well as for total angular momentum $\v{J}=\v{L}+\v{S}$.

In the spherical basis for $J=1$, the three states that form a complete basis are denoted as $\ket{+1} \equiv \ket{J=1, m=1}_z$, $\ket{-1} \equiv \ket{J=1, m=-1}_z$, and $\ket{0} \equiv \ket{J=1, m=0}_z$, where the subscript indicates that these spherical states are projections along the $\v{\hat{z}}$ axis. Although less common, it is also possible to use spherical basis states with respect to the $\v{\hat{x}}$ or $\v{\hat{y}}$ axes. We define a set of Cartesian basis states that are linear combinations of the $\v{\hat{z}}$ projection states:
\begin{align}
    \ket{J_x} &\equiv \ket{J=1, m=0}_x = \frac{-\ket{+1} + \ket{-1}}{\sqrt{2}} \label{Eq:Jx}\\
    \ket{J_y} &\equiv \ket{J=1, m=0}_y = \frac{\ket{+1} + \ket{-1}}{-i\sqrt{2}}\\
    \ket{J_z} &\equiv \ket{J=1, m=0}_z = \ket{0}\label{Eq:Jz}
\end{align}
As indicated, the $\ket{J_x}$ and $\ket{J_y}$ basis states correspond to zero projection of angular momentum along the $\v{\hat{x}}$ and $\v{\hat{y}}$ axes, analogous to $\ket{J=1, m=0}_z$ in the spherical basis. These $\ket{J_n}$ Cartesian states are symmetric with respect to the $\v{\hat{x}}$, $\v{\hat{y}}$, $\v{\hat{z}}$ axes and form a complete basis for the $J=1$ states that is complementary to the spherical basis:  $\sum_{n=1}^{3} \ket{J_n}\bra{J_n} = \sum_{m=-1}^{+1} \ket{m}\bra{m} = 1$.
The Cartesian states arise naturally when an atom is driven by linearly polarized light, since in this case an atom initially in an $L=0$ state $\ketJ{\ell}{0}{0}$ will transition to a $\ket{J_n}$ state, as we will see below. To transform between these two bases, we compute the amplitudes $\braket{m | J_n}$, which may be represented in matrix form as
\begin{align}
    U_{mn} &\equiv \braket{m | J_n} \doteq 
    \begin{pmatrix}
        \tfrac{-1}{\sqrt{2}} & \tfrac{i}{\sqrt{2}} & 0 \\
        \tfrac{1}{\sqrt{2}} & \tfrac{i}{\sqrt{2}} & 0  \\
        0 & 0 & 1
    \end{pmatrix}
\end{align}
where we use the following conventions for the state representations: $\bra{+1}\doteq (1,0,0)$, $\bra{-1}\doteq (0,1,0)$, $\bra{0}\doteq (0,0,1)$, for the $\ket{m}$ states and $\ket{J_x}\doteq (1,0,0)^\intercal$, $\ket{J_y}\doteq (0,1,0)^\intercal$, $\ket{J_z}\doteq (0,0,1)^\intercal$ for the Cartesian states.  Note that the unconventional ordering of the $\ket{0}$ and $\ket{-1}$ states in this matrix representation is chosen so that the $\ket{J_z}$ and $\ket{0}$ states agree.  Also, as expected, the basis transformation $U_{mn}$ is unitary with $U^\dagger U=1$.

We can now express the three-photon coupling $\Omega_\text{eff}$ in the Cartesian basis.  To begin, the detuning denominator $\Delta_{1,m_1}$ in \Cref{Eq:ThreePhotonCouplingSpherical} can be written as a diagonal matrix in the spherical basis:
\begin{widetext}
\begin{align}
    \Omega_\text{eff} &= \sum_{m_1,m_1',m_2}\frac{\bra{\threePzero{}}(\v{\op{\mu}}\cdot\v{\widetilde{E}}_3) \ket{\threeSone{},m_2} \bra{\threeSone{},m_2}(\v{\op{\mu}}\cdot\v{\widetilde{E}}_2^*)\ket{\threePone{},m_1'}\bra{m_1'}\op{T}\ket{m_1}\bra{\threePone{},m_1}(\v{\op{\mu}}\cdot\v{\widetilde{E}}_1^*) \ket{\oneSzero{}}}{4\Delta_2\hbar^3} \nonumber
\end{align}
\end{widetext}
where we define the matrix $\bra{m_1'}\op{T}\ket{m_1}$ as the inverse of the detunings of the $\threePone{}$ state.  Explicitly we have $T_{m_1',m_1}\equiv\bra{m_1'}\op{T}\ket{m_1}$ and
\begin{align}
    T_{m_1',m_1} &\doteq
    \begingroup
    \setlength\arraycolsep{1.5pt}
    \begin{pmatrix}
        \tfrac{1}{\Delta_{1,1}} & 0 & 0 \\
        0 & \tfrac{1}{\Delta_{1,-1}} & 0  \\
        0 & 0 & \tfrac{1}{\Delta_{1,0}}
    \end{pmatrix} = 
    \begin{pmatrix}
        \tfrac{1}{\Delta_1 - \delta\omega_B} & 0 & 0 \\
        0 & \tfrac{1}{\Delta_1 +\delta\omega_B} & 0  \\
        0 & 0 & \tfrac{1}{\Delta_1}
    \end{pmatrix} \nonumber
    \endgroup
\end{align}
where, as before, $\Delta_1$ is the detuning of $\v{E}_1$ from $\ket{\threePone{},m_1=0}$ while $\pm\delta\omega_B=\pm g_J\mu_B B/\hbar$ is the Zeeman detuning of the $m_1=\pm 1$ states, which is linearly proportional to the applied magnetic field. For \threePone{}, the Land\'e $g$ factor $g_J=3/2$.

Since the $\ket{J_n}$ form a complete basis, the sum over $m_2$ in $\Omega_\text{eff}$ may be written as $\sum_{m_2} \ket{\threeSone{},m_2} \bra{\threeSone{},m_2} = \sum_{l=1}^{3} \ket{\threeSone{},J_l} \bra{\threeSone{},J_l}$.
Similarly, we can express the $m_1$ and $m_1'$ sums in the Cartesian basis by inserting the identity
\begin{widetext}
\begin{align}
    \sum_{m_1,m_1'} \ket{\threePone{},m_1'}T_{m_1',m_1}\bra{\threePone{},m_1} &= \sum_{m_1,m_1'} \left(\sum_{j=1}^{3} \ket{J_j}\bra{J_j}\right)\ket{\threePone{},m_1'}T_{m_1',m_1}\bra{\threePone{},m_1}\left(\sum_{n=1}^{3} \ket{J_n}\bra{J_n}\right) \\ &= \sum_{j,n} \ket{\threePone{},J_j} D_{jn} \bra{\threePone{},J_n}
\end{align}
where $D_{jn} \equiv \sum_{m_1,m_1'} \braket{J_j | m_1'} T_{m_1',m_1}\braket{m_1 | J_n} = \sum_{m_1,m_1'} (U_{m_1',j})^\dagger T_{m_1',m_1} U_{m_1,n}$. 
Explicitly, we find
\begin{align}
    D_{jn}& \doteq 
    \begin{pmatrix}
        \dfrac{\strut \Delta_1}{\strut \Delta_1^2 - \delta\omega_B^2} & \dfrac{\strut -i\,\delta\omega_B}{\strut \Delta_1^2 - \delta\omega_B^2} & 0 \\
        \dfrac{\strut i\,\delta\omega_B}{\strut \Delta_1^2 - \delta\omega_B^2} & \dfrac{\strut \Delta_1}{\strut \Delta_1^2 - \delta\omega_B^2} & 0  \\
        0 & 0 & \dfrac{\strut 1}{\strut \Delta_1}
    \end{pmatrix}\label{Eq:DMatrix}
\end{align}
This matrix encodes the magnetic-field-induced mixing of the Cartesian orbital described in the main text.  In particular, the off-diagonal entries of this matrix are responsible for allowing the three-photon transition to proceed using collinear light, as further described below.  With these substitutions, the three-photon coupling becomes
\begin{align}
    \Omega_\text{eff} &= \sum_{l, j, n}\frac{\bra{\threePzero{}} \v{\op{\mu}}\cdot\v{\widetilde{E}}_3 \ket{\threeSone{},J_l} \bra{\threeSone{},J_l} \v{\op{\mu}}\cdot\v{\widetilde{E}}_2^* \ket{\threePone{},J_j}D_{jn}\bra{\threePone{},J_n} \v{\op{\mu}}\cdot\v{\widetilde{E}}_1^* \ket{\oneSzero{}}}{4\Delta_2\hbar^3} \label{Eq:ThreePhotonCouplingCartesian}
\end{align}
\end{widetext}

\subsection{Decomposition of the states into L and S}

Next we consider each of the four energy levels that enter into the three-photon coupling and express the states in terms of $\v{L}$ and $\v{S}$ angular momentum components.  

\subsubsection*{\texorpdfstring{$\bm{\oneSzero{}}$}{1S0}}

\noindent The ground state $\ket{\oneSzero{}}$ has $L=0$ and $S=0$, which we write as
\begin{align}
    \ket{\oneSzero{}}=\ket{|\oneSzero{}|}\ketJ{\ell}{0}{0}\ketJ{s}{0}{0}
\end{align}
where $\ket{|\oneSzero{}|}$ denotes the radial part of the \oneSzero{} state, i.e.\ the nonangular part. Here $\ketJ{J}{j}{m}$ is a spherical basis state with angular momentum $j$ and $z$ projection $m$, and the $J$ subscript indicates either orbital $J=\ell$ or spin $J=s$.  This notation will be used for the other states as well.

\subsubsection*{\texorpdfstring{$\bm{\threePone{}}$}{3P1}}

\noindent Because of the $LS$ coupling, the $\ket{\threePone{}}$ state is not an eigenstate of $\v{S}$, but consists of a mixture of two bare states (before $LS$ mixing), each with $J=1$.  In terms of the bare states:
\begin{align}\label{Eq:3P1Decomposition}
    \ket{\threePone{}} &= a\,\ket{\threePoneZero{}} + b\,\ket{\onePoneZero{}} 
\end{align}
where $a$ and $b$ are the amplitudes of the $\ket{\threePoneZero{}}$ (triplet) and $\ket{\onePoneZero{}}$ (singlet) bare-state contributions, respectively. First, $\ket{\onePoneZero{}}$ has $L=1$ and $S=0$, so in the Cartesian basis we have
\begin{align}
    \ket{\onePoneZero{}, J_n}  &=  \ket{\big|\onePoneZero{}\big|} \ket{L_n} \ket{0,0}_s \label{Eq:1P1LS}
\end{align}
Now consider the triplet component $\ket{\threePoneZero{},m}$.  In the spherical basis, we can use Clebsch-Gordan coefficients to express the angular momentum state in terms of $L=1$ and $S=1$ states and their projections along $\v{\hat{z}}$:
\begin{align}
    \ket{J=1,m=1}_z &= \frac{\ketLS{1}{1}{1}{0}-\ketLS{1}{0}{1}{1}}{\sqrt{2}} \label{eq:J1SsphericalLS}\\
    \ket{J=1,m=-1}_z &= \frac{-\ketLS{1}{-1}{1}{0}+\ketLS{1}{0}{1}{-1}}{\sqrt{2}} \\
    \ket{J=1,m=0}_z &= \frac{\ketLS{1}{1}{1}{-1}-\ketLS{1}{-1}{1}{1}}{\sqrt{2}}
\end{align}
Using \Crefrange{Eq:Jx}{Eq:Jz}, we can convert these states to the Cartesian basis, using $\ket{1,\pm 1}_{\ell} = \frac{1}{\sqrt{2}}(\mp\ket{L_x} - i \ket{L_y})$, $\ket{1,0}_{\ell} = \ket{L_z}$ and 
$\ket{1,\pm 1}_s = \frac{1}{\sqrt{2}}(\mp\ket{S_x} - i \ket{S_y})$, $\ket{1,0}_s = \ket{S_z}$. Thus the $J=1$ state for the triplet component can be written as
\begin{align}
    \ket{J_n} &= \frac{i}{\sqrt{2}} \epsilon_{npq} \ket{L_p}\ket{S_q} \label{Eq:JLcrossS}
\end{align}
where $\epsilon_{npq}$ is the Levi-Civita symbol and the sums over $p$ and $q$ are implied.  Overall this yields
\begin{align}
    \ket{\threePone{}, J_n} &= a\ket{\big|\threePone{}^{(0)}\big|} \frac{i}{\sqrt{2}} \epsilon_{npq} \ket{L_p}\ket{S_q} \nonumber \\ & \qquad + b \ket{\big|\onePone{}^{(0)}\big|} \ket{L_n} \ket{0,0}_s
\end{align}

\subsubsection*{\texorpdfstring{$\bm{\threeSone{}}$}{3S1}}

\noindent The $\ket{\threeSone{}}$ state has $L=0$ and $S=1$. We can decompose the state in terms of the $\v{L}$ and $\v{S}$ states in the spherical $\v{\hat{z}}$ basis as
\begin{align}
    \ket{\threeSone{}, m} &= \ket{|\threeSone{}|} \ketJ{\ell}{0}{0}\ketJ{s}{1}{m}
\end{align}
In terms of the Cartesian basis the state is then
\begin{align}
    \ket{\threeSone{}, J_n} &= \ket{|\threeSone{}|} \ketJ{\ell}{0}{0}\ket{S_n}\label{eq:Cartesian3S1}
\end{align}

\subsubsection*{\texorpdfstring{$\bm{\threePzero{}}$}{3P0}}

\noindent The $\ket{\threePzero{}}$ state has $L=1$, $S=1$, and $J=0$. In the spherical basis we have
\begin{align}
    \ket{J=0,m=0} &= \frac{1}{\sqrt{3}}\Big(\! \ketJ{\ell}{1}{1}\ketJ{s}{1}{-1}-\ketJ{\ell}{1}{0}\ketJ{s}{1}{0} \nonumber\\ & \quad \qquad + \ketJ{\ell}{1}{-1}\ketJ{s}{1}{1} \!\Big) \label{eq:J00sphericalLS}
\end{align}
which in the Cartesian basis is
\begin{align}
    \ket{J=0,m=0} &= \frac{-1}{\sqrt{3}}\sum_n \ket{L_n}\ket{S_n}
\end{align}
The complete state in the Cartesian basis is then
\begin{align}
    \ket{\threePzero{}, J=0, m = 0} &= \ket{|\threePzero{}|} \ket{J=0,m=0} \nonumber\\ &= \frac{-1}{\sqrt{3}}\ket{|\threePzero{}|} \sum_n \ket{L_n}\ket{S_n}\label{Eq:3P0}
\end{align}

\subsection{Dipole matrix elements}

Since the dipole operator $\v{\op{\mu}}$ does not operate on $\v{S}$ states, only components of the two states with the same $\v{S}$ state can contribute to the dipole matrix element.  In evaluating the matrix elements below, we will take advantage of the fact that the $\ket{S_i}$ states with $S=1$ are an orthonormal set and are also orthogonal to the $S=0$ state:
\begin{align}
    \braket{S_i | S_j} &= \delta_{ij} \\
    {}_{s}\!\braket{0,0 | S_i} &= 0
\end{align}
The matrix element for the $\v{E}_1$ coupling is
\begin{widetext}
\begin{align}
    \bra{\threePone{},J_n}\v{\op{\mu}}\cdot\v{\widetilde{E}}_1^* \ket{\oneSzero{}} &= \left( a^*\bra{\big|\threePoneZero{}\big|} \frac{-i}{\sqrt{2}} \epsilon_{npq} \bra{L_p}\bra{S_q} + b^* \bra{\big|\onePoneZero{}\big|} \bra{L_n} \bra{0,0}_s \right) \v{\op{\mu}}\cdot\v{\widetilde{E}}_1^* \ket{|\oneSzero{}|}\ketJ{\ell}{0}{0}\ketJ{s}{0}{0} \nonumber \\
    &= b^* \bra{\big|\onePoneZero{}\big|} \bra{L_n} \v{\op{\mu}}\cdot\v{\widetilde{E}}_1^* \ket{|\oneSzero{}|}\ketJ{\ell}{0}{0}
\end{align}
The matrix element for $\v{E}_2$ is
\begin{align}
    \bra{\threeSone{},J_l}\v{\op{\mu}}\cdot\v{\widetilde{E}}_2^*\ket{\threePone{},J_j} &= \bra{|\threeSone{}|} \braJ{\ell}{0}{0}\bra{S_l}\v{\op{\mu}}\cdot\v{\widetilde{E}}_2^* \left(a\ket{\big|\threePoneZero{}\big|} \frac{i}{\sqrt{2}} \epsilon_{jkq} \ket{L_k}\ket{S_q} + b \ket{\big|\onePoneZero{}\big|} \ket{L_j} \ket{0,0}_s\right) \nonumber \\
    &=\frac{i a}{\sqrt{2}}\epsilon_{jkq}\bra{|\threeSone{}|} \braJ{\ell}{0}{0}\v{\op{\mu}}\cdot\v{\widetilde{E}}_2^* \ket{\big|\threePoneZero{}\big|} \ket{L_k}\braket{S_l | S_q} \nonumber \\
    &=\frac{i a}{\sqrt{2}}\epsilon_{jkl}\bra{|\threeSone{}|} \braJ{\ell}{0}{0}\v{\op{\mu}}\cdot\v{\widetilde{E}}_2^* \ket{\big|\threePoneZero{}\big|} \ket{L_k}
\end{align}
where we first used $\braket{S_l | S_q} = \delta_{lq}$ and then performed the sum over $q$. Similarly, the matrix element for $\v{E}_3$ is
\begin{align}
    \bra{\threePzero{}}\v{\op{\mu}}\cdot\v{\widetilde{E}}_3 \ket{\threeSone{},J_l} &= \left(\frac{-1}{\sqrt{3}}\bra{|\threePzero{}|} \sum_n \bra{L_n}\bra{S_n}\right) \v{\op{\mu}}\cdot\v{\widetilde{E}}_3 \ket{|\threeSone{}|} \ketJ{\ell}{0}{0}\ket{S_l} \nonumber \\
    &=\frac{-1}{\sqrt{3}}\bra{|\threePzero{}|} \bra{L_l}\v{\op{\mu}}\cdot\v{\widetilde{E}}_3 \ket{|\threeSone{}|} \ketJ{\ell}{0}{0}
\end{align}
Collecting these results:
\begin{align}
    \bra{\threePone{},J_n}\v{\op{\mu}}\cdot\v{\widetilde{E}}_1^* \ket{\oneSzero{}} &= b^* \bra{|\onePoneZero{}|} \bra{L_n} \v{\op{\mu}}\cdot\v{\widetilde{E}}_1^* \ket{|\oneSzero{}|}\ketJ{\ell}{0}{0} \\
    \bra{\threeSone{},J_l}\v{\op{\mu}}\cdot\v{\widetilde{E}}_2^*\ket{\threePone{},J_j} &= \frac{i a}{\sqrt{2}}\epsilon_{jkl}\bra{|\threeSone{}|} \braJ{\ell}{0}{0}\v{\op{\mu}}\cdot\v{\widetilde{E}}_2^* \ket{|\threePoneZero{}|} \ket{L_k} \\
    \bra{\threePzero{}}\v{\op{\mu}}\cdot\v{\widetilde{E}}_3 \ket{\threeSone{},J_i} &= \frac{-1}{\sqrt{3}}\bra{|\threePzero{}|} \bra{L_i}\v{\op{\mu}}\cdot\v{\widetilde{E}}_3 \ket{|\threeSone{}|} \ketJ{\ell}{0}{0}
\end{align}
\end{widetext}
Next, we define polarization unit vectors $\v{e}^{(\alpha)}$, where $\alpha=1,2,3$ corresponds to the three laser fields such that $\v{\widetilde{E}}_\alpha \equiv E^{(\alpha)} \v{e}^{(\alpha)}$, where $E^{(\alpha)}$ is the field magnitude.  The polarization unit vectors are normalized so that $(\v{e}^{(\alpha)})^*\cdot\v{e}^{(\alpha)} = 1$. The dipole matrix operator for laser field $\alpha$ is then $\v{\op{\mu}}\cdot\v{\widetilde{E}}_\alpha = E^{(\alpha)} \sum_n \op{\mu}_n  e^{(\alpha)}_n$, where the sum over $n$ is over the Cartesian components $e^{(\alpha)}_n$ of the polarization vector of laser $\alpha$, and $\op{\mu}_n$ are the Cartesian components of the dipole operator.  Note that while the components $e^{(\alpha)}_n$ may be complex for general elliptically polarized light, they are real for linear polarized light.

The Cartesian components of the dipole operator couple the $L=0$ state to $L=1$ states in the Cartesian basis: 
\begin{align}\label{Eq:DipoleOn0}
\bra{L_k} \op{\mu}_n \ketJ{\ell}{0}{0} = \op{\mu} \, \delta_{kn}    
\end{align}
where $\op{\mu} \equiv |\v{\op{\mu}}|$ is the magnitude of the dipole operator, which encodes the radial degree of freedom of the atom's internal state.
When evaluating each of the matrix elements we therefore have
\begin{align}
    \bra{L_k} \v{\op{\mu}}\cdot\v{\widetilde{E}}_\alpha \ketJ{\ell}{0}{0} &= E^{(\alpha)}\sum_n\bra{L_k} \op{\mu}_n  e^{(\alpha)}_n \ketJ{\ell}{0}{0} \nonumber \\
    &= E^{(\alpha)}\sum_n \op{\mu} \, \delta_{kn}  e^{(\alpha)}_n = E^{(\alpha)} \op{\mu} \, e^{(\alpha)}_k \nonumber
\end{align}
which implies that a linearly polarized light field couples the $L=0$ state to the $L=1$ Cartesian state that is aligned with the polarization direction.
With this, the complete dipole matrix elements are
\begin{align}
\begin{split}
    \bra{\threePone{},J_n}\v{\op{\mu}}\cdot\v{\widetilde{E}}_1^* \ket{\oneSzero{}} & = b^* E^{(1)}\! \bra{\big|\onePoneZero{}\big|} \op{\mu} \ket{ \vphantom{\Big|}|\oneSzero{}|} (e^{(1)}_n)^* \phantom{\sum_k} \\
    \bra{\threeSone{},J_l}\v{\op{\mu}}\cdot\v{\widetilde{E}}_2^* \ket{\threePone{},J_j} & =  \\
    & \mkern-40mu \frac{i a E^{(2)}}{\sqrt{2}}\sum_k \epsilon_{jkl}\bra{\vphantom{\Big|}|\threeSone{}|} \op{\mu} \ket{\big|\threePoneZero{}\big|} (e^{(2)}_k)^* \\
    \bra{\threePzero{}}\v{\op{\mu}}\cdot\v{\widetilde{E}}_3 \ket{\threeSone{},J_l} & = \frac{-E^{(3)}}{\sqrt{3}}\bra{|\threePzero{}|} \op{\mu} \ket{|\threeSone{}|} e^{(3)}_l  \phantom{\sum_k} 
\end{split}\label{eq:DipoleMatrixElementsE123}
\end{align}

Up to this point, we allow for general $\v{\widetilde{E}}_\alpha$ with arbitrary polarization $\v{e}^{(\alpha)}$.  However, in our definition of the single-photon Rabi couplings $\Omega_i$ between the intermediate levels, it is convenient to choose a normalization convention that is consistent with the specific polarizations chosen experimentally.  In particular, in the main text we focus on the polarization sequence $\v{\hat{x}}\,\mhyphen\,\v{\hat{x}}\,\mhyphen\,\v{\hat{z}}$ for the three photons.  In this case we have $\v{\widetilde{E}}_1 = E^{(1)} \v{\hat{x}} = \v{E}_1^{(+)} + \v{E}_1^{(-)}$ and $\v{\widetilde{E}}_2 = E^{(2)} \v{\hat{x}} = \v{E}_2^{(+)} + \v{E}_2^{(-)}$, where we decompose the linear polarization into right and left circular polarization using the spherical basis: $\v{\hat{x}} = -\frac{1}{\sqrt{2}} \v{\hat{\sigma}}^{+} + \frac{1}{\sqrt{2}} \v{\hat{\sigma}}^{-}$.  Likewise, we have $\v{\widetilde{E}}_3 = E^{(3)} \v{\hat{z}} = \v{E}_3^{(\pi)}$ with  $\v{\hat{z}} = \v{\hat{\pi}}$.  Therefore, the spherical basis components of the fields used in the main text are
\begin{align}
    \v{E}_1^{(\pm)} &= \mp\frac{1}{\sqrt{2}} E^{(1)} \v{\hat{\sigma}}^{\pm} \\
    \v{E}_2^{(\pm)} &= \mp\frac{1}{\sqrt{2}} E^{(2)} \v{\hat{\sigma}}^{\pm} \\
    \v{E}_3^{(\pi)} &= E^{(3)} \v{\hat{\pi}}
\end{align}
Furthermore, we define $\Omega_i$ as the couplings between \textit{spherical basis} states, as shown in \figref[b]{Fig:LevelDiagrams} in the main text,
\begin{align}
    \Omega_1 &\equiv \frac{1}{\hbar} \bra{\threePone{}, m=1}\v{\op{\mu}}\cdot\v{E}_1^{(+)} \ket{\oneSzero{}} \\
    \Omega_2 &\equiv \frac{1}{\hbar} \bra{\threeSone{},m=0}\v{\op{\mu}}\cdot\v{E}_2^{(-)}\ket{\threePone{},m=1} \\
    \Omega_3 &\equiv \frac{1}{\hbar} \bra{\threePzero{}}\v{\op{\mu}}\cdot\v{E}_3^{(\pi)} \ket{\threeSone{},m=0} 
\end{align}
The states represented in the spherical basis are
\begin{align}
    \ket{\threePone{}, m=1} &= a\ket{\big|\threePoneZero{}\big|} \ket{J=1,m=1}_z  \nonumber \\ & \qquad\qquad + b \ket{\big|\onePoneZero{}\big|} \ketLS{1}{1}{0}{0}  \nonumber \\
    \phantom{\ket{\big|\threePoneZero{}\big|}} \ket{\threeSone{}, m=0} &= \ket{|\threeSone{}|} \ketJ{\ell}{0}{0}\ketJ{s}{1}{0} \nonumber \\
    \phantom{\ket{\big|\threePoneZero{}\big|}} \ket{\threePzero{}} &= \ket{|\threePzero{}|} \ket{J=0,m=0} \nonumber
\end{align}
and for transitions between these states, we have $\braL{1}{\pm 1} \v{\op{\mu}}\cdot\v{\hat{\sigma}}^{\pm} \ketJ{\ell}{0}{0} = \op{\mu}$ and $\braL{1}{0} \v{\op{\mu}}\cdot\v{\hat{\pi}} \ketJ{\ell}{0}{0} = \op{\mu}$, for circularly polarized $\sigma^{\pm}$ and $\pi$ light, respectively.  Evaluating the first single-photon coupling in the spherical basis yields 
\begin{align}
    \Omega_1 &= \frac{b^*}{\hbar} \bra{\big|\onePoneZero{}\big|} \braLS{1}{1}{0}{0} \v{\op{\mu}}\cdot\v{E}_1^{(+)} \ket{\vphantom{\Big|}|\oneSzero{}|}\ketJ{\ell}{0}{0}\ketJ{s}{0}{0}\nonumber \\
    &= \frac{b^*}{\hbar} \left(-\tfrac{1}{\sqrt{2}} E^{(1)}\right) \bra{\big|\onePoneZero{}\big|} \braL{1}{1} \v{\op{\mu}}\cdot\v{\hat{\sigma}}^{+}  \ketJ{\ell}{0}{0}\ket{\vphantom{\Big|}|\oneSzero{}|}\nonumber \\
    &= -\frac{b^* E^{(1)}}{\sqrt{2} \hbar} \bra{\big|\onePoneZero{}\big|} \op{\mu} \ket{\vphantom{\Big|}|\oneSzero{}|} 
\end{align}
Likewise, for the second coupling we have 
\begin{align}
    \Omega_2 &= \frac{a}{\hbar} \bra{|\threeSone{}|} \braJ{\ell}{0}{0}\braJ{s}{1}{0} \! \v{\op{\mu}}\cdot\v{E}_2^{(-)}\! \ket{\big|\threePoneZero{}\big|} \!\ket{ J=1,m=1}_z\nonumber \\
    &= \frac{a E^{(2)}}{2\hbar} \bra{|\threeSone{}|} \braJ{\ell}{0}{0}\braJ{s}{1}{0} \v{\op{\mu}}\cdot\v{\hat{\sigma}}^{-}\!\ketLS{1}{1}{1}{0} \ket{\big|\threePoneZero{}\big|}\nonumber \\
    &= \frac{a E^{(2)}}{2\hbar} \bra{\vphantom{\Big|}\big|\threeSone{}\big|} \op{\mu} \ket{\big|\threePoneZero{}\big|}\label{Eq:Omega2ReducedMatrix}
\end{align}
where in the second line we used \Cref{eq:J1SsphericalLS}. Finally,
\begin{align}
    \Omega_3 &= \frac{1}{\hbar} \bra{|\threePzero{}|} \bra{J=0,m=0} \v{\op{\mu}}\cdot\v{E}_3^{(\pi)} \ket{|\threeSone{}|} \ketJ{\ell}{0}{0}\ketJ{s}{1}{0}\nonumber \\
    &= \frac{-E^{(3)}}{\sqrt{3}\hbar}  \bra{|\threePzero{}|}\braJ{\ell}{1}{0}\braJ{s}{1}{0}\v{\op{\mu}}\cdot\v{\hat{\pi}}  \ketJ{\ell}{0}{0}\ketJ{s}{1}{0} \ket{|\threeSone{}|}\nonumber \\
    &= -\frac{E^{(3)}}{\sqrt{3}\hbar} \bra{|\threePzero{}|} \op{\mu} \ket{|\threeSone{}|}
\end{align}
where in the second line we used \Cref{eq:J00sphericalLS}.
We can now write the matrix elements between the Cartesian states (\Cref{eq:DipoleMatrixElementsE123}) in terms of the single-photon Rabi couplings:
\begin{align}
   \phantom{\sum_k} \bra{\threePone{},J_n}\v{\op{\mu}}\cdot\v{\widetilde{E}}_1^* \ket{\oneSzero{}} &= -\sqrt{2} \hbar \, \Omega_1 (e^{(1)}_n)^* \label{Eq:Omega1Def} \\
    \bra{\threeSone{},J_l}\v{\op{\mu}}\cdot\v{\widetilde{E}}_2^* \ket{\threePone{},J_j} &= i\sqrt{2} \hbar\, \Omega_2 \sum_k\epsilon_{jkl} (e^{(2)}_k)^* \label{Eq:Omega2Def}\\
   \phantom{\sum_k} \bra{\threePzero{}}\v{\op{\mu}}\cdot\v{\widetilde{E}}_3 \ket{\threeSone{},J_l} &= \hbar\, \Omega_3 e^{(3)}_l \label{Eq:Omega3Def}
\end{align}

\subsection{The effective coupling in the Cartesian basis}\label{Sec:EffCoupling}
Substituting the above matrix elements into \Cref{Eq:ThreePhotonCouplingCartesian} yields the main result:
\begin{equation}
    \Omega_\text{eff} = \,\frac{\Omega_1 \Omega_2 \Omega_3}{2\Delta_2} \sum_{n, j, k, l}  \epsilon_{jkl} D_{jn} (e^{(1)}_n)^* (e^{(2)}_k)^* e^{(3)}_l\label{Eq:RabiDTripleProduct}
\end{equation}
where here and in the main text we drop the overall phase of the three-photon coupling and report only the magnitude. This reproduces \Cref{eq:ThreePhotonCoupling} in the main text, assuming linearly polarized (real) $e^{(\alpha)}_i$.
In this work, we use $\v{e}^{(1)} = \v{e}^{(2)} = \v{\hat{x}}$ and $\v{e}^{(3)} = \v{\hat{z}}$, so the three-photon Rabi frequency is
\begin{align}
    \Omega_\text{eff} &= \frac{\Omega_1 \Omega_2 \Omega_3}{2\Delta_2} \epsilon_{213} D_{21} \nonumber\\
    &=\frac{\Omega_1 \Omega_2 \Omega_3}{2\Delta_2}\frac{\delta\omega_B}{\Delta_1^2-\delta\omega_B^2}\nonumber\\
    &=\frac{\Omega_1 \Omega_2 \Omega_3}{4\Delta_2}\left(\frac{1}{\Delta_1-\delta\omega_B}-\frac{1}{\Delta_1+\delta\omega_B}\right)\label{Eq:ThreePhotonCoupling}
\end{align}
where on the second line we once again drop the overall phase. This matches \Cref{eq:DestInterference} in the main text. In the absence of a magnetic field, the Zeeman shift $\delta\omega_B$ is zero and the matrix $D_{jn}$ is diagonal.  Specifically, we have $ D_{jn} = \frac{1}{\Delta_1}\delta_{jn}$ which gives
\begin{align}
    \Omega_\text{eff}({\v{B}=0}) &= \frac{\Omega_1 \Omega_2 \Omega_3}{2\Delta_1\Delta_2} \sum_{j, k, l}  \epsilon_{jkl} (e^{(1)}_j)^* (e^{(2)}_k)^* e^{(3)}_l \nonumber\\ &= \frac{\Omega_1 \Omega_2 \Omega_3}{2\Delta_1\Delta_2} \left[(\v{e}^{(2)})^*\times \v{e}^{(3)}\right] \cdot (\v{e}^{(1)})^*
\end{align}
In the case of linear polarization, we recover \Cref{eq:TripleProduct} from the main text:
\begin{align}
    \Omega_\text{eff}({\v{B}=0}) &= \frac{\Omega_1 \Omega_2 \Omega_3}{2\Delta_1\Delta_2} \,\v{e}^{(1)} \cdot \left(\v{e}^{(2)}\times \v{e}^{(3)}\right)
\end{align}
This shows that with no magnetic field, $\Omega_\text{eff}$ is proportional to the volume spanned by the three $\v{e}^{(i)}$ vectors. When all three vectors are mutually orthogonal, the coupling is $\Omega_0\equiv\Omega_1\Omega_2\Omega_3/(2\Delta_1\Delta_2)$, which corresponds to the maximum Rabi frequency. In this limit, it is not possible to have all three light fields collinear, because in that case the volume spanned is zero.

\subsection{Effective wave function of the \texorpdfstring{$\bm{\threePone{}}$}{3P1} state}\label{Sec:PsiTheta}
In \figref{Fig:TransitionPaths} in the main text, we visualize the three-photon coupling in terms of the three sequential single-photon couplings linking \oneSzero{}, \threePone{}, \threeSone{}, and \threePzero{}.  
The effect of the magnetic field on the coupling can be understood by defining an effective wavefunction that couples to $\ket{\threeSone{}}$ as it appears in \Cref{Eq:ThreePhotonCouplingCartesian}
\begin{align}
    \ket{\psi (\theta_B)}&\equiv\frac{1}{N}\sum_{j,n}\ket{\threePone{},J_j}D_{jn}\bra{\threePone{},J_n}\v{\hat{\mu}}\cdot\v{\widetilde{E}}_1^*\ket{\oneSzero{}} \nonumber \\
     &= \frac{\sqrt{2}\hbar\Omega_1}{N}\sum_{j}\ket{\threePone{},J_j}D_{j1} \label{eq:EffectiveWavefunctionDefinition}
\end{align}
where $N$ is a normalization factor and we have substituted in the definition of $\Omega_1$ from \Cref{Eq:Omega1Def} with $\v{e}^{(1)}=\v{\hat{x}}$.
With this definition, \Cref{Eq:ThreePhotonCouplingCartesian} has the form of a two-photon transition from $\ket{\psi(\theta_B)}$ to $\ket{\threePzero{}}$, i.e.
\begin{align}
    \Omega_{\text{eff}}&=N\sum_{l}\frac{\bra{\threePzero{}}\v{\hat{\mu}}\cdot\v{\widetilde{E}}_3\ket{\threeSone{},J_l}\bra{\threeSone{},J_l}\v{\hat{\mu}}\cdot\v{\widetilde{E}}_2^*\ket{\psi(\theta_B)}}{4\Delta_2\hbar^3}\label{Eq:EffectiveTwoPhoton}
\end{align}
To obtain an explicit expression for $\ket{\psi(\theta_B)}$, we define the dimensionless parameter $\beta\equiv\delta\omega_B/\Delta_1$ so that $D_{jn}$ from \Cref{Eq:DMatrix} becomes
\begin{align}
    D_{jn} &\doteq  \dfrac{1}{\Delta_1} 
     \begin{pmatrix}
        \dfrac{\strut 1}{1-\beta^2} & -\dfrac{i\,\beta}{1-\beta^2} & 0 \\
        \dfrac{\strut i\,\beta}{1-\beta^2} & \dfrac{1}{1-\beta^2} & 0  \\[4mm]
         0 & 0 & 1
    \end{pmatrix}\label{Eq:DjnBeta}
\end{align}
Although $\Omega_{\text{eff}}$ is nonzero for $j=n=3$, this case requires noncollinear light fields and gives a coupling strength that is identical to the $\v{B}=0$ case, independent of the magnetic field. Thus, with minimal loss of generality, we neglect this case and consider only the upper left $2\times2$ part of the matrix:
\begin{gather}
\centering
D_{jn} \doteq  \frac{1}{\Delta_1}
\left(
\begin{array}{c c|c}
    \multicolumn{2}{c|}{\multirow{2}{*}{$d_{jn}$}} & 0 \\ 
    \multicolumn{2}{c|}{} & 0 \\ \hline
    0 & 0 &  1 \\
\end{array}
\right)
 \quad   d_{jn} \doteq 
\begin{pmatrix}
    \frac{1}{1-\beta^2} & -\frac{i\beta}{1-\beta^2} \\
    \frac{i\beta}{1-\beta^2} & \frac{1}{1-\beta^2}
\end{pmatrix}
\end{gather}
Further defining $\theta_B\equiv \arctan{\beta}$, the matrix $d_{jn}$ becomes
\begin{align}\label{Eq:djnThetaB}
    d_{jn}&\doteq \frac{\sqrt{1+\beta^2}}{1-\beta^2}
\begin{pmatrix}
    \cos\theta_B & -i\sin\theta_B \\
    i\sin\theta_B & \cos\theta_B
\end{pmatrix}
\end{align}

After we impose the above restriction to the interesting case where the polarization $\v{e}^{(1)}$ lies in the plane perpendicular to $\v{B}$, we are free to orient our coordinate system such that the first laser is polarized along $\v{\hat{x}}$.  Thus, without further loss of generality we choose $\v{e}^{(1)} = \v{\hat{x}}$.  With this, the sum in \Cref{eq:EffectiveWavefunctionDefinition} reduces to a sum over $d_{j1}$, and for the normalization factor $N$ we find
\begin{align}
    N &= \frac{\sqrt{2}\hbar\Omega_1}{\Delta_1}\frac{\sqrt{1+\beta^2}}{1-\beta^2}
\end{align}
The effective wave function is then
\begin{align}
    \ket{\psi(\theta_B)} &= a\ket{\big|\threePoneZero{}\big|}\Big(\cos\theta_B\ket{J_x}+i\sin\theta_B\ket{J_y} \Big)  \nonumber\\ &+ b \ket{\big|\onePoneZero{}\big|} \Big(\cos\theta_B\ket{L_x}+i\sin\theta_B\ket{L_y} \Big) \ket{0,0}_s \nonumber
\end{align}
where we used \Cref{Eq:3P1Decomposition} and \Cref{Eq:1P1LS} to express the result in terms of the triplet and singlet parts.  Here $\ket{\big|\threePoneZero{}\big|}$ and $\ket{\big|\onePoneZero{}\big|}$ are the radial parts of the triplet and singlet components, respectively. As we will see below, only the triplet component of $\ket{\psi(\theta_B)}$ ultimately contributes to the three-photon coupling since the other atomic states are orthogonal to $\ket{0,0}_s$.  We define the angular parts of the effective wave function as
\begin{align}
    \ket{\threePone{},\theta_B} &\equiv \cos\theta_B\ket{J_x}+i\sin\theta_B\ket{J_y} \\
    \ket{\onePone{},\theta_B} &\equiv (\cos\theta_B\ket{L_x}+i\sin\theta_B\ket{L_y})\ket{0,0}_s
\end{align}
resulting in
\begin{align}
    \ket{\psi(\theta_B)} &= a\ket{\big|\threePoneZero{}\big|}\ket{\threePone{},\theta_B} + b \ket{\big|\onePoneZero{}\big|} \ket{\onePone{},\theta_B}  \label{Eq:PsiThetaNormalized}
\end{align}

\noindent Focusing on the triplet component, we decompose $\ket{\threePone{},\theta_B}$ in terms of $\v{L}$ and $\v{S}$ using \Cref{Eq:JLcrossS}:
\begin{align}
    \ket{\threePone{},\theta_B} &= \frac{1}{\sqrt{2}}\Big(\sin{\theta_B}\ket{L_x} + i\cos{\theta_B}\ket{L_y}\Big) \ket{S_z} \nonumber\\ &-\frac{1}{\sqrt{2}} \ket{L_z} \Big(\sin{\theta_B} \ket{S_x} +  i \cos{\theta_B} \ket{S_y}\Big) \label{Eq:JThetaB}
\end{align}
This corresponds to \Cref{Eq:PsiTheta} in the main text. 

In \figref{Fig:TransitionPaths} in the main text, we show the angular wave function of each of the atomic states that determine the sequential single-photon couplings in \Cref{Eq:ThreePhotonCouplingCartesian}, with color indicating the spin state.  For a particular state $\ket{\psi}$, we plot the probability density $f_n(\theta,\phi)=|\braket{S_n,\v{r} | \psi}|^2$ for each spin projection.
We choose to represent the different spin states using color according to $f_0(\theta,\phi)=|\braket{S=0,\v{r} | \psi}|^2$ (gray) for $S=0$, and $\{f_x~\text{(red)},~f_y~\text{(green)},~f_z~\text{(blue)}\}$ for $S=1$.  The position-space representations of the Cartesian basis states $\braket{\v{r} | L_n}$ follow from \Crefrange{Eq:Jx}{Eq:Jz}: 
\vspace{-4pt}
\begin{align}
    \braket{\v{r} | L=0} &= Y_{0,0} \nonumber \\
    \braket{\v{r} | L_x}&=-\tfrac{1}{\sqrt{2}}\left(Y_{1,+1}-Y_{1,-1}\right) \nonumber\\
    \braket{\v{r} | L_y}&=-\tfrac{1}{i\sqrt{2}}\left(Y_{1,+1}+Y_{1,-1}\right) \nonumber\\
    \braket{\v{r} | L_z}&=Y_{1,0} \nonumber
\end{align}
where $Y_{\ell,m}(\theta,\phi)$ are the spherical harmonics.

The \oneSzero{} state has $L=0$ and $S=0$, resulting in a spherically symmetric $s$ orbital colored gray according to our convention. Next, the angular wave function in \Cref{Eq:JThetaB} describes the triplet component of the $\threePone{}$ state (assuming $\v{e}^{(1)} = \v{\hat{x}}$,  as above) as a function of $\theta_B$, resulting in the following spin components
\begin{align}
    f_x(\theta,\phi)&=\tfrac{1}{2}\sin^2\theta_B|Y_{1,0}|^2 \nonumber \\
    f_y(\theta,\phi)&=\tfrac{1}{2}\cos^2\theta_B|Y_{1,0}|^2 \nonumber\\
    f_z(\theta,\phi)&=\tfrac{1}{4}\Big(\sin^2\theta_B|Y_{1,+1}-Y_{1,-1}|^2 \nonumber \\ & \qquad +\cos^2\theta_B|Y_{1,+1}+Y_{1,-1}|^2\Big) \nonumber
\end{align}
Note that the $f_x$ and $f_y$ components here both have the shape of a $p_z$ orbital, and so the color interpolates between green and red as a function of $\theta_B$ in our visualization.  \figureref{Fig:TransitionPaths} in the main text omits the singlet component of \threePone{} that appears in \Cref{Eq:PsiThetaNormalized} for the practical reason that $|b|^2\ll|a|^2$, and also because only the triplet component is coupled by the 688~nm transition.
The \threeSone{} state has $L=0$ and $S=1$, and so it appears as a spherically symmetric $s$ orbital whose color depends on the $\ket{S_n}$ state of the excitation pathway.  Finally, the Cartesian basis representation of $\ket{\threePzero{}}$ is given by \Cref{Eq:3P0}, resulting in spin components
\begin{align}
    f_x(\theta,\phi)&=\tfrac{1}{6}|Y_{1,+1}-Y_{1,-1}|^2 \nonumber\\
    f_y(\theta,\phi)&=\tfrac{1}{6}|Y_{1,+1}+Y_{1,-1}|^2 \nonumber\\
    f_z(\theta,\phi)&=\tfrac{1}{3}|Y_{1,0}|^2 \nonumber
\end{align}
which have the shape of a $p_x$, $p_y$, and $p_z$ orbital, respectively.

\subsection{Parameterizations of the three-photon coupling}\label{Sec:Metaplots}

In terms of the dimensionless detuning $\beta\equiv\delta\omega_B/\Delta_1$, the three-photon coupling in \Cref{Eq:ThreePhotonCoupling} is
\begin{align}
    \Omega_{\text{eff}}=\Omega_0\,\frac{\beta}{1-\beta^2}
\end{align}
matching \Cref{eq:meta1} in the main text.
This can also be written in terms of the projection angle $\theta_B$.  Starting from \Cref{Eq:EffectiveTwoPhoton}, we use \Cref{Eq:PsiThetaNormalized} to express the coupling in terms of the angular part of the effective \threePone{} wave function:
\begin{widetext}
\begin{align}
    \Omega_{\text{eff}}&=N a \sum_{l}\frac{\bra{\threePzero{}}\v{\hat{\mu}}\cdot\v{\widetilde{E}}_3\ket{\threeSone{},J_l}\bra{\threeSone{},J_l}\v{\hat{\mu}}\cdot\v{\widetilde{E}}_2^*\ket{\big|\threePoneZero{}\big|}\ket{\threePone{},\theta_B}}{4\Delta_2\hbar^3}
\end{align}
\end{widetext}
where only the triplet component of the effective wave function contributes since $\bra{\threeSone{},J_l}$ is orthogonal to the singlet part of $\ket{\psi(\theta_B)}$.
In the experiment, we used polarizations $\v{\hat{x}}-\v{\hat{x}}-\v{\hat{z}}$. Assuming $\v{e}^{(3)} = \v{\hat{z}}$, the sum over $l$ collapses to $l=3$ and picks out the $\ket{J_z}$ term:
\begin{align}
    \Omega_{\text{eff}}&= \frac{N a\, \Omega_3}{4\Delta_2\hbar^2} \bra{\vphantom{\Big|} \threeSone{},J_z}\v{\hat{\mu}}\cdot\v{\widetilde{E}}_2^*\ket{\big|\threePoneZero{}\big|}\ket{\threePone{},\theta_B}
\end{align}
where we use $\Omega_3$ as defined in \Cref{Eq:Omega3Def}.  For $\v{E}_2$, we allow a more general elliptical polarization:
\begin{align}
    \v{e}^{(2)}=\sin\theta \,\v{\hat{x}}+i\cos\theta\,\v{\hat{y}}
\end{align}
where $\theta$ is the ellipticity angle and we have restricted the polarization to the $xy$-plane in order to be orthogonal to $\v{e}^{(3)}=\v{\hat{z}}$, as is required to maximize the coupling. As noted in the main text, $\theta=\pi/2$ is the only choice for $\v{e}^{(2)}$ that is compatible with collinear propagation along $\v{\hat{y}}$. In the more general case, we therefore have
\begin{align}
    \v{\hat{\mu}}\cdot\v{\widetilde{E}}_2^* &= E^{(2)}\Big( \sin\theta \,\hat{\mu}_x - i\cos\theta\,\hat{\mu}_y\Big)
\end{align}
which then gives
\begin{widetext}
\begin{align}
    \Omega_{\text{eff}}&= \frac{N a\, E^{(2)} \Omega_3}{4\Delta_2\hbar^2} \bra{\vphantom{\Big|}\big|\threeSone{}\big|} \bra{S_z}\braJ{\ell}{0}{0}\Big( \sin\theta \,\hat{\mu}_x - i\cos\theta\,\hat{\mu}_y\Big)\ket{\big|\threePoneZero{}\big|}\ket{\threePone{},\theta_B}
\end{align}
where we used \Cref{eq:Cartesian3S1}.  From \Cref{Eq:DipoleOn0}, we have $\op{\mu}_n\ketJ{\ell}{0}{0} = \op{\mu} \ket{L_n}$ in general, which implies
\begin{align}
    \Omega_{\text{eff}}&= \frac{N a\, E^{(2)} \Omega_3}{4\Delta_2\hbar^2} \bra{\vphantom{\Big|}\big|\threeSone{}\big|} \op{\mu} \ket{\big|\threePoneZero{}\big|} \bra{S_z}\Big( \sin\theta \,\bra{L_x} - i\cos\theta\,\bra{L_y}\Big) \ket{\threePone{},\theta_B} .
\end{align}
\end{widetext} 
Using \Cref{Eq:Omega2ReducedMatrix} for the dipole matrix element,
\begin{align}
    \Omega_{\text{eff}} &= \frac{N \Omega_2 \Omega_3}{2\Delta_2\hbar} \bra{S_z}\bra{L,\theta} \ket{\threePone{},\theta_B}
\end{align}
where we have defined
\begin{align}
    \bra{L,\theta} &\equiv \sin\theta \,\bra{L_x} - i\cos\theta\,\bra{L_y}.
\end{align}
Inserting the normalization factor $N$, we find
\begin{align}
    \Omega_{\text{eff}} &=\sqrt{2} \, \Omega_0 \frac{\sqrt{1+\beta^2}}{1-\beta^2} \bra{S_z}\bra{L,\theta} \ket{\threePone{},\theta_B}\nonumber\\
     &=\sqrt{2} \, \Omega_B \bra{S_z}\bra{L,\theta} \ket{\threePone{},\theta_B}
\end{align}
where $\Omega_B\equiv\Omega_0\sqrt{1+\beta^2}/(1-\beta^2)$. Thus we see that the three-photon coupling is proportional to the projection of the effective wave function $\ket{\threePone{},\theta_B}$ onto a vector $\bra{S_z}\bra{L,\theta}$ determined by the light polarization ellipticity angle $\theta$.

\pagebreak[4]\noindent
Using \Cref{Eq:JThetaB}, 
\begin{align}
    \bra{S_z}\bra{L,\theta} \ket{\threePone{},\theta_B} &= \frac{1}{\sqrt{2}}\left(\sin{\theta} \sin{\theta_B} + \cos{\theta} \cos{\theta_B}\right) \nonumber
\end{align}
and so we have
\begin{equation}
    \Omega_{\text{eff}}=\Omega_B \, \cos{(\theta_B-\theta)} .
\end{equation}

This shows that the coupling depends on the relative alignment of the atom's effective dipole moment (as determined by $\theta_B$) with the polarization of the light (as set by $\theta$).  The magnetic field tunes this dipole moment in much the same way as a wave plate rotates the polarization of light. 
Finally, when we restrict to the case $\theta=\pi/2$ as we assume in the main text, we have 
\begin{equation}
    \Omega_{\text{eff}}=\Omega_B \, \sin\theta_B
\end{equation}
which reproduces \Cref{eq:meta2} in the main text.

\bibliography{bibliography}

\begin{thebibliography}{70}%
\makeatletter
\providecommand \@ifxundefined [1]{%
 \@ifx{#1\undefined}
}%
\providecommand \@ifnum [1]{%
 \ifnum #1\expandafter \@firstoftwo
 \else \expandafter \@secondoftwo
 \fi
}%
\providecommand \@ifx [1]{%
 \ifx #1\expandafter \@firstoftwo
 \else \expandafter \@secondoftwo
 \fi
}%
\providecommand \natexlab [1]{#1}%
\providecommand \enquote  [1]{``#1''}%
\providecommand \bibnamefont  [1]{#1}%
\providecommand \bibfnamefont [1]{#1}%
\providecommand \citenamefont [1]{#1}%
\providecommand \href@noop [0]{\@secondoftwo}%
\providecommand \href [0]{\begingroup \@sanitize@url \@href}%
\providecommand \@href[1]{\@@startlink{#1}\@@href}%
\providecommand \@@href[1]{\endgroup#1\@@endlink}%
\providecommand \@sanitize@url [0]{\catcode `\\12\catcode `\$12\catcode `\&12\catcode `\#12\catcode `\^12\catcode `\_12\catcode `\%12\relax}%
\providecommand \@@startlink[1]{}%
\providecommand \@@endlink[0]{}%
\providecommand \url  [0]{\begingroup\@sanitize@url \@url }%
\providecommand \@url [1]{\endgroup\@href {#1}{\urlprefix }}%
\providecommand \urlprefix  [0]{URL }%
\providecommand \Eprint [0]{\href }%
\providecommand \doibase [0]{https://doi.org/}%
\providecommand \selectlanguage [0]{\@gobble}%
\providecommand \bibinfo  [0]{\@secondoftwo}%
\providecommand \bibfield  [0]{\@secondoftwo}%
\providecommand \translation [1]{[#1]}%
\providecommand \BibitemOpen [0]{}%
\providecommand \bibitemStop [0]{}%
\providecommand \bibitemNoStop [0]{.\EOS\space}%
\providecommand \EOS [0]{\spacefactor3000\relax}%
\providecommand \BibitemShut  [1]{\csname bibitem#1\endcsname}%
\let\auto@bib@innerbib\@empty
\bibitem [{\citenamefont {Katori}(2011)}]{Katori2011}%
  \BibitemOpen
  \bibfield  {author} {\bibinfo {author} {\bibfnamefont {H.}~\bibnamefont {Katori}},\ }\bibfield  {title} {\emph {\bibinfo {title} {Optical lattice clocks and quantum metrology}},\ }\href {https://doi.org/10.1038/nphoton.2011.45} {\bibfield  {journal} {\bibinfo  {journal} {Nat. Photon.}\ }\textbf {\bibinfo {volume} {5}},\ \bibinfo {pages} {203--210} (\bibinfo {year} {2011})}\BibitemShut {NoStop}%
\bibitem [{\citenamefont {Kozlov}\ \emph {et~al.}(2018)\citenamefont {Kozlov}, \citenamefont {Safronova}, \citenamefont {Crespo L\'opez-Urrutia},\ and\ \citenamefont {Schmidt}}]{Kozlov2018}%
  \BibitemOpen
  \bibfield  {author} {\bibinfo {author} {\bibfnamefont {M.~G.}\ \bibnamefont {Kozlov}}, \bibinfo {author} {\bibfnamefont {M.~S.}\ \bibnamefont {Safronova}}, \bibinfo {author} {\bibfnamefont {J.~R.}\ \bibnamefont {Crespo L\'opez-Urrutia}},\ and\ \bibinfo {author} {\bibfnamefont {P.~O.}\ \bibnamefont {Schmidt}},\ }\bibfield  {title} {\emph {\bibinfo {title} {Highly charged ions: Optical clocks and applications in fundamental physics}},\ }\href {https://doi.org/10.1103/RevModPhys.90.045005} {\bibfield  {journal} {\bibinfo  {journal} {Rev. Mod. Phys.}\ }\textbf {\bibinfo {volume} {90}},\ \bibinfo {pages} {045005} (\bibinfo {year} {2018})}\BibitemShut {NoStop}%
\bibitem [{\citenamefont {Grotti}\ \emph {et~al.}(2018)\citenamefont {Grotti}, \citenamefont {Koller}, \citenamefont {Vogt}, \citenamefont {Haefner}, \citenamefont {Sterr}, \citenamefont {Lisdat}, \citenamefont {Denker}, \citenamefont {Voigt}, \citenamefont {Timmen}, \citenamefont {Rolland}, \citenamefont {Baynes}, \citenamefont {Margolis}, \citenamefont {Zampaolo}, \citenamefont {Thoumany}, \citenamefont {Pizzocaro}, \citenamefont {Rauf}, \citenamefont {Bregolin}, \citenamefont {Tampellini}, \citenamefont {Barbieri},\ and\ \citenamefont {Calonico}}]{Grotti2018}%
  \BibitemOpen
  \bibfield  {author} {\bibinfo {author} {\bibfnamefont {J.}~\bibnamefont {Grotti}}, \bibinfo {author} {\bibfnamefont {S.}~\bibnamefont {Koller}}, \bibinfo {author} {\bibfnamefont {S.}~\bibnamefont {Vogt}}, \bibinfo {author} {\bibfnamefont {S.}~\bibnamefont {Haefner}}, \bibinfo {author} {\bibfnamefont {U.}~\bibnamefont {Sterr}}, \bibinfo {author} {\bibfnamefont {C.}~\bibnamefont {Lisdat}}, \bibinfo {author} {\bibfnamefont {H.}~\bibnamefont {Denker}}, \bibinfo {author} {\bibfnamefont {C.}~\bibnamefont {Voigt}}, \bibinfo {author} {\bibfnamefont {L.}~\bibnamefont {Timmen}}, \bibinfo {author} {\bibfnamefont {A.}~\bibnamefont {Rolland}}, \bibinfo {author} {\bibfnamefont {F.}~\bibnamefont {Baynes}}, \bibinfo {author} {\bibfnamefont {H.}~\bibnamefont {Margolis}}, \bibinfo {author} {\bibfnamefont {M.}~\bibnamefont {Zampaolo}}, \bibinfo {author} {\bibfnamefont {P.}~\bibnamefont {Thoumany}}, \bibinfo {author} {\bibfnamefont {M.}~\bibnamefont {Pizzocaro}}, \bibinfo {author} {\bibfnamefont {B.}~\bibnamefont {Rauf}},
  \bibinfo {author} {\bibfnamefont {F.}~\bibnamefont {Bregolin}}, \bibinfo {author} {\bibfnamefont {A.}~\bibnamefont {Tampellini}}, \bibinfo {author} {\bibfnamefont {P.}~\bibnamefont {Barbieri}},\ and\ \bibinfo {author} {\bibfnamefont {D.}~\bibnamefont {Calonico}},\ }\bibfield  {title} {\emph {\bibinfo {title} {Geodesy and metrology with a transportable optical clock}},\ }\href {https://doi.org/10.1038/s41567-017-0042-3} {\bibfield  {journal} {\bibinfo  {journal} {Nat. Phys.}\ }\textbf {\bibinfo {volume} {14}},\ \bibinfo {pages} {437--441} (\bibinfo {year} {2018})}\BibitemShut {NoStop}%
\bibitem [{\citenamefont {Bloom}\ \emph {et~al.}(2014)\citenamefont {Bloom}, \citenamefont {Nicholson}, \citenamefont {Williams}, \citenamefont {Campbell}, \citenamefont {Bishof}, \citenamefont {Zhang}, \citenamefont {Zhang}, \citenamefont {Bromley},\ and\ \citenamefont {Ye}}]{Bloom2014}%
  \BibitemOpen
  \bibfield  {author} {\bibinfo {author} {\bibfnamefont {B.~J.}\ \bibnamefont {Bloom}}, \bibinfo {author} {\bibfnamefont {T.~L.}\ \bibnamefont {Nicholson}}, \bibinfo {author} {\bibfnamefont {J.~R.}\ \bibnamefont {Williams}}, \bibinfo {author} {\bibfnamefont {S.~L.}\ \bibnamefont {Campbell}}, \bibinfo {author} {\bibfnamefont {M.}~\bibnamefont {Bishof}}, \bibinfo {author} {\bibfnamefont {X.}~\bibnamefont {Zhang}}, \bibinfo {author} {\bibfnamefont {W.}~\bibnamefont {Zhang}}, \bibinfo {author} {\bibfnamefont {S.~L.}\ \bibnamefont {Bromley}},\ and\ \bibinfo {author} {\bibfnamefont {J.}~\bibnamefont {Ye}},\ }\bibfield  {title} {\emph {\bibinfo {title} {An optical lattice clock with accuracy and stability at the $10^{-18}$ level}},\ }\href {https://doi.org/10.1038/nature12941} {\bibfield  {journal} {\bibinfo  {journal} {Nature}\ }\textbf {\bibinfo {volume} {506}},\ \bibinfo {pages} {71} (\bibinfo {year} {2014})}\BibitemShut {NoStop}%
\bibitem [{\citenamefont {Ludlow}\ \emph {et~al.}(2015)\citenamefont {Ludlow}, \citenamefont {Boyd}, \citenamefont {Ye}, \citenamefont {Peik},\ and\ \citenamefont {Schmidt}}]{Ludlow2015}%
  \BibitemOpen
  \bibfield  {author} {\bibinfo {author} {\bibfnamefont {A.~D.}\ \bibnamefont {Ludlow}}, \bibinfo {author} {\bibfnamefont {M.~M.}\ \bibnamefont {Boyd}}, \bibinfo {author} {\bibfnamefont {J.}~\bibnamefont {Ye}}, \bibinfo {author} {\bibfnamefont {E.}~\bibnamefont {Peik}},\ and\ \bibinfo {author} {\bibfnamefont {P.~O.}\ \bibnamefont {Schmidt}},\ }\bibfield  {title} {\emph {\bibinfo {title} {Optical atomic clocks}},\ }\href {https://doi.org/10.1103/RevModPhys.87.637} {\bibfield  {journal} {\bibinfo  {journal} {Rev. Mod. Phys.}\ }\textbf {\bibinfo {volume} {87}},\ \bibinfo {pages} {637--701} (\bibinfo {year} {2015})}\BibitemShut {NoStop}%
\bibitem [{\citenamefont {Marti}\ \emph {et~al.}(2018)\citenamefont {Marti}, \citenamefont {Hutson}, \citenamefont {Goban}, \citenamefont {Campbell}, \citenamefont {Poli},\ and\ \citenamefont {Ye}}]{Marti2018}%
  \BibitemOpen
  \bibfield  {author} {\bibinfo {author} {\bibfnamefont {G.~E.}\ \bibnamefont {Marti}}, \bibinfo {author} {\bibfnamefont {R.~B.}\ \bibnamefont {Hutson}}, \bibinfo {author} {\bibfnamefont {A.}~\bibnamefont {Goban}}, \bibinfo {author} {\bibfnamefont {S.~L.}\ \bibnamefont {Campbell}}, \bibinfo {author} {\bibfnamefont {N.}~\bibnamefont {Poli}},\ and\ \bibinfo {author} {\bibfnamefont {J.}~\bibnamefont {Ye}},\ }\bibfield  {title} {\emph {\bibinfo {title} {{Imaging Optical Frequencies with $100~\ensuremath{\micro}\mathrm{Hz}$ Precision and $1.1~\ensuremath{\micro}\mathrm{m}$ Resolution}}},\ }\href {https://doi.org/10.1103/PhysRevLett.120.103201} {\bibfield  {journal} {\bibinfo  {journal} {Phys. Rev. Lett.}\ }\textbf {\bibinfo {volume} {120}},\ \bibinfo {pages} {103201} (\bibinfo {year} {2018})}\BibitemShut {NoStop}%
\bibitem [{\citenamefont {Blatt}\ \emph {et~al.}(2008)\citenamefont {Blatt}, \citenamefont {Ludlow}, \citenamefont {Campbell}, \citenamefont {Thomsen}, \citenamefont {Zelevinsky}, \citenamefont {Boyd}, \citenamefont {Ye}, \citenamefont {Baillard}, \citenamefont {Fouch\'e}, \citenamefont {Le~Targat}, \citenamefont {Brusch}, \citenamefont {Lemonde}, \citenamefont {Takamoto}, \citenamefont {Hong}, \citenamefont {Katori},\ and\ \citenamefont {Flambaum}}]{Blatt2008}%
  \BibitemOpen
  \bibfield  {author} {\bibinfo {author} {\bibfnamefont {S.}~\bibnamefont {Blatt}}, \bibinfo {author} {\bibfnamefont {A.~D.}\ \bibnamefont {Ludlow}}, \bibinfo {author} {\bibfnamefont {G.~K.}\ \bibnamefont {Campbell}}, \bibinfo {author} {\bibfnamefont {J.~W.}\ \bibnamefont {Thomsen}}, \bibinfo {author} {\bibfnamefont {T.}~\bibnamefont {Zelevinsky}}, \bibinfo {author} {\bibfnamefont {M.~M.}\ \bibnamefont {Boyd}}, \bibinfo {author} {\bibfnamefont {J.}~\bibnamefont {Ye}}, \bibinfo {author} {\bibfnamefont {X.}~\bibnamefont {Baillard}}, \bibinfo {author} {\bibfnamefont {M.}~\bibnamefont {Fouch\'e}}, \bibinfo {author} {\bibfnamefont {R.}~\bibnamefont {Le~Targat}}, \bibinfo {author} {\bibfnamefont {A.}~\bibnamefont {Brusch}}, \bibinfo {author} {\bibfnamefont {P.}~\bibnamefont {Lemonde}}, \bibinfo {author} {\bibfnamefont {M.}~\bibnamefont {Takamoto}}, \bibinfo {author} {\bibfnamefont {F.-L.}\ \bibnamefont {Hong}}, \bibinfo {author} {\bibfnamefont {H.}~\bibnamefont {Katori}},\ and\ \bibinfo {author} {\bibfnamefont
  {V.~V.}\ \bibnamefont {Flambaum}},\ }\bibfield  {title} {\emph {\bibinfo {title} {{New Limits on Coupling of Fundamental Constants to Gravity Using $^{87}\mathrm{Sr}$ Optical Lattice Clocks}}},\ }\href {https://doi.org/10.1103/PhysRevLett.100.140801} {\bibfield  {journal} {\bibinfo  {journal} {Phys. Rev. Lett.}\ }\textbf {\bibinfo {volume} {100}},\ \bibinfo {pages} {140801} (\bibinfo {year} {2008})}\BibitemShut {NoStop}%
\bibitem [{\citenamefont {Godun}\ \emph {et~al.}(2014)\citenamefont {Godun}, \citenamefont {Nisbet-Jones}, \citenamefont {Jones}, \citenamefont {King}, \citenamefont {Johnson}, \citenamefont {Margolis}, \citenamefont {Szymaniec}, \citenamefont {Lea}, \citenamefont {Bongs},\ and\ \citenamefont {Gill}}]{Godun2014}%
  \BibitemOpen
  \bibfield  {author} {\bibinfo {author} {\bibfnamefont {R.~M.}\ \bibnamefont {Godun}}, \bibinfo {author} {\bibfnamefont {P.~B.~R.}\ \bibnamefont {Nisbet-Jones}}, \bibinfo {author} {\bibfnamefont {J.~M.}\ \bibnamefont {Jones}}, \bibinfo {author} {\bibfnamefont {S.~A.}\ \bibnamefont {King}}, \bibinfo {author} {\bibfnamefont {L.~A.~M.}\ \bibnamefont {Johnson}}, \bibinfo {author} {\bibfnamefont {H.~S.}\ \bibnamefont {Margolis}}, \bibinfo {author} {\bibfnamefont {K.}~\bibnamefont {Szymaniec}}, \bibinfo {author} {\bibfnamefont {S.~N.}\ \bibnamefont {Lea}}, \bibinfo {author} {\bibfnamefont {K.}~\bibnamefont {Bongs}},\ and\ \bibinfo {author} {\bibfnamefont {P.}~\bibnamefont {Gill}},\ }\bibfield  {title} {\emph {\bibinfo {title} {{Frequency Ratio of Two Optical Clock Transitions in $^{171}{\mathrm{Yb}}^{+}$ and Constraints on the Time Variation of Fundamental Constants}}},\ }\href {https://doi.org/10.1103/PhysRevLett.113.210801} {\bibfield  {journal} {\bibinfo  {journal} {Phys. Rev. Lett.}\ }\textbf {\bibinfo
  {volume} {113}},\ \bibinfo {pages} {210801} (\bibinfo {year} {2014})}\BibitemShut {NoStop}%
\bibitem [{\citenamefont {{Takamoto}}\ \emph {et~al.}(2020)\citenamefont {{Takamoto}}, \citenamefont {{Ushijima}}, \citenamefont {{Ohmae}}, \citenamefont {{Yahagi}}, \citenamefont {{Kokado}}, \citenamefont {{Shinkai}},\ and\ \citenamefont {{Katori}}}]{Takamoto2020}%
  \BibitemOpen
  \bibfield  {author} {\bibinfo {author} {\bibfnamefont {M.}~\bibnamefont {{Takamoto}}}, \bibinfo {author} {\bibfnamefont {I.}~\bibnamefont {{Ushijima}}}, \bibinfo {author} {\bibfnamefont {N.}~\bibnamefont {{Ohmae}}}, \bibinfo {author} {\bibfnamefont {T.}~\bibnamefont {{Yahagi}}}, \bibinfo {author} {\bibfnamefont {K.}~\bibnamefont {{Kokado}}}, \bibinfo {author} {\bibfnamefont {H.}~\bibnamefont {{Shinkai}}},\ and\ \bibinfo {author} {\bibfnamefont {H.}~\bibnamefont {{Katori}}},\ }\bibfield  {title} {\emph {\bibinfo {title} {{Test of general relativity by a pair of transportable optical lattice clocks}}},\ }\href {https://doi.org/10.1038/s41566-020-0619-8} {\bibfield  {journal} {\bibinfo  {journal} {Nat. Photon.}\ }\textbf {\bibinfo {volume} {14}},\ \bibinfo {pages} {411--415} (\bibinfo {year} {2020})}\BibitemShut {NoStop}%
\bibitem [{\citenamefont {Bothwell}\ \emph {et~al.}(2022)\citenamefont {Bothwell}, \citenamefont {Kennedy}, \citenamefont {Aeppli}, \citenamefont {Kedar}, \citenamefont {Robinson}, \citenamefont {Oelker}, \citenamefont {Staron},\ and\ \citenamefont {Ye}}]{Bothwell2022}%
  \BibitemOpen
  \bibfield  {author} {\bibinfo {author} {\bibfnamefont {T.}~\bibnamefont {Bothwell}}, \bibinfo {author} {\bibfnamefont {C.~J.}\ \bibnamefont {Kennedy}}, \bibinfo {author} {\bibfnamefont {A.}~\bibnamefont {Aeppli}}, \bibinfo {author} {\bibfnamefont {D.}~\bibnamefont {Kedar}}, \bibinfo {author} {\bibfnamefont {J.~M.}\ \bibnamefont {Robinson}}, \bibinfo {author} {\bibfnamefont {E.}~\bibnamefont {Oelker}}, \bibinfo {author} {\bibfnamefont {A.}~\bibnamefont {Staron}},\ and\ \bibinfo {author} {\bibfnamefont {J.}~\bibnamefont {Ye}},\ }\bibfield  {title} {\emph {\bibinfo {title} {Resolving the gravitational redshift across a millimetre-scale atomic sample}},\ }\href {https://doi.org/10.1038/s41586-021-04349-7} {\bibfield  {journal} {\bibinfo  {journal} {Nature}\ }\textbf {\bibinfo {volume} {602}},\ \bibinfo {pages} {420--424} (\bibinfo {year} {2022})}\BibitemShut {NoStop}%
\bibitem [{\citenamefont {Bothwell}\ \emph {et~al.}(2019)\citenamefont {Bothwell}, \citenamefont {Kedar}, \citenamefont {Oelker}, \citenamefont {Robinson}, \citenamefont {Bromley}, \citenamefont {Tew}, \citenamefont {Ye},\ and\ \citenamefont {Kennedy}}]{Bothwell2019}%
  \BibitemOpen
  \bibfield  {author} {\bibinfo {author} {\bibfnamefont {T.}~\bibnamefont {Bothwell}}, \bibinfo {author} {\bibfnamefont {D.}~\bibnamefont {Kedar}}, \bibinfo {author} {\bibfnamefont {E.}~\bibnamefont {Oelker}}, \bibinfo {author} {\bibfnamefont {J.}~\bibnamefont {Robinson}}, \bibinfo {author} {\bibfnamefont {S.}~\bibnamefont {Bromley}}, \bibinfo {author} {\bibfnamefont {W.}~\bibnamefont {Tew}}, \bibinfo {author} {\bibfnamefont {J.}~\bibnamefont {Ye}},\ and\ \bibinfo {author} {\bibfnamefont {C.}~\bibnamefont {Kennedy}},\ }\bibfield  {title} {\emph {\bibinfo {title} {{JILA $\mathrm{Sr~I}$ optical lattice clock with uncertainty of $2.0\times10^{-18}$}}},\ }\href {https://doi.org/10.1088/1681-7575/ab4089} {\bibfield  {journal} {\bibinfo  {journal} {Metrologia}\ }\textbf {\bibinfo {volume} {56}} (\bibinfo {year} {2019})}\BibitemShut {NoStop}%
\bibitem [{\citenamefont {McGrew}\ \emph {et~al.}(2018)\citenamefont {McGrew}, \citenamefont {Zhang}, \citenamefont {Fasano}, \citenamefont {Sch{\"a}ffer}, \citenamefont {Beloy}, \citenamefont {Nicolodi}, \citenamefont {Brown}, \citenamefont {Hinkley}, \citenamefont {Milani}, \citenamefont {Schioppo}, \citenamefont {Yoon},\ and\ \citenamefont {Ludlow}}]{McGrew2018}%
  \BibitemOpen
  \bibfield  {author} {\bibinfo {author} {\bibfnamefont {W.~F.}\ \bibnamefont {McGrew}}, \bibinfo {author} {\bibfnamefont {X.}~\bibnamefont {Zhang}}, \bibinfo {author} {\bibfnamefont {R.~J.}\ \bibnamefont {Fasano}}, \bibinfo {author} {\bibfnamefont {S.~A.}\ \bibnamefont {Sch{\"a}ffer}}, \bibinfo {author} {\bibfnamefont {K.}~\bibnamefont {Beloy}}, \bibinfo {author} {\bibfnamefont {D.}~\bibnamefont {Nicolodi}}, \bibinfo {author} {\bibfnamefont {R.~C.}\ \bibnamefont {Brown}}, \bibinfo {author} {\bibfnamefont {N.}~\bibnamefont {Hinkley}}, \bibinfo {author} {\bibfnamefont {G.}~\bibnamefont {Milani}}, \bibinfo {author} {\bibfnamefont {M.}~\bibnamefont {Schioppo}}, \bibinfo {author} {\bibfnamefont {T.~H.}\ \bibnamefont {Yoon}},\ and\ \bibinfo {author} {\bibfnamefont {A.~D.}\ \bibnamefont {Ludlow}},\ }\bibfield  {title} {\emph {\bibinfo {title} {Atomic clock performance enabling geodesy below the centimetre level}},\ }\href {https://doi.org/10.1038/s41586-018-0738-2} {\bibfield  {journal} {\bibinfo  {journal}
  {Nature}\ }\textbf {\bibinfo {volume} {564}},\ \bibinfo {pages} {87--90} (\bibinfo {year} {2018})}\BibitemShut {NoStop}%
\bibitem [{\citenamefont {Kulosa}\ \emph {et~al.}(2015)\citenamefont {Kulosa}, \citenamefont {Fim}, \citenamefont {Zipfel}, \citenamefont {R\"uhmann}, \citenamefont {Sauer}, \citenamefont {Jha}, \citenamefont {Gibble}, \citenamefont {Ertmer}, \citenamefont {Rasel}, \citenamefont {Safronova}, \citenamefont {Safronova},\ and\ \citenamefont {Porsev}}]{Kulosa2015}%
  \BibitemOpen
  \bibfield  {author} {\bibinfo {author} {\bibfnamefont {A.~P.}\ \bibnamefont {Kulosa}}, \bibinfo {author} {\bibfnamefont {D.}~\bibnamefont {Fim}}, \bibinfo {author} {\bibfnamefont {K.~H.}\ \bibnamefont {Zipfel}}, \bibinfo {author} {\bibfnamefont {S.}~\bibnamefont {R\"uhmann}}, \bibinfo {author} {\bibfnamefont {S.}~\bibnamefont {Sauer}}, \bibinfo {author} {\bibfnamefont {N.}~\bibnamefont {Jha}}, \bibinfo {author} {\bibfnamefont {K.}~\bibnamefont {Gibble}}, \bibinfo {author} {\bibfnamefont {W.}~\bibnamefont {Ertmer}}, \bibinfo {author} {\bibfnamefont {E.~M.}\ \bibnamefont {Rasel}}, \bibinfo {author} {\bibfnamefont {M.~S.}\ \bibnamefont {Safronova}}, \bibinfo {author} {\bibfnamefont {U.~I.}\ \bibnamefont {Safronova}},\ and\ \bibinfo {author} {\bibfnamefont {S.~G.}\ \bibnamefont {Porsev}},\ }\bibfield  {title} {\emph {\bibinfo {title} {{Towards a $\mathrm{Mg}$ Lattice Clock: Observation of the $^{1}{\mathrm{S}}_{0}-^{3}{\mathrm{P}}_{0}$ Transition and Determination of the Magic Wavelength}}},\ }\href
  {https://doi.org/10.1103/PhysRevLett.115.240801} {\bibfield  {journal} {\bibinfo  {journal} {Phys. Rev. Lett.}\ }\textbf {\bibinfo {volume} {115}},\ \bibinfo {pages} {240801} (\bibinfo {year} {2015})}\BibitemShut {NoStop}%
\bibitem [{\citenamefont {Brewer}\ \emph {et~al.}(2019)\citenamefont {Brewer}, \citenamefont {Chen}, \citenamefont {Hankin}, \citenamefont {Clements}, \citenamefont {Chou}, \citenamefont {Wineland}, \citenamefont {Hume},\ and\ \citenamefont {Leibrandt}}]{Brewer2019}%
  \BibitemOpen
  \bibfield  {author} {\bibinfo {author} {\bibfnamefont {S.~M.}\ \bibnamefont {Brewer}}, \bibinfo {author} {\bibfnamefont {J.-S.}\ \bibnamefont {Chen}}, \bibinfo {author} {\bibfnamefont {A.~M.}\ \bibnamefont {Hankin}}, \bibinfo {author} {\bibfnamefont {E.~R.}\ \bibnamefont {Clements}}, \bibinfo {author} {\bibfnamefont {C.~W.}\ \bibnamefont {Chou}}, \bibinfo {author} {\bibfnamefont {D.~J.}\ \bibnamefont {Wineland}}, \bibinfo {author} {\bibfnamefont {D.~B.}\ \bibnamefont {Hume}},\ and\ \bibinfo {author} {\bibfnamefont {D.~R.}\ \bibnamefont {Leibrandt}},\ }\bibfield  {title} {\emph {\bibinfo {title} {{$^{27}{\mathrm{Al}}^{+}$ Quantum-Logic Clock with a Systematic Uncertainty below ${10}^{-18}$}}},\ }\href {https://doi.org/10.1103/PhysRevLett.123.033201} {\bibfield  {journal} {\bibinfo  {journal} {Phys. Rev. Lett.}\ }\textbf {\bibinfo {volume} {123}},\ \bibinfo {pages} {033201} (\bibinfo {year} {2019})}\BibitemShut {NoStop}%
\bibitem [{\citenamefont {Taichenachev}\ \emph {et~al.}(2006)\citenamefont {Taichenachev}, \citenamefont {Yudin}, \citenamefont {Oates}, \citenamefont {Hoyt}, \citenamefont {Barber},\ and\ \citenamefont {Hollberg}}]{Taichenachev2006}%
  \BibitemOpen
  \bibfield  {author} {\bibinfo {author} {\bibfnamefont {A.~V.}\ \bibnamefont {Taichenachev}}, \bibinfo {author} {\bibfnamefont {V.~I.}\ \bibnamefont {Yudin}}, \bibinfo {author} {\bibfnamefont {C.~W.}\ \bibnamefont {Oates}}, \bibinfo {author} {\bibfnamefont {C.~W.}\ \bibnamefont {Hoyt}}, \bibinfo {author} {\bibfnamefont {Z.~W.}\ \bibnamefont {Barber}},\ and\ \bibinfo {author} {\bibfnamefont {L.}~\bibnamefont {Hollberg}},\ }\bibfield  {title} {\emph {\bibinfo {title} {{Magnetic Field-Induced Spectroscopy of Forbidden Optical Transitions with Application to Lattice-Based Optical Atomic Clocks}}},\ }\href {https://doi.org/10.1103/PhysRevLett.96.083001} {\bibfield  {journal} {\bibinfo  {journal} {Phys. Rev. Lett.}\ }\textbf {\bibinfo {volume} {96}},\ \bibinfo {pages} {083001} (\bibinfo {year} {2006})}\BibitemShut {NoStop}%
\bibitem [{\citenamefont {Barber}\ \emph {et~al.}(2006)\citenamefont {Barber}, \citenamefont {Hoyt}, \citenamefont {Oates}, \citenamefont {Hollberg}, \citenamefont {Taichenachev},\ and\ \citenamefont {Yudin}}]{Barber2006}%
  \BibitemOpen
  \bibfield  {author} {\bibinfo {author} {\bibfnamefont {Z.~W.}\ \bibnamefont {Barber}}, \bibinfo {author} {\bibfnamefont {C.~W.}\ \bibnamefont {Hoyt}}, \bibinfo {author} {\bibfnamefont {C.~W.}\ \bibnamefont {Oates}}, \bibinfo {author} {\bibfnamefont {L.}~\bibnamefont {Hollberg}}, \bibinfo {author} {\bibfnamefont {A.~V.}\ \bibnamefont {Taichenachev}},\ and\ \bibinfo {author} {\bibfnamefont {V.~I.}\ \bibnamefont {Yudin}},\ }\bibfield  {title} {\emph {\bibinfo {title} {{Direct excitation of the forbidden clock transition in neutral $^{174}\mathrm{Yb}$ atoms confined to an optical lattice}}},\ }\href {https://doi.org/10.1103/PhysRevLett.96.083002} {\bibfield  {journal} {\bibinfo  {journal} {Phys. Rev. Lett.}\ }\textbf {\bibinfo {volume} {96}},\ \bibinfo {pages} {083002} (\bibinfo {year} {2006})}\BibitemShut {NoStop}%
\bibitem [{\citenamefont {Akatsuka}\ \emph {et~al.}(2008)\citenamefont {Akatsuka}, \citenamefont {Takamoto},\ and\ \citenamefont {Katori}}]{Akatsuka2008}%
  \BibitemOpen
  \bibfield  {author} {\bibinfo {author} {\bibfnamefont {T.}~\bibnamefont {Akatsuka}}, \bibinfo {author} {\bibfnamefont {M.}~\bibnamefont {Takamoto}},\ and\ \bibinfo {author} {\bibfnamefont {H.}~\bibnamefont {Katori}},\ }\bibfield  {title} {\emph {\bibinfo {title} {Optical lattice clocks with non-interacting bosons and fermions}},\ }\href {https://doi.org/10.1038/nphys1108} {\bibfield  {journal} {\bibinfo  {journal} {Nat. Phys.}\ }\textbf {\bibinfo {volume} {4}},\ \bibinfo {pages} {954--959} (\bibinfo {year} {2008})}\BibitemShut {NoStop}%
\bibitem [{\citenamefont {Poli}\ \emph {et~al.}(2014)\citenamefont {Poli}, \citenamefont {Schioppo}, \citenamefont {Vogt}, \citenamefont {Falke}, \citenamefont {Sterr}, \citenamefont {Lisdat},\ and\ \citenamefont {Tino}}]{Poli2014}%
  \BibitemOpen
  \bibfield  {author} {\bibinfo {author} {\bibfnamefont {N.}~\bibnamefont {Poli}}, \bibinfo {author} {\bibfnamefont {M.}~\bibnamefont {Schioppo}}, \bibinfo {author} {\bibfnamefont {S.}~\bibnamefont {Vogt}}, \bibinfo {author} {\bibfnamefont {S.}~\bibnamefont {Falke}}, \bibinfo {author} {\bibfnamefont {U.}~\bibnamefont {Sterr}}, \bibinfo {author} {\bibfnamefont {C.}~\bibnamefont {Lisdat}},\ and\ \bibinfo {author} {\bibfnamefont {G.~M.}\ \bibnamefont {Tino}},\ }\bibfield  {title} {\emph {\bibinfo {title} {{A transportable strontium optical lattice clock}}},\ }\href {https://doi.org/10.1007/s00340-014-5932-9} {\bibfield  {journal} {\bibinfo  {journal} {Appl. Phys. B.}\ }\textbf {\bibinfo {volume} {117}},\ \bibinfo {pages} {1107 -- 1116} (\bibinfo {year} {2014})}\BibitemShut {NoStop}%
\bibitem [{\citenamefont {Santra}\ \emph {et~al.}(2005)\citenamefont {Santra}, \citenamefont {Arimondo}, \citenamefont {Ido}, \citenamefont {Greene},\ and\ \citenamefont {Ye}}]{Santra2005}%
  \BibitemOpen
  \bibfield  {author} {\bibinfo {author} {\bibfnamefont {R.}~\bibnamefont {Santra}}, \bibinfo {author} {\bibfnamefont {E.}~\bibnamefont {Arimondo}}, \bibinfo {author} {\bibfnamefont {T.}~\bibnamefont {Ido}}, \bibinfo {author} {\bibfnamefont {C.~H.}\ \bibnamefont {Greene}},\ and\ \bibinfo {author} {\bibfnamefont {J.}~\bibnamefont {Ye}},\ }\bibfield  {title} {\emph {\bibinfo {title} {{High-Accuracy Optical Clock via Three-Level Coherence in Neutral Bosonic $^{88}\mathrm{Sr}$}}},\ }\href {https://doi.org/10.1103/PhysRevLett.94.173002} {\bibfield  {journal} {\bibinfo  {journal} {Phys. Rev. Lett.}\ }\textbf {\bibinfo {volume} {94}},\ \bibinfo {pages} {173002} (\bibinfo {year} {2005})}\BibitemShut {NoStop}%
\bibitem [{\citenamefont {Hong}\ \emph {et~al.}(2005)\citenamefont {Hong}, \citenamefont {Cramer}, \citenamefont {Nagourney},\ and\ \citenamefont {Fortson}}]{Hong2005}%
  \BibitemOpen
  \bibfield  {author} {\bibinfo {author} {\bibfnamefont {T.}~\bibnamefont {Hong}}, \bibinfo {author} {\bibfnamefont {C.}~\bibnamefont {Cramer}}, \bibinfo {author} {\bibfnamefont {W.}~\bibnamefont {Nagourney}},\ and\ \bibinfo {author} {\bibfnamefont {E.~N.}\ \bibnamefont {Fortson}},\ }\bibfield  {title} {\emph {\bibinfo {title} {{Optical Clocks Based on Ultranarrow Three-Photon Resonances in Alkaline Earth Atoms}}},\ }\href {https://doi.org/10.1103/PhysRevLett.94.050801} {\bibfield  {journal} {\bibinfo  {journal} {Phys. Rev. Lett.}\ }\textbf {\bibinfo {volume} {94}},\ \bibinfo {pages} {050801} (\bibinfo {year} {2005})}\BibitemShut {NoStop}%
\bibitem [{\citenamefont {Ovsiannikov}\ \emph {et~al.}(2007)\citenamefont {Ovsiannikov}, \citenamefont {Pal'chikov}, \citenamefont {Taichenachev}, \citenamefont {Yudin}, \citenamefont {Katori},\ and\ \citenamefont {Takamoto}}]{Vitaly2007}%
  \BibitemOpen
  \bibfield  {author} {\bibinfo {author} {\bibfnamefont {V.~D.}\ \bibnamefont {Ovsiannikov}}, \bibinfo {author} {\bibfnamefont {V.~G.}\ \bibnamefont {Pal'chikov}}, \bibinfo {author} {\bibfnamefont {A.~V.}\ \bibnamefont {Taichenachev}}, \bibinfo {author} {\bibfnamefont {V.~I.}\ \bibnamefont {Yudin}}, \bibinfo {author} {\bibfnamefont {H.}~\bibnamefont {Katori}},\ and\ \bibinfo {author} {\bibfnamefont {M.}~\bibnamefont {Takamoto}},\ }\bibfield  {title} {\emph {\bibinfo {title} {{Magic-wave-induced $^{1}\mathrm{S}_{0}-^{3}\mathrm{P}_{0}$ transition in even isotopes of alkaline-earth-metal-like atoms}}},\ }\href {https://doi.org/10.1103/PhysRevA.75.020501} {\bibfield  {journal} {\bibinfo  {journal} {Phys. Rev. A}\ }\textbf {\bibinfo {volume} {75}},\ \bibinfo {pages} {020501(R)} (\bibinfo {year} {2007})}\BibitemShut {NoStop}%
\bibitem [{\citenamefont {Alden}\ \emph {et~al.}(2014)\citenamefont {Alden}, \citenamefont {Moore},\ and\ \citenamefont {Leanhardt}}]{Alden2014}%
  \BibitemOpen
  \bibfield  {author} {\bibinfo {author} {\bibfnamefont {E.~A.}\ \bibnamefont {Alden}}, \bibinfo {author} {\bibfnamefont {K.~R.}\ \bibnamefont {Moore}},\ and\ \bibinfo {author} {\bibfnamefont {A.~E.}\ \bibnamefont {Leanhardt}},\ }\bibfield  {title} {\emph {\bibinfo {title} {{Two-photon $E1$-$M1$ optical clock}}},\ }\href {https://doi.org/10.1103/PhysRevA.90.012523} {\bibfield  {journal} {\bibinfo  {journal} {Phys. Rev. A}\ }\textbf {\bibinfo {volume} {90}},\ \bibinfo {pages} {012523} (\bibinfo {year} {2014})}\BibitemShut {NoStop}%
\bibitem [{\citenamefont {Barker}\ \emph {et~al.}(2016)\citenamefont {Barker}, \citenamefont {Pisenti}, \citenamefont {Reschovsky},\ and\ \citenamefont {Campbell}}]{Barker2016}%
  \BibitemOpen
  \bibfield  {author} {\bibinfo {author} {\bibfnamefont {D.~S.}\ \bibnamefont {Barker}}, \bibinfo {author} {\bibfnamefont {N.~C.}\ \bibnamefont {Pisenti}}, \bibinfo {author} {\bibfnamefont {B.~J.}\ \bibnamefont {Reschovsky}},\ and\ \bibinfo {author} {\bibfnamefont {G.~K.}\ \bibnamefont {Campbell}},\ }\bibfield  {title} {\emph {\bibinfo {title} {Three-photon process for producing a degenerate gas of metastable alkaline-earth-metal atoms}},\ }\href {https://doi.org/10.1103/PhysRevA.93.053417} {\bibfield  {journal} {\bibinfo  {journal} {Phys. Rev. A}\ }\textbf {\bibinfo {volume} {93}},\ \bibinfo {pages} {053417} (\bibinfo {year} {2016})}\BibitemShut {NoStop}%
\bibitem [{\citenamefont {Grynberg}\ \emph {et~al.}(1980)\citenamefont {Grynberg}, \citenamefont {Cagnac},\ and\ \citenamefont {Biraben}}]{Grynberg1980}%
  \BibitemOpen
  \bibfield  {author} {\bibinfo {author} {\bibfnamefont {G.}~\bibnamefont {Grynberg}}, \bibinfo {author} {\bibfnamefont {B.}~\bibnamefont {Cagnac}},\ and\ \bibinfo {author} {\bibfnamefont {F.}~\bibnamefont {Biraben}},\ }\bibfield  {title} {\emph {\bibinfo {title} {{Multiphoton Resonant Processes in Atoms}}},\ }in\ \href {https://doi.org/10.1007/978-3-642-81495-2_4} {\emph {\bibinfo {booktitle} {Coherent Nonlinear Optics: Recent Advances}}}\ (\bibinfo  {publisher} {Springer},\ \bibinfo {year} {1980})\ pp.\ \bibinfo {pages} {111--164}\BibitemShut {NoStop}%
\bibitem [{\citenamefont {Panelli}\ \emph {et~al.}(2025)\citenamefont {Panelli}, \citenamefont {Burd}, \citenamefont {Porter},\ and\ \citenamefont {Kasevich}}]{Panelli2024}%
  \BibitemOpen
  \bibfield  {author} {\bibinfo {author} {\bibfnamefont {G.}~\bibnamefont {Panelli}}, \bibinfo {author} {\bibfnamefont {S.~C.}\ \bibnamefont {Burd}}, \bibinfo {author} {\bibfnamefont {E.~J.}\ \bibnamefont {Porter}},\ and\ \bibinfo {author} {\bibfnamefont {M.}~\bibnamefont {Kasevich}},\ }\bibfield  {title} {\emph {\bibinfo {title} {Doppler-free three-photon spectroscopy on narrow-line optical transitions}},\ }\href {https://doi.org/10.1103/PhysRevA.111.033112} {\bibfield  {journal} {\bibinfo  {journal} {Phys. Rev. A}\ }\textbf {\bibinfo {volume} {111}},\ \bibinfo {pages} {033112} (\bibinfo {year} {2025})}\BibitemShut {NoStop}%
\bibitem [{\citenamefont {Madjarov}\ \emph {et~al.}(2020)\citenamefont {Madjarov}, \citenamefont {Covey}, \citenamefont {Shaw}, \citenamefont {Choi}, \citenamefont {Kale}, \citenamefont {Cooper}, \citenamefont {Pichler}, \citenamefont {Schkolnik}, \citenamefont {Williams},\ and\ \citenamefont {Endres}}]{Madjarov2020}%
  \BibitemOpen
  \bibfield  {author} {\bibinfo {author} {\bibfnamefont {I.~S.}\ \bibnamefont {Madjarov}}, \bibinfo {author} {\bibfnamefont {J.~P.}\ \bibnamefont {Covey}}, \bibinfo {author} {\bibfnamefont {A.~L.}\ \bibnamefont {Shaw}}, \bibinfo {author} {\bibfnamefont {J.}~\bibnamefont {Choi}}, \bibinfo {author} {\bibfnamefont {A.}~\bibnamefont {Kale}}, \bibinfo {author} {\bibfnamefont {A.}~\bibnamefont {Cooper}}, \bibinfo {author} {\bibfnamefont {H.}~\bibnamefont {Pichler}}, \bibinfo {author} {\bibfnamefont {V.}~\bibnamefont {Schkolnik}}, \bibinfo {author} {\bibfnamefont {J.~R.}\ \bibnamefont {Williams}},\ and\ \bibinfo {author} {\bibfnamefont {M.}~\bibnamefont {Endres}},\ }\bibfield  {title} {\emph {\bibinfo {title} {{High-fidelity entanglement and detection of alkaline-earth Rydberg atoms}}},\ }\href {https://doi.org/10.1038/s41567-020-0903-z} {\bibfield  {journal} {\bibinfo  {journal} {Nat. Phys.}\ }\textbf {\bibinfo {volume} {16}},\ \bibinfo {pages} {857} (\bibinfo {year} {2020})}\BibitemShut {NoStop}%
\bibitem [{\citenamefont {Schine}\ \emph {et~al.}(2022)\citenamefont {Schine}, \citenamefont {Young}, \citenamefont {Eckner}, \citenamefont {Martin},\ and\ \citenamefont {Kaufman}}]{Schine2022}%
  \BibitemOpen
  \bibfield  {author} {\bibinfo {author} {\bibfnamefont {N.}~\bibnamefont {Schine}}, \bibinfo {author} {\bibfnamefont {A.}~\bibnamefont {Young}}, \bibinfo {author} {\bibfnamefont {W.}~\bibnamefont {Eckner}}, \bibinfo {author} {\bibfnamefont {M.}~\bibnamefont {Martin}},\ and\ \bibinfo {author} {\bibfnamefont {A.}~\bibnamefont {Kaufman}},\ }\bibfield  {title} {\emph {\bibinfo {title} {{Long-lived Bell states in an array of optical clock qubits}}},\ }\href {https://doi.org/10.1038/s41567-022-01678-w} {\bibfield  {journal} {\bibinfo  {journal} {Nat. Phys.}\ }\textbf {\bibinfo {volume} {18}} (\bibinfo {year} {2022})}\BibitemShut {NoStop}%
\bibitem [{\citenamefont {Eckner}\ \emph {et~al.}(2023)\citenamefont {Eckner}, \citenamefont {Oppong}, \citenamefont {Cao}, \citenamefont {Young}, \citenamefont {Milner}, \citenamefont {Robinson}, \citenamefont {Ye},\ and\ \citenamefont {Kaufman}}]{Eckner2023}%
  \BibitemOpen
  \bibfield  {author} {\bibinfo {author} {\bibfnamefont {W.}~\bibnamefont {Eckner}}, \bibinfo {author} {\bibfnamefont {N.}~\bibnamefont {Oppong}}, \bibinfo {author} {\bibfnamefont {A.}~\bibnamefont {Cao}}, \bibinfo {author} {\bibfnamefont {A.}~\bibnamefont {Young}}, \bibinfo {author} {\bibfnamefont {W.}~\bibnamefont {Milner}}, \bibinfo {author} {\bibfnamefont {J.}~\bibnamefont {Robinson}}, \bibinfo {author} {\bibfnamefont {J.}~\bibnamefont {Ye}},\ and\ \bibinfo {author} {\bibfnamefont {A.}~\bibnamefont {Kaufman}},\ }\bibfield  {title} {\emph {\bibinfo {title} {{Realizing spin squeezing with Rydberg interactions in an optical clock}}},\ }\href {https://doi.org/10.1038/s41586-023-06360-6} {\bibfield  {journal} {\bibinfo  {journal} {Nature}\ }\textbf {\bibinfo {volume} {621}},\ \bibinfo {pages} {734--739} (\bibinfo {year} {2023})}\BibitemShut {NoStop}%
\bibitem [{\citenamefont {Cao}\ \emph {et~al.}(2024)\citenamefont {Cao}, \citenamefont {Eckner}, \citenamefont {Yelin}, \citenamefont {Young}, \citenamefont {Jandura}, \citenamefont {Yan}, \citenamefont {Kim}, \citenamefont {Pupillo}, \citenamefont {Ye}, \citenamefont {Oppong},\ and\ \citenamefont {Kaufman}}]{Cao2024}%
  \BibitemOpen
  \bibfield  {author} {\bibinfo {author} {\bibfnamefont {A.}~\bibnamefont {Cao}}, \bibinfo {author} {\bibfnamefont {W.~J.}\ \bibnamefont {Eckner}}, \bibinfo {author} {\bibfnamefont {T.~L.}\ \bibnamefont {Yelin}}, \bibinfo {author} {\bibfnamefont {A.~W.}\ \bibnamefont {Young}}, \bibinfo {author} {\bibfnamefont {S.}~\bibnamefont {Jandura}}, \bibinfo {author} {\bibfnamefont {L.}~\bibnamefont {Yan}}, \bibinfo {author} {\bibfnamefont {K.}~\bibnamefont {Kim}}, \bibinfo {author} {\bibfnamefont {G.}~\bibnamefont {Pupillo}}, \bibinfo {author} {\bibfnamefont {J.}~\bibnamefont {Ye}}, \bibinfo {author} {\bibfnamefont {N.~D.}\ \bibnamefont {Oppong}},\ and\ \bibinfo {author} {\bibfnamefont {A.~M.}\ \bibnamefont {Kaufman}},\ }\bibfield  {title} {\emph {\bibinfo {title} {{Multi-qubit gates and 'Schr\"odinger cat' states in an optical clock}}},\ }\href {https://doi.org/10.1038/s41586-024-07913-z} {\bibfield  {journal} {\bibinfo  {journal} {Nature}\ }\textbf {\bibinfo {volume} {634}},\ \bibinfo {pages} {315--320} (\bibinfo
  {year} {2024})}\BibitemShut {NoStop}%
\bibitem [{\citenamefont {Hinkley}\ \emph {et~al.}(2013)\citenamefont {Hinkley}, \citenamefont {Sherman}, \citenamefont {Phillips}, \citenamefont {Schioppo}, \citenamefont {Lemke}, \citenamefont {Beloy}, \citenamefont {Pizzocaro}, \citenamefont {Oates},\ and\ \citenamefont {Ludlow}}]{Hinkley2013}%
  \BibitemOpen
  \bibfield  {author} {\bibinfo {author} {\bibfnamefont {N.}~\bibnamefont {Hinkley}}, \bibinfo {author} {\bibfnamefont {J.~A.}\ \bibnamefont {Sherman}}, \bibinfo {author} {\bibfnamefont {N.~B.}\ \bibnamefont {Phillips}}, \bibinfo {author} {\bibfnamefont {M.}~\bibnamefont {Schioppo}}, \bibinfo {author} {\bibfnamefont {N.~D.}\ \bibnamefont {Lemke}}, \bibinfo {author} {\bibfnamefont {K.}~\bibnamefont {Beloy}}, \bibinfo {author} {\bibfnamefont {M.}~\bibnamefont {Pizzocaro}}, \bibinfo {author} {\bibfnamefont {C.~W.}\ \bibnamefont {Oates}},\ and\ \bibinfo {author} {\bibfnamefont {A.~D.}\ \bibnamefont {Ludlow}},\ }\bibfield  {title} {\emph {\bibinfo {title} {{An Atomic Clock with $10^{-18}$ Instability}}},\ }\href {https://doi.org/10.1126/science.1240420} {\bibfield  {journal} {\bibinfo  {journal} {Science}\ }\textbf {\bibinfo {volume} {341}},\ \bibinfo {pages} {1215--1218} (\bibinfo {year} {2013})}\BibitemShut {NoStop}%
\bibitem [{\citenamefont {Kolkowitz}\ \emph {et~al.}(2016)\citenamefont {Kolkowitz}, \citenamefont {Pikovski}, \citenamefont {Langellier}, \citenamefont {Lukin}, \citenamefont {Walsworth},\ and\ \citenamefont {Ye}}]{Kolkowitz2016}%
  \BibitemOpen
  \bibfield  {author} {\bibinfo {author} {\bibfnamefont {S.}~\bibnamefont {Kolkowitz}}, \bibinfo {author} {\bibfnamefont {I.}~\bibnamefont {Pikovski}}, \bibinfo {author} {\bibfnamefont {N.}~\bibnamefont {Langellier}}, \bibinfo {author} {\bibfnamefont {M.~D.}\ \bibnamefont {Lukin}}, \bibinfo {author} {\bibfnamefont {R.~L.}\ \bibnamefont {Walsworth}},\ and\ \bibinfo {author} {\bibfnamefont {J.}~\bibnamefont {Ye}},\ }\bibfield  {title} {\emph {\bibinfo {title} {{Gravitational wave detection with optical lattice atomic clocks}}},\ }\href {https://doi.org/10.1103/PhysRevD.94.124043} {\bibfield  {journal} {\bibinfo  {journal} {Phys. Rev. D}\ }\textbf {\bibinfo {volume} {94}},\ \bibinfo {pages} {124043} (\bibinfo {year} {2016})}\BibitemShut {NoStop}%
\bibitem [{\citenamefont {Schioppo}\ \emph {et~al.}(2016)\citenamefont {Schioppo}, \citenamefont {Brown}, \citenamefont {McGrew}, \citenamefont {Hinkley}, \citenamefont {Fasano}, \citenamefont {Beloy}, \citenamefont {Yoon}, \citenamefont {Milani}, \citenamefont {Nicolodi}, \citenamefont {Sherman}, \citenamefont {Phillips}, \citenamefont {Oates},\ and\ \citenamefont {Ludlow}}]{Schioppo2016}%
  \BibitemOpen
  \bibfield  {author} {\bibinfo {author} {\bibfnamefont {M.}~\bibnamefont {Schioppo}}, \bibinfo {author} {\bibfnamefont {R.}~\bibnamefont {Brown}}, \bibinfo {author} {\bibfnamefont {W.}~\bibnamefont {McGrew}}, \bibinfo {author} {\bibfnamefont {N.}~\bibnamefont {Hinkley}}, \bibinfo {author} {\bibfnamefont {R.}~\bibnamefont {Fasano}}, \bibinfo {author} {\bibfnamefont {K.}~\bibnamefont {Beloy}}, \bibinfo {author} {\bibfnamefont {T.~H.}\ \bibnamefont {Yoon}}, \bibinfo {author} {\bibfnamefont {G.}~\bibnamefont {Milani}}, \bibinfo {author} {\bibfnamefont {D.}~\bibnamefont {Nicolodi}}, \bibinfo {author} {\bibfnamefont {J.}~\bibnamefont {Sherman}}, \bibinfo {author} {\bibfnamefont {N.}~\bibnamefont {Phillips}}, \bibinfo {author} {\bibfnamefont {C.}~\bibnamefont {Oates}},\ and\ \bibinfo {author} {\bibfnamefont {A.}~\bibnamefont {Ludlow}},\ }\bibfield  {title} {\emph {\bibinfo {title} {{Ultra-stable optical clock with two cold-atom ensembles}}},\ }\href {https://doi.org/10.1038/nphoton.2016.231} {\bibfield  {journal}
  {\bibinfo  {journal} {Nat. Photon.}\ }\textbf {\bibinfo {volume} {11}},\ \bibinfo {pages} {48--52} (\bibinfo {year} {2016})}\BibitemShut {NoStop}%
\bibitem [{\citenamefont {Oelker}\ \emph {et~al.}(2019)\citenamefont {Oelker}, \citenamefont {Hutson}, \citenamefont {Kennedy}, \citenamefont {Sonderhouse}, \citenamefont {Bothwell}, \citenamefont {Goban}, \citenamefont {Kedar}, \citenamefont {Sanner}, \citenamefont {Robinson}, \citenamefont {Marti}, \citenamefont {Matei}, \citenamefont {Legero}, \citenamefont {Giunta}, \citenamefont {Holzwarth}, \citenamefont {Riehle}, \citenamefont {Sterr},\ and\ \citenamefont {Ye}}]{Oelker2019}%
  \BibitemOpen
  \bibfield  {author} {\bibinfo {author} {\bibfnamefont {E.}~\bibnamefont {Oelker}}, \bibinfo {author} {\bibfnamefont {R.}~\bibnamefont {Hutson}}, \bibinfo {author} {\bibfnamefont {C.}~\bibnamefont {Kennedy}}, \bibinfo {author} {\bibfnamefont {L.}~\bibnamefont {Sonderhouse}}, \bibinfo {author} {\bibfnamefont {T.}~\bibnamefont {Bothwell}}, \bibinfo {author} {\bibfnamefont {A.}~\bibnamefont {Goban}}, \bibinfo {author} {\bibfnamefont {D.}~\bibnamefont {Kedar}}, \bibinfo {author} {\bibfnamefont {C.}~\bibnamefont {Sanner}}, \bibinfo {author} {\bibfnamefont {J.}~\bibnamefont {Robinson}}, \bibinfo {author} {\bibfnamefont {G.}~\bibnamefont {Marti}}, \bibinfo {author} {\bibfnamefont {D.}~\bibnamefont {Matei}}, \bibinfo {author} {\bibfnamefont {T.}~\bibnamefont {Legero}}, \bibinfo {author} {\bibfnamefont {M.}~\bibnamefont {Giunta}}, \bibinfo {author} {\bibfnamefont {R.}~\bibnamefont {Holzwarth}}, \bibinfo {author} {\bibfnamefont {F.}~\bibnamefont {Riehle}}, \bibinfo {author} {\bibfnamefont {U.}~\bibnamefont {Sterr}},\
  and\ \bibinfo {author} {\bibfnamefont {J.}~\bibnamefont {Ye}},\ }\bibfield  {title} {\emph {\bibinfo {title} {{Demonstration of $4.8 \times 10^{-17}$ stability at $1~\mathrm{s}$ for two independent optical clocks}}},\ }\href {https://doi.org/10.1038/s41566-019-0493-4} {\bibfield  {journal} {\bibinfo  {journal} {Nat. Photon.}\ }\textbf {\bibinfo {volume} {13}} (\bibinfo {year} {2019})}\BibitemShut {NoStop}%
\bibitem [{\citenamefont {{Boulder Atomic Clock Optical Network (BACON) Collaboration}}(2021)}]{BACON2021}%
  \BibitemOpen
  \bibfield  {author} {\bibinfo {author} {\bibnamefont {{Boulder Atomic Clock Optical Network (BACON) Collaboration}}},\ }\bibfield  {title} {\emph {\bibinfo {title} {Frequency ratio measurements at 18-digit accuracy using an optical clock network}},\ }\href {https://doi.org/10.1038/s41586-021-03253-4} {\bibfield  {journal} {\bibinfo  {journal} {Nature}\ }\textbf {\bibinfo {volume} {591}},\ \bibinfo {pages} {564--569} (\bibinfo {year} {2021})}\BibitemShut {NoStop}%
\bibitem [{\citenamefont {Zheng}\ \emph {et~al.}(2022)\citenamefont {Zheng}, \citenamefont {Dolde}, \citenamefont {Lochab}, \citenamefont {Merriman}, \citenamefont {Li},\ and\ \citenamefont {Kolkowitz}}]{Zheng2022}%
  \BibitemOpen
  \bibfield  {author} {\bibinfo {author} {\bibfnamefont {X.}~\bibnamefont {Zheng}}, \bibinfo {author} {\bibfnamefont {J.}~\bibnamefont {Dolde}}, \bibinfo {author} {\bibfnamefont {V.}~\bibnamefont {Lochab}}, \bibinfo {author} {\bibfnamefont {B.}~\bibnamefont {Merriman}}, \bibinfo {author} {\bibfnamefont {H.}~\bibnamefont {Li}},\ and\ \bibinfo {author} {\bibfnamefont {S.}~\bibnamefont {Kolkowitz}},\ }\bibfield  {title} {\emph {\bibinfo {title} {{Differential clock comparisons with a multiplexed optical lattice clock}}},\ }\href {https://doi.org/10.1038/s41586-021-04344-y} {\bibfield  {journal} {\bibinfo  {journal} {Nature}\ }\textbf {\bibinfo {volume} {602}},\ \bibinfo {pages} {425--430} (\bibinfo {year} {2022})}\BibitemShut {NoStop}%
\bibitem [{\citenamefont {Hu}\ \emph {et~al.}(2017)\citenamefont {Hu}, \citenamefont {Poli}, \citenamefont {Salvi},\ and\ \citenamefont {Tino}}]{Hu2017}%
  \BibitemOpen
  \bibfield  {author} {\bibinfo {author} {\bibfnamefont {L.}~\bibnamefont {Hu}}, \bibinfo {author} {\bibfnamefont {N.}~\bibnamefont {Poli}}, \bibinfo {author} {\bibfnamefont {L.}~\bibnamefont {Salvi}},\ and\ \bibinfo {author} {\bibfnamefont {G.~M.}\ \bibnamefont {Tino}},\ }\bibfield  {title} {\emph {\bibinfo {title} {{Atom Interferometry with the $\mathrm{Sr}$ Optical Clock Transition}}},\ }\href {https://doi.org/10.1103/PhysRevLett.119.263601} {\bibfield  {journal} {\bibinfo  {journal} {Phys. Rev. Lett.}\ }\textbf {\bibinfo {volume} {119}},\ \bibinfo {pages} {263601} (\bibinfo {year} {2017})}\BibitemShut {NoStop}%
\bibitem [{\citenamefont {Abe}\ \emph {et~al.}(2021)\citenamefont {Abe}, \citenamefont {Adamson}, \citenamefont {Borcean}, \citenamefont {Bortoletto}, \citenamefont {Bridges}, \citenamefont {Carman}, \citenamefont {Chattopadhyay}, \citenamefont {Coleman}, \citenamefont {Curfman}, \citenamefont {DeRose}, \citenamefont {Deshpande}, \citenamefont {Dimopoulos}, \citenamefont {Foot}, \citenamefont {Frisch}, \citenamefont {Garber}, \citenamefont {Geer}, \citenamefont {Gibson}, \citenamefont {Glick}, \citenamefont {Graham}, \citenamefont {Hahn}, \citenamefont {Harnik}, \citenamefont {Hawkins}, \citenamefont {Hindley}, \citenamefont {Hogan}, \citenamefont {Jiang}, \citenamefont {Kasevich}, \citenamefont {Kellett}, \citenamefont {Kiburg}, \citenamefont {Kovachy}, \citenamefont {Lykken}, \citenamefont {March-Russell}, \citenamefont {Mitchell}, \citenamefont {Murphy}, \citenamefont {Nantel}, \citenamefont {Nobrega}, \citenamefont {Plunkett}, \citenamefont {Rajendran}, \citenamefont {Rudolph}, \citenamefont {Sachdeva},
  \citenamefont {Safdari}, \citenamefont {Santucci}, \citenamefont {Schwartzman}, \citenamefont {Shipsey}, \citenamefont {Swan}, \citenamefont {Valerio}, \citenamefont {Vasonis}, \citenamefont {Wang},\ and\ \citenamefont {Wilkason}}]{Abe2021}%
  \BibitemOpen
  \bibfield  {author} {\bibinfo {author} {\bibfnamefont {M.}~\bibnamefont {Abe}}, \bibinfo {author} {\bibfnamefont {P.}~\bibnamefont {Adamson}}, \bibinfo {author} {\bibfnamefont {M.}~\bibnamefont {Borcean}}, \bibinfo {author} {\bibfnamefont {D.}~\bibnamefont {Bortoletto}}, \bibinfo {author} {\bibfnamefont {K.}~\bibnamefont {Bridges}}, \bibinfo {author} {\bibfnamefont {S.~P.}\ \bibnamefont {Carman}}, \bibinfo {author} {\bibfnamefont {S.}~\bibnamefont {Chattopadhyay}}, \bibinfo {author} {\bibfnamefont {J.}~\bibnamefont {Coleman}}, \bibinfo {author} {\bibfnamefont {N.~M.}\ \bibnamefont {Curfman}}, \bibinfo {author} {\bibfnamefont {K.}~\bibnamefont {DeRose}}, \bibinfo {author} {\bibfnamefont {T.}~\bibnamefont {Deshpande}}, \bibinfo {author} {\bibfnamefont {S.}~\bibnamefont {Dimopoulos}}, \bibinfo {author} {\bibfnamefont {C.~J.}\ \bibnamefont {Foot}}, \bibinfo {author} {\bibfnamefont {J.~C.}\ \bibnamefont {Frisch}}, \bibinfo {author} {\bibfnamefont {B.~E.}\ \bibnamefont {Garber}}, \bibinfo {author} {\bibfnamefont
  {S.}~\bibnamefont {Geer}}, \bibinfo {author} {\bibfnamefont {V.}~\bibnamefont {Gibson}}, \bibinfo {author} {\bibfnamefont {J.}~\bibnamefont {Glick}}, \bibinfo {author} {\bibfnamefont {P.~W.}\ \bibnamefont {Graham}}, \bibinfo {author} {\bibfnamefont {S.~R.}\ \bibnamefont {Hahn}}, \bibinfo {author} {\bibfnamefont {R.}~\bibnamefont {Harnik}}, \bibinfo {author} {\bibfnamefont {L.}~\bibnamefont {Hawkins}}, \bibinfo {author} {\bibfnamefont {S.}~\bibnamefont {Hindley}}, \bibinfo {author} {\bibfnamefont {J.~M.}\ \bibnamefont {Hogan}}, \bibinfo {author} {\bibfnamefont {Y.}~\bibnamefont {Jiang}}, \bibinfo {author} {\bibfnamefont {M.~A.}\ \bibnamefont {Kasevich}}, \bibinfo {author} {\bibfnamefont {R.~J.}\ \bibnamefont {Kellett}}, \bibinfo {author} {\bibfnamefont {M.}~\bibnamefont {Kiburg}}, \bibinfo {author} {\bibfnamefont {T.}~\bibnamefont {Kovachy}}, \bibinfo {author} {\bibfnamefont {J.~D.}\ \bibnamefont {Lykken}}, \bibinfo {author} {\bibfnamefont {J.}~\bibnamefont {March-Russell}}, \bibinfo {author} {\bibfnamefont
  {J.}~\bibnamefont {Mitchell}}, \bibinfo {author} {\bibfnamefont {M.}~\bibnamefont {Murphy}}, \bibinfo {author} {\bibfnamefont {M.}~\bibnamefont {Nantel}}, \bibinfo {author} {\bibfnamefont {L.~E.}\ \bibnamefont {Nobrega}}, \bibinfo {author} {\bibfnamefont {R.~K.}\ \bibnamefont {Plunkett}}, \bibinfo {author} {\bibfnamefont {S.}~\bibnamefont {Rajendran}}, \bibinfo {author} {\bibfnamefont {J.}~\bibnamefont {Rudolph}}, \bibinfo {author} {\bibfnamefont {N.}~\bibnamefont {Sachdeva}}, \bibinfo {author} {\bibfnamefont {M.}~\bibnamefont {Safdari}}, \bibinfo {author} {\bibfnamefont {J.~K.}\ \bibnamefont {Santucci}}, \bibinfo {author} {\bibfnamefont {A.~G.}\ \bibnamefont {Schwartzman}}, \bibinfo {author} {\bibfnamefont {I.}~\bibnamefont {Shipsey}}, \bibinfo {author} {\bibfnamefont {H.}~\bibnamefont {Swan}}, \bibinfo {author} {\bibfnamefont {L.~R.}\ \bibnamefont {Valerio}}, \bibinfo {author} {\bibfnamefont {A.}~\bibnamefont {Vasonis}}, \bibinfo {author} {\bibfnamefont {Y.}~\bibnamefont {Wang}},\ and\ \bibinfo {author}
  {\bibfnamefont {T.}~\bibnamefont {Wilkason}},\ }\bibfield  {title} {\emph {\bibinfo {title} {{Matter-wave Atomic Gradiometer Interferometric Sensor (MAGIS-100)}}},\ }\href {https://doi.org/10.1088/2058-9565/abf719} {\bibfield  {journal} {\bibinfo  {journal} {Quantum Sci. Technol.}\ }\textbf {\bibinfo {volume} {6}},\ \bibinfo {pages} {044003} (\bibinfo {year} {2021})}\BibitemShut {NoStop}%
\bibitem [{\citenamefont {Graham}\ \emph {et~al.}(2013)\citenamefont {Graham}, \citenamefont {Hogan}, \citenamefont {Kasevich},\ and\ \citenamefont {Rajendran}}]{Graham2013}%
  \BibitemOpen
  \bibfield  {author} {\bibinfo {author} {\bibfnamefont {P.~W.}\ \bibnamefont {Graham}}, \bibinfo {author} {\bibfnamefont {J.~M.}\ \bibnamefont {Hogan}}, \bibinfo {author} {\bibfnamefont {M.~A.}\ \bibnamefont {Kasevich}},\ and\ \bibinfo {author} {\bibfnamefont {S.}~\bibnamefont {Rajendran}},\ }\bibfield  {title} {\emph {\bibinfo {title} {{New Method for Gravitational Wave Detection with Atomic Sensors}}},\ }\href {https://doi.org/10.1103/PhysRevLett.110.171102} {\bibfield  {journal} {\bibinfo  {journal} {Phys. Rev. Lett.}\ }\textbf {\bibinfo {volume} {110}},\ \bibinfo {pages} {171102} (\bibinfo {year} {2013})}\BibitemShut {NoStop}%
\bibitem [{Note1()}]{Note1}%
  \BibitemOpen
  \bibinfo {note} {Though we only show the case where the first photon is $\protect \ensuremath {\protect \bm {\protect \mathrm {\protect \hat {x}}}}$ polarized, the requirement for noncoplanar polarizations is independent of the choice of the first polarization for any $\protect \ensuremath {\protect \bm {\protect \mathrm {e}}}^{(1)}\perp \protect \ensuremath {\protect \bm {\protect \mathrm {B}}}$.}\BibitemShut {Stop}%
\bibitem [{\citenamefont {{Kasevich~}}\ and\ \citenamefont {{Chu}}(1991)}]{Kasevich1991}%
  \BibitemOpen
  \bibfield  {author} {\bibinfo {author} {\bibfnamefont {M.~A.}\ \bibnamefont {{Kasevich~}}}\ and\ \bibinfo {author} {\bibfnamefont {S.}~\bibnamefont {{Chu}}},\ }\bibfield  {title} {\emph {\bibinfo {title} {Atomic interferometry using stimulated {Raman} transitions}},\ }\href {https://doi.org/10.1103/PhysRevLett.67.181} {\bibfield  {journal} {\bibinfo  {journal} {Phys. Rev. Lett.}\ }\textbf {\bibinfo {volume} {67}},\ \bibinfo {pages} {181--184} (\bibinfo {year} {1991})}\BibitemShut {NoStop}%
\bibitem [{\citenamefont {Hu}\ \emph {et~al.}(2019)\citenamefont {Hu}, \citenamefont {Wang}, \citenamefont {Salvi}, \citenamefont {Tinsley}, \citenamefont {Tino},\ and\ \citenamefont {Poli}}]{Hu2019}%
  \BibitemOpen
  \bibfield  {author} {\bibinfo {author} {\bibfnamefont {L.}~\bibnamefont {Hu}}, \bibinfo {author} {\bibfnamefont {E.}~\bibnamefont {Wang}}, \bibinfo {author} {\bibfnamefont {L.}~\bibnamefont {Salvi}}, \bibinfo {author} {\bibfnamefont {J.~N.}\ \bibnamefont {Tinsley}}, \bibinfo {author} {\bibfnamefont {G.~M.}\ \bibnamefont {Tino}},\ and\ \bibinfo {author} {\bibfnamefont {N.}~\bibnamefont {Poli}},\ }\bibfield  {title} {\emph {\bibinfo {title} {{$\mathrm{Sr}$ atom interferometry with the optical clock transition as a gravimeter and a gravity gradiometer}}},\ }\href {https://doi.org/10.1088/1361-6382/ab4d18} {\bibfield  {journal} {\bibinfo  {journal} {Class. Quantum Grav.}\ }\textbf {\bibinfo {volume} {37}},\ \bibinfo {pages} {014001} (\bibinfo {year} {2019})}\BibitemShut {NoStop}%
\bibitem [{\citenamefont {Hartwig}\ \emph {et~al.}(2015)\citenamefont {Hartwig}, \citenamefont {Abend}, \citenamefont {Schubert}, \citenamefont {Schlippert}, \citenamefont {Ahlers}, \citenamefont {Posso-Trujillo}, \citenamefont {Gaaloul}, \citenamefont {Ertmer},\ and\ \citenamefont {Rasel}}]{Hartwig2015}%
  \BibitemOpen
  \bibfield  {author} {\bibinfo {author} {\bibfnamefont {J.}~\bibnamefont {Hartwig}}, \bibinfo {author} {\bibfnamefont {S.}~\bibnamefont {Abend}}, \bibinfo {author} {\bibfnamefont {C.}~\bibnamefont {Schubert}}, \bibinfo {author} {\bibfnamefont {D.}~\bibnamefont {Schlippert}}, \bibinfo {author} {\bibfnamefont {H.}~\bibnamefont {Ahlers}}, \bibinfo {author} {\bibfnamefont {K.}~\bibnamefont {Posso-Trujillo}}, \bibinfo {author} {\bibfnamefont {N.}~\bibnamefont {Gaaloul}}, \bibinfo {author} {\bibfnamefont {W.}~\bibnamefont {Ertmer}},\ and\ \bibinfo {author} {\bibfnamefont {E.~M.}\ \bibnamefont {Rasel}},\ }\bibfield  {title} {\emph {\bibinfo {title} {Testing the universality of free fall with rubidium and ytterbium in a very large baseline atom interferometer}},\ }\href {https://doi.org/10.1088/1367-2630/17/3/035011} {\bibfield  {journal} {\bibinfo  {journal} {New J. Phys.}\ }\textbf {\bibinfo {volume} {17}},\ \bibinfo {pages} {035011} (\bibinfo {year} {2015})}\BibitemShut {NoStop}%
\bibitem [{\citenamefont {Kovachy}\ \emph {et~al.}(2015)\citenamefont {Kovachy}, \citenamefont {Asenbaum}, \citenamefont {Overstreet}, \citenamefont {Donnelly}, \citenamefont {Dickerson}, \citenamefont {Sugarbaker}, \citenamefont {Hogan},\ and\ \citenamefont {Kasevich}}]{Kovachy2015}%
  \BibitemOpen
  \bibfield  {author} {\bibinfo {author} {\bibfnamefont {T.}~\bibnamefont {Kovachy}}, \bibinfo {author} {\bibfnamefont {P.}~\bibnamefont {Asenbaum}}, \bibinfo {author} {\bibfnamefont {C.}~\bibnamefont {Overstreet}}, \bibinfo {author} {\bibfnamefont {C.~A.}\ \bibnamefont {Donnelly}}, \bibinfo {author} {\bibfnamefont {S.~M.}\ \bibnamefont {Dickerson}}, \bibinfo {author} {\bibfnamefont {A.}~\bibnamefont {Sugarbaker}}, \bibinfo {author} {\bibfnamefont {J.~M.}\ \bibnamefont {Hogan}},\ and\ \bibinfo {author} {\bibfnamefont {M.~A.}\ \bibnamefont {Kasevich}},\ }\bibfield  {title} {\emph {\bibinfo {title} {Quantum superposition at the half-metre scale}},\ }\href {https://www.nature.com/articles/nature16155.pdf} {\bibfield  {journal} {\bibinfo  {journal} {Nature}\ }\textbf {\bibinfo {volume} {528}},\ \bibinfo {pages} {530} (\bibinfo {year} {2015})}\BibitemShut {NoStop}%
\bibitem [{\citenamefont {Muniz}\ \emph {et~al.}(2021)\citenamefont {Muniz}, \citenamefont {Young}, \citenamefont {Cline},\ and\ \citenamefont {Thompson}}]{Muniz2021}%
  \BibitemOpen
  \bibfield  {author} {\bibinfo {author} {\bibfnamefont {J.~A.}\ \bibnamefont {Muniz}}, \bibinfo {author} {\bibfnamefont {D.~J.}\ \bibnamefont {Young}}, \bibinfo {author} {\bibfnamefont {J.~R.~K.}\ \bibnamefont {Cline}},\ and\ \bibinfo {author} {\bibfnamefont {J.~K.}\ \bibnamefont {Thompson}},\ }\bibfield  {title} {\emph {\bibinfo {title} {{Cavity-QED measurements of the $^{87}\mathrm{Sr}$ millihertz optical clock transition and determination of its natural linewidth}}},\ }\href {https://doi.org/10.1103/PhysRevResearch.3.023152} {\bibfield  {journal} {\bibinfo  {journal} {Phys. Rev. Res.}\ }\textbf {\bibinfo {volume} {3}},\ \bibinfo {pages} {023152} (\bibinfo {year} {2021})}\BibitemShut {NoStop}%
\bibitem [{\citenamefont {Xu}\ \emph {et~al.}(2019)\citenamefont {Xu}, \citenamefont {Jaffe}, \citenamefont {Panda}, \citenamefont {Kristensen}, \citenamefont {Clark},\ and\ \citenamefont {M{\"u}ller}}]{Xu2019}%
  \BibitemOpen
  \bibfield  {author} {\bibinfo {author} {\bibfnamefont {V.}~\bibnamefont {Xu}}, \bibinfo {author} {\bibfnamefont {M.}~\bibnamefont {Jaffe}}, \bibinfo {author} {\bibfnamefont {C.~D.}\ \bibnamefont {Panda}}, \bibinfo {author} {\bibfnamefont {S.~L.}\ \bibnamefont {Kristensen}}, \bibinfo {author} {\bibfnamefont {L.~W.}\ \bibnamefont {Clark}},\ and\ \bibinfo {author} {\bibfnamefont {H.}~\bibnamefont {M{\"u}ller}},\ }\bibfield  {title} {\emph {\bibinfo {title} {{Probing gravity by holding atoms for 20 seconds}}},\ }\href {https://doi.org/10.1126/science.aay6428} {\bibfield  {journal} {\bibinfo  {journal} {Science}\ }\textbf {\bibinfo {volume} {366}},\ \bibinfo {pages} {745--749} (\bibinfo {year} {2019})}\BibitemShut {NoStop}%
\bibitem [{\citenamefont {Wilkason}\ \emph {et~al.}(2022)\citenamefont {Wilkason}, \citenamefont {Nantel}, \citenamefont {Rudolph}, \citenamefont {Jiang}, \citenamefont {Garber}, \citenamefont {Swan}, \citenamefont {Carman}, \citenamefont {Abe},\ and\ \citenamefont {Hogan}}]{Wilkason2022}%
  \BibitemOpen
  \bibfield  {author} {\bibinfo {author} {\bibfnamefont {T.}~\bibnamefont {Wilkason}}, \bibinfo {author} {\bibfnamefont {M.}~\bibnamefont {Nantel}}, \bibinfo {author} {\bibfnamefont {J.}~\bibnamefont {Rudolph}}, \bibinfo {author} {\bibfnamefont {Y.}~\bibnamefont {Jiang}}, \bibinfo {author} {\bibfnamefont {B.~E.}\ \bibnamefont {Garber}}, \bibinfo {author} {\bibfnamefont {H.}~\bibnamefont {Swan}}, \bibinfo {author} {\bibfnamefont {S.~P.}\ \bibnamefont {Carman}}, \bibinfo {author} {\bibfnamefont {M.}~\bibnamefont {Abe}},\ and\ \bibinfo {author} {\bibfnamefont {J.~M.}\ \bibnamefont {Hogan}},\ }\bibfield  {title} {\emph {\bibinfo {title} {{Atom Interferometry with Floquet Atom Optics}}},\ }\href {https://doi.org/10.1103/PhysRevLett.129.183202} {\bibfield  {journal} {\bibinfo  {journal} {Phys. Rev. Lett.}\ }\textbf {\bibinfo {volume} {129}},\ \bibinfo {pages} {183202} (\bibinfo {year} {2022})}\BibitemShut {NoStop}%
\bibitem [{\citenamefont {Rudolph}\ \emph {et~al.}(2020)\citenamefont {Rudolph}, \citenamefont {Wilkason}, \citenamefont {Nantel}, \citenamefont {Swan}, \citenamefont {Holland}, \citenamefont {Jiang}, \citenamefont {Garber}, \citenamefont {Carman},\ and\ \citenamefont {Hogan}}]{Rudolph2020}%
  \BibitemOpen
  \bibfield  {author} {\bibinfo {author} {\bibfnamefont {J.}~\bibnamefont {Rudolph}}, \bibinfo {author} {\bibfnamefont {T.}~\bibnamefont {Wilkason}}, \bibinfo {author} {\bibfnamefont {M.}~\bibnamefont {Nantel}}, \bibinfo {author} {\bibfnamefont {H.}~\bibnamefont {Swan}}, \bibinfo {author} {\bibfnamefont {C.~M.}\ \bibnamefont {Holland}}, \bibinfo {author} {\bibfnamefont {Y.}~\bibnamefont {Jiang}}, \bibinfo {author} {\bibfnamefont {B.~E.}\ \bibnamefont {Garber}}, \bibinfo {author} {\bibfnamefont {S.~P.}\ \bibnamefont {Carman}},\ and\ \bibinfo {author} {\bibfnamefont {J.~M.}\ \bibnamefont {Hogan}},\ }\bibfield  {title} {\emph {\bibinfo {title} {{Large Momentum Transfer Clock Atom Interferometry on the $\mathrm{689~nm}$ Intercombination Line of Strontium}}},\ }\href {https://doi.org/10.1103/PhysRevLett.124.083604} {\bibfield  {journal} {\bibinfo  {journal} {Phys. Rev. Lett.}\ }\textbf {\bibinfo {volume} {124}},\ \bibinfo {pages} {083604} (\bibinfo {year} {2020})}\BibitemShut {NoStop}%
\bibitem [{\citenamefont {Szigeti}\ \emph {et~al.}(2012)\citenamefont {Szigeti}, \citenamefont {Debs}, \citenamefont {Hope}, \citenamefont {Robins},\ and\ \citenamefont {Close}}]{Szigeti2012}%
  \BibitemOpen
  \bibfield  {author} {\bibinfo {author} {\bibfnamefont {S.~S.}\ \bibnamefont {Szigeti}}, \bibinfo {author} {\bibfnamefont {J.~E.}\ \bibnamefont {Debs}}, \bibinfo {author} {\bibfnamefont {J.~J.}\ \bibnamefont {Hope}}, \bibinfo {author} {\bibfnamefont {N.~P.}\ \bibnamefont {Robins}},\ and\ \bibinfo {author} {\bibfnamefont {J.~D.}\ \bibnamefont {Close}},\ }\bibfield  {title} {\emph {\bibinfo {title} {Why momentum width matters for atom interferometry with {Bragg} pulses}},\ }\href {https://doi.org/10.1088/1367-2630/14/2/023009} {\bibfield  {journal} {\bibinfo  {journal} {New J. Phys.}\ }\textbf {\bibinfo {volume} {14}},\ \bibinfo {pages} {023009} (\bibinfo {year} {2012})}\BibitemShut {NoStop}%
\bibitem [{\citenamefont {Cronin}\ \emph {et~al.}(2009)\citenamefont {Cronin}, \citenamefont {Schmiedmayer},\ and\ \citenamefont {Pritchard}}]{Cronin2009}%
  \BibitemOpen
  \bibfield  {author} {\bibinfo {author} {\bibfnamefont {A.~D.}\ \bibnamefont {Cronin}}, \bibinfo {author} {\bibfnamefont {J.}~\bibnamefont {Schmiedmayer}},\ and\ \bibinfo {author} {\bibfnamefont {D.~E.}\ \bibnamefont {Pritchard}},\ }\bibfield  {title} {\emph {\bibinfo {title} {Optics and interferometry with atoms and molecules}},\ }\href {https://doi.org/10.1103/RevModPhys.81.1051} {\bibfield  {journal} {\bibinfo  {journal} {Rev. Mod. Phys.}\ }\textbf {\bibinfo {volume} {81}},\ \bibinfo {pages} {1051--1129} (\bibinfo {year} {2009})}\BibitemShut {NoStop}%
\bibitem [{\citenamefont {Badurina}\ \emph {et~al.}(2020)\citenamefont {Badurina}, \citenamefont {Bentine}, \citenamefont {Blas}, \citenamefont {Bongs}, \citenamefont {Bortoletto}, \citenamefont {Bowcock}, \citenamefont {Bridges}, \citenamefont {Bowden}, \citenamefont {Buchmueller}, \citenamefont {Burrage}, \citenamefont {Coleman}, \citenamefont {Elertas}, \citenamefont {Ellis}, \citenamefont {Foot}, \citenamefont {Gibson}, \citenamefont {Haehnelt}, \citenamefont {Harte}, \citenamefont {Hedges}, \citenamefont {Hobson}, \citenamefont {Holynski}, \citenamefont {Jones}, \citenamefont {Langlois}, \citenamefont {Lellouch}, \citenamefont {Lewicki}, \citenamefont {Maiolino}, \citenamefont {Majewski}, \citenamefont {Malik}, \citenamefont {March-Russell}, \citenamefont {McCabe}, \citenamefont {Newbold}, \citenamefont {Sauer}, \citenamefont {Schneider}, \citenamefont {Shipsey}, \citenamefont {Singh}, \citenamefont {Uchida}, \citenamefont {Valenzuela}, \citenamefont {van~der Grinten}, \citenamefont {Vaskonen},
  \citenamefont {Vossebeld}, \citenamefont {Weatherill},\ and\ \citenamefont {Wilmut}}]{Badurina2020}%
  \BibitemOpen
  \bibfield  {author} {\bibinfo {author} {\bibfnamefont {L.}~\bibnamefont {Badurina}}, \bibinfo {author} {\bibfnamefont {E.}~\bibnamefont {Bentine}}, \bibinfo {author} {\bibfnamefont {D.}~\bibnamefont {Blas}}, \bibinfo {author} {\bibfnamefont {K.}~\bibnamefont {Bongs}}, \bibinfo {author} {\bibfnamefont {D.}~\bibnamefont {Bortoletto}}, \bibinfo {author} {\bibfnamefont {T.}~\bibnamefont {Bowcock}}, \bibinfo {author} {\bibfnamefont {K.}~\bibnamefont {Bridges}}, \bibinfo {author} {\bibfnamefont {W.}~\bibnamefont {Bowden}}, \bibinfo {author} {\bibfnamefont {O.}~\bibnamefont {Buchmueller}}, \bibinfo {author} {\bibfnamefont {C.}~\bibnamefont {Burrage}}, \bibinfo {author} {\bibfnamefont {J.}~\bibnamefont {Coleman}}, \bibinfo {author} {\bibfnamefont {G.}~\bibnamefont {Elertas}}, \bibinfo {author} {\bibfnamefont {J.}~\bibnamefont {Ellis}}, \bibinfo {author} {\bibfnamefont {C.}~\bibnamefont {Foot}}, \bibinfo {author} {\bibfnamefont {V.}~\bibnamefont {Gibson}}, \bibinfo {author} {\bibfnamefont {M.}~\bibnamefont
  {Haehnelt}}, \bibinfo {author} {\bibfnamefont {T.}~\bibnamefont {Harte}}, \bibinfo {author} {\bibfnamefont {S.}~\bibnamefont {Hedges}}, \bibinfo {author} {\bibfnamefont {R.}~\bibnamefont {Hobson}}, \bibinfo {author} {\bibfnamefont {M.}~\bibnamefont {Holynski}}, \bibinfo {author} {\bibfnamefont {T.}~\bibnamefont {Jones}}, \bibinfo {author} {\bibfnamefont {M.}~\bibnamefont {Langlois}}, \bibinfo {author} {\bibfnamefont {S.}~\bibnamefont {Lellouch}}, \bibinfo {author} {\bibfnamefont {M.}~\bibnamefont {Lewicki}}, \bibinfo {author} {\bibfnamefont {R.}~\bibnamefont {Maiolino}}, \bibinfo {author} {\bibfnamefont {P.}~\bibnamefont {Majewski}}, \bibinfo {author} {\bibfnamefont {S.}~\bibnamefont {Malik}}, \bibinfo {author} {\bibfnamefont {J.}~\bibnamefont {March-Russell}}, \bibinfo {author} {\bibfnamefont {C.}~\bibnamefont {McCabe}}, \bibinfo {author} {\bibfnamefont {D.}~\bibnamefont {Newbold}}, \bibinfo {author} {\bibfnamefont {B.}~\bibnamefont {Sauer}}, \bibinfo {author} {\bibfnamefont {U.}~\bibnamefont {Schneider}},
  \bibinfo {author} {\bibfnamefont {I.}~\bibnamefont {Shipsey}}, \bibinfo {author} {\bibfnamefont {Y.}~\bibnamefont {Singh}}, \bibinfo {author} {\bibfnamefont {M.}~\bibnamefont {Uchida}}, \bibinfo {author} {\bibfnamefont {T.}~\bibnamefont {Valenzuela}}, \bibinfo {author} {\bibfnamefont {M.}~\bibnamefont {van~der Grinten}}, \bibinfo {author} {\bibfnamefont {V.}~\bibnamefont {Vaskonen}}, \bibinfo {author} {\bibfnamefont {J.}~\bibnamefont {Vossebeld}}, \bibinfo {author} {\bibfnamefont {D.}~\bibnamefont {Weatherill}},\ and\ \bibinfo {author} {\bibfnamefont {I.}~\bibnamefont {Wilmut}},\ }\bibfield  {title} {\emph {\bibinfo {title} {{AION}: an atom interferometer observatory and network}},\ }\href {https://doi.org/10.1088/1475-7516/2020/05/011} {\bibfield  {journal} {\bibinfo  {journal} {J. Cosmol. Astropart. Phys.}\ }\textbf {\bibinfo {volume} {2020}}\bibinfo  {number} { (05)},\ \bibinfo {pages} {011}}\BibitemShut {NoStop}%
\bibitem [{\citenamefont {B\'eguin}\ \emph {et~al.}(2023)\citenamefont {B\'eguin}, \citenamefont {Rodzinka}, \citenamefont {Calmels}, \citenamefont {Allard},\ and\ \citenamefont {Gauguet}}]{Beguin2023}%
  \BibitemOpen
\bibfield  {number} {  }\bibfield  {author} {\bibinfo {author} {\bibfnamefont {A.}~\bibnamefont {B\'eguin}}, \bibinfo {author} {\bibfnamefont {T.}~\bibnamefont {Rodzinka}}, \bibinfo {author} {\bibfnamefont {L.}~\bibnamefont {Calmels}}, \bibinfo {author} {\bibfnamefont {B.}~\bibnamefont {Allard}},\ and\ \bibinfo {author} {\bibfnamefont {A.}~\bibnamefont {Gauguet}},\ }\bibfield  {title} {\emph {\bibinfo {title} {{Atom Interferometry with Coherent Enhancement of Bragg Pulse Sequences}}},\ }\href {https://doi.org/10.1103/PhysRevLett.131.143401} {\bibfield  {journal} {\bibinfo  {journal} {Phys. Rev. Lett.}\ }\textbf {\bibinfo {volume} {131}},\ \bibinfo {pages} {143401} (\bibinfo {year} {2023})}\BibitemShut {NoStop}%
\bibitem [{\citenamefont {Dunning}\ \emph {et~al.}(2014)\citenamefont {Dunning}, \citenamefont {Gregory}, \citenamefont {Bateman}, \citenamefont {Cooper}, \citenamefont {Himsworth}, \citenamefont {Jones},\ and\ \citenamefont {Freegarde}}]{Dunning2014}%
  \BibitemOpen
  \bibfield  {author} {\bibinfo {author} {\bibfnamefont {A.}~\bibnamefont {Dunning}}, \bibinfo {author} {\bibfnamefont {R.}~\bibnamefont {Gregory}}, \bibinfo {author} {\bibfnamefont {J.}~\bibnamefont {Bateman}}, \bibinfo {author} {\bibfnamefont {N.}~\bibnamefont {Cooper}}, \bibinfo {author} {\bibfnamefont {M.}~\bibnamefont {Himsworth}}, \bibinfo {author} {\bibfnamefont {J.~A.}\ \bibnamefont {Jones}},\ and\ \bibinfo {author} {\bibfnamefont {T.}~\bibnamefont {Freegarde}},\ }\bibfield  {title} {\emph {\bibinfo {title} {Composite pulses for interferometry in a thermal cold atom cloud}},\ }\href {https://doi.org/10.1103/PhysRevA.90.033608} {\bibfield  {journal} {\bibinfo  {journal} {Phys. Rev. A}\ }\textbf {\bibinfo {volume} {90}},\ \bibinfo {pages} {033608} (\bibinfo {year} {2014})}\BibitemShut {NoStop}%
\bibitem [{\citenamefont {Baillard}\ \emph {et~al.}(2007)\citenamefont {Baillard}, \citenamefont {Fouch\'e}, \citenamefont {Targat}, \citenamefont {Westergaard}, \citenamefont {Lecallier}, \citenamefont {Le~Coq}, \citenamefont {Rovera}, \citenamefont {Bize},\ and\ \citenamefont {Lemonde}}]{Baillard2007}%
  \BibitemOpen
  \bibfield  {author} {\bibinfo {author} {\bibfnamefont {X.}~\bibnamefont {Baillard}}, \bibinfo {author} {\bibfnamefont {M.}~\bibnamefont {Fouch\'e}}, \bibinfo {author} {\bibfnamefont {R.}~\bibnamefont {Targat}}, \bibinfo {author} {\bibfnamefont {P.}~\bibnamefont {Westergaard}}, \bibinfo {author} {\bibfnamefont {A.}~\bibnamefont {Lecallier}}, \bibinfo {author} {\bibfnamefont {Y.}~\bibnamefont {Le~Coq}}, \bibinfo {author} {\bibfnamefont {G.}~\bibnamefont {Rovera}}, \bibinfo {author} {\bibfnamefont {S.}~\bibnamefont {Bize}},\ and\ \bibinfo {author} {\bibfnamefont {P.}~\bibnamefont {Lemonde}},\ }\bibfield  {title} {\emph {\bibinfo {title} {{Accuracy Evaluation of an Optical Lattice Clock with Bosonic Atoms}}},\ }\href {https://doi.org/10.1364/OL.32.001812} {\bibfield  {journal} {\bibinfo  {journal} {Opt. Lett.}\ }\textbf {\bibinfo {volume} {32}},\ \bibinfo {pages} {1812--4} (\bibinfo {year} {2007})}\BibitemShut {NoStop}%
\bibitem [{\citenamefont {Akatsuka}\ \emph {et~al.}(2010)\citenamefont {Akatsuka}, \citenamefont {Takamoto},\ and\ \citenamefont {Katori}}]{Akatsuka2010}%
  \BibitemOpen
  \bibfield  {author} {\bibinfo {author} {\bibfnamefont {T.}~\bibnamefont {Akatsuka}}, \bibinfo {author} {\bibfnamefont {M.}~\bibnamefont {Takamoto}},\ and\ \bibinfo {author} {\bibfnamefont {H.}~\bibnamefont {Katori}},\ }\bibfield  {title} {\emph {\bibinfo {title} {Three-dimensional optical lattice clock with bosonic $^{88}\mathrm{Sr}$ atoms}},\ }\href {https://doi.org/10.1103/PhysRevA.81.023402} {\bibfield  {journal} {\bibinfo  {journal} {Phys. Rev. A}\ }\textbf {\bibinfo {volume} {81}},\ \bibinfo {pages} {023402} (\bibinfo {year} {2010})}\BibitemShut {NoStop}%
\bibitem [{\citenamefont {Origlia}\ \emph {et~al.}(2018)\citenamefont {Origlia}, \citenamefont {Pramod}, \citenamefont {Schiller}, \citenamefont {Singh}, \citenamefont {Bongs}, \citenamefont {Schwarz}, \citenamefont {Al-Masoudi}, \citenamefont {D\"orscher}, \citenamefont {Herbers}, \citenamefont {H\"afner}, \citenamefont {Sterr},\ and\ \citenamefont {Lisdat}}]{Origlia2018}%
  \BibitemOpen
  \bibfield  {author} {\bibinfo {author} {\bibfnamefont {S.}~\bibnamefont {Origlia}}, \bibinfo {author} {\bibfnamefont {M.~S.}\ \bibnamefont {Pramod}}, \bibinfo {author} {\bibfnamefont {S.}~\bibnamefont {Schiller}}, \bibinfo {author} {\bibfnamefont {Y.}~\bibnamefont {Singh}}, \bibinfo {author} {\bibfnamefont {K.}~\bibnamefont {Bongs}}, \bibinfo {author} {\bibfnamefont {R.}~\bibnamefont {Schwarz}}, \bibinfo {author} {\bibfnamefont {A.}~\bibnamefont {Al-Masoudi}}, \bibinfo {author} {\bibfnamefont {S.}~\bibnamefont {D\"orscher}}, \bibinfo {author} {\bibfnamefont {S.}~\bibnamefont {Herbers}}, \bibinfo {author} {\bibfnamefont {S.}~\bibnamefont {H\"afner}}, \bibinfo {author} {\bibfnamefont {U.}~\bibnamefont {Sterr}},\ and\ \bibinfo {author} {\bibfnamefont {C.}~\bibnamefont {Lisdat}},\ }\bibfield  {title} {\emph {\bibinfo {title} {{Towards an optical clock for space: Compact, high-performance optical lattice clock based on bosonic atoms}}},\ }\href {https://doi.org/10.1103/PhysRevA.98.053443} {\bibfield  {journal}
  {\bibinfo  {journal} {Phys. Rev. A}\ }\textbf {\bibinfo {volume} {98}},\ \bibinfo {pages} {053443} (\bibinfo {year} {2018})}\BibitemShut {NoStop}%
\bibitem [{\citenamefont {Norcia}\ \emph {et~al.}(2019)\citenamefont {Norcia}, \citenamefont {Young}, \citenamefont {Eckner}, \citenamefont {Oelker}, \citenamefont {Ye},\ and\ \citenamefont {Kaufman}}]{Norcia2019}%
  \BibitemOpen
  \bibfield  {author} {\bibinfo {author} {\bibfnamefont {M.~A.}\ \bibnamefont {Norcia}}, \bibinfo {author} {\bibfnamefont {A.~W.}\ \bibnamefont {Young}}, \bibinfo {author} {\bibfnamefont {W.~J.}\ \bibnamefont {Eckner}}, \bibinfo {author} {\bibfnamefont {E.}~\bibnamefont {Oelker}}, \bibinfo {author} {\bibfnamefont {J.}~\bibnamefont {Ye}},\ and\ \bibinfo {author} {\bibfnamefont {A.~M.}\ \bibnamefont {Kaufman}},\ }\bibfield  {title} {\emph {\bibinfo {title} {{Seconds-scale coherence on an optical clock transition in a tweezer array}}},\ }\href {https://doi.org/10.1126/science.aay0644} {\bibfield  {journal} {\bibinfo  {journal} {Science}\ }\textbf {\bibinfo {volume} {366}},\ \bibinfo {pages} {93--97} (\bibinfo {year} {2019})}\BibitemShut {NoStop}%
\bibitem [{\citenamefont {Stock}\ \emph {et~al.}(2008)\citenamefont {Stock}, \citenamefont {Babcock}, \citenamefont {Raizen},\ and\ \citenamefont {Sanders}}]{Stock2008}%
  \BibitemOpen
  \bibfield  {author} {\bibinfo {author} {\bibfnamefont {R.}~\bibnamefont {Stock}}, \bibinfo {author} {\bibfnamefont {N.~S.}\ \bibnamefont {Babcock}}, \bibinfo {author} {\bibfnamefont {M.~G.}\ \bibnamefont {Raizen}},\ and\ \bibinfo {author} {\bibfnamefont {B.~C.}\ \bibnamefont {Sanders}},\ }\bibfield  {title} {\emph {\bibinfo {title} {{Entanglement of group-II-like atoms with fast measurement for quantum information processing}}},\ }\href {https://doi.org/10.1103/PhysRevA.78.022301} {\bibfield  {journal} {\bibinfo  {journal} {Phys. Rev. A}\ }\textbf {\bibinfo {volume} {78}},\ \bibinfo {pages} {022301} (\bibinfo {year} {2008})}\BibitemShut {NoStop}%
\bibitem [{\citenamefont {Young}\ \emph {et~al.}(2020)\citenamefont {Young}, \citenamefont {Eckner}, \citenamefont {Milner}, \citenamefont {Kedar}, \citenamefont {Norcia}, \citenamefont {Oelker}, \citenamefont {Schine}, \citenamefont {Ye},\ and\ \citenamefont {Kaufman}}]{Young2020}%
  \BibitemOpen
  \bibfield  {author} {\bibinfo {author} {\bibfnamefont {A.}~\bibnamefont {Young}}, \bibinfo {author} {\bibfnamefont {W.}~\bibnamefont {Eckner}}, \bibinfo {author} {\bibfnamefont {W.}~\bibnamefont {Milner}}, \bibinfo {author} {\bibfnamefont {D.}~\bibnamefont {Kedar}}, \bibinfo {author} {\bibfnamefont {M.}~\bibnamefont {Norcia}}, \bibinfo {author} {\bibfnamefont {E.}~\bibnamefont {Oelker}}, \bibinfo {author} {\bibfnamefont {N.}~\bibnamefont {Schine}}, \bibinfo {author} {\bibfnamefont {J.}~\bibnamefont {Ye}},\ and\ \bibinfo {author} {\bibfnamefont {A.}~\bibnamefont {Kaufman}},\ }\bibfield  {title} {\emph {\bibinfo {title} {{Half-minute-scale atomic coherence and high relative stability in a tweezer clock}}},\ }\href {https://doi.org/10.1038/s41586-020-3009-y} {\bibfield  {journal} {\bibinfo  {journal} {Nature}\ }\textbf {\bibinfo {volume} {588}},\ \bibinfo {pages} {408--413} (\bibinfo {year} {2020})}\BibitemShut {NoStop}%
\bibitem [{\citenamefont {Pagano}\ \emph {et~al.}(2022)\citenamefont {Pagano}, \citenamefont {Weber}, \citenamefont {Jaschke}, \citenamefont {Pfau}, \citenamefont {Meinert}, \citenamefont {Montangero},\ and\ \citenamefont {B\"uchler}}]{Pagano2022}%
  \BibitemOpen
  \bibfield  {author} {\bibinfo {author} {\bibfnamefont {A.}~\bibnamefont {Pagano}}, \bibinfo {author} {\bibfnamefont {S.}~\bibnamefont {Weber}}, \bibinfo {author} {\bibfnamefont {D.}~\bibnamefont {Jaschke}}, \bibinfo {author} {\bibfnamefont {T.}~\bibnamefont {Pfau}}, \bibinfo {author} {\bibfnamefont {F.}~\bibnamefont {Meinert}}, \bibinfo {author} {\bibfnamefont {S.}~\bibnamefont {Montangero}},\ and\ \bibinfo {author} {\bibfnamefont {H.~P.}\ \bibnamefont {B\"uchler}},\ }\bibfield  {title} {\emph {\bibinfo {title} {{Error budgeting for a controlled-phase gate with strontium-88 Rydberg atoms}}},\ }\href {https://doi.org/10.1103/PhysRevResearch.4.033019} {\bibfield  {journal} {\bibinfo  {journal} {Phys. Rev. Res.}\ }\textbf {\bibinfo {volume} {4}},\ \bibinfo {pages} {033019} (\bibinfo {year} {2022})}\BibitemShut {NoStop}%
\bibitem [{\citenamefont {Okuno}\ \emph {et~al.}(2022)\citenamefont {Okuno}, \citenamefont {Nakamura}, \citenamefont {Kusano}, \citenamefont {Takasu}, \citenamefont {Takei}, \citenamefont {Konishi},\ and\ \citenamefont {Takahashi}}]{Okuno2022}%
  \BibitemOpen
  \bibfield  {author} {\bibinfo {author} {\bibfnamefont {D.}~\bibnamefont {Okuno}}, \bibinfo {author} {\bibfnamefont {Y.}~\bibnamefont {Nakamura}}, \bibinfo {author} {\bibfnamefont {T.}~\bibnamefont {Kusano}}, \bibinfo {author} {\bibfnamefont {Y.}~\bibnamefont {Takasu}}, \bibinfo {author} {\bibfnamefont {N.}~\bibnamefont {Takei}}, \bibinfo {author} {\bibfnamefont {H.}~\bibnamefont {Konishi}},\ and\ \bibinfo {author} {\bibfnamefont {Y.}~\bibnamefont {Takahashi}},\ }\bibfield  {title} {\emph {\bibinfo {title} {{High-resolution Spectroscopy and Single-photon Rydberg Excitation of Reconfigurable Ytterbium Atom Tweezer Arrays Utilizing a Metastable State}}},\ }\href {https://doi.org/10.7566/JPSJ.91.084301} {\bibfield  {journal} {\bibinfo  {journal} {J. Phys. Soc. Jpn.}\ }\textbf {\bibinfo {volume} {91}},\ \bibinfo {pages} {084301} (\bibinfo {year} {2022})}\BibitemShut {NoStop}%
\bibitem [{\citenamefont {Trautmann}\ \emph {et~al.}(2023)\citenamefont {Trautmann}, \citenamefont {Yankelev}, \citenamefont {Kl\"usener}, \citenamefont {Park}, \citenamefont {Bloch},\ and\ \citenamefont {Blatt}}]{Trautmann2023}%
  \BibitemOpen
  \bibfield  {author} {\bibinfo {author} {\bibfnamefont {J.}~\bibnamefont {Trautmann}}, \bibinfo {author} {\bibfnamefont {D.}~\bibnamefont {Yankelev}}, \bibinfo {author} {\bibfnamefont {V.}~\bibnamefont {Kl\"usener}}, \bibinfo {author} {\bibfnamefont {A.~J.}\ \bibnamefont {Park}}, \bibinfo {author} {\bibfnamefont {I.}~\bibnamefont {Bloch}},\ and\ \bibinfo {author} {\bibfnamefont {S.}~\bibnamefont {Blatt}},\ }\bibfield  {title} {\emph {\bibinfo {title} {{$^{1}\mathrm{S}_{0}\text{\ensuremath{-}}^{3}\mathrm{P}_{2}$ magnetic quadrupole transition in neutral strontium}}},\ }\href {https://doi.org/10.1103/PhysRevResearch.5.013219} {\bibfield  {journal} {\bibinfo  {journal} {Phys. Rev. Res.}\ }\textbf {\bibinfo {volume} {5}},\ \bibinfo {pages} {013219} (\bibinfo {year} {2023})}\BibitemShut {NoStop}%
\bibitem [{\citenamefont {Pucher}\ \emph {et~al.}(2024)\citenamefont {Pucher}, \citenamefont {Kl\"usener}, \citenamefont {Spriestersbach}, \citenamefont {Geiger}, \citenamefont {Schindewolf}, \citenamefont {Bloch},\ and\ \citenamefont {Blatt}}]{Pucher2024}%
  \BibitemOpen
  \bibfield  {author} {\bibinfo {author} {\bibfnamefont {S.}~\bibnamefont {Pucher}}, \bibinfo {author} {\bibfnamefont {V.}~\bibnamefont {Kl\"usener}}, \bibinfo {author} {\bibfnamefont {F.}~\bibnamefont {Spriestersbach}}, \bibinfo {author} {\bibfnamefont {J.}~\bibnamefont {Geiger}}, \bibinfo {author} {\bibfnamefont {A.}~\bibnamefont {Schindewolf}}, \bibinfo {author} {\bibfnamefont {I.}~\bibnamefont {Bloch}},\ and\ \bibinfo {author} {\bibfnamefont {S.}~\bibnamefont {Blatt}},\ }\bibfield  {title} {\emph {\bibinfo {title} {{Fine-Structure Qubit Encoded in Metastable Strontium Trapped in an Optical Lattice}}},\ }\href {https://doi.org/10.1103/PhysRevLett.132.150605} {\bibfield  {journal} {\bibinfo  {journal} {Phys. Rev. Lett.}\ }\textbf {\bibinfo {volume} {132}},\ \bibinfo {pages} {150605} (\bibinfo {year} {2024})}\BibitemShut {NoStop}%
\bibitem [{\citenamefont {Kl\"usener}\ \emph {et~al.}(2024)\citenamefont {Kl\"usener}, \citenamefont {Pucher}, \citenamefont {Yankelev}, \citenamefont {Trautmann}, \citenamefont {Spriestersbach}, \citenamefont {Filin}, \citenamefont {Porsev}, \citenamefont {Safronova}, \citenamefont {Bloch},\ and\ \citenamefont {Blatt}}]{Klusener2024}%
  \BibitemOpen
  \bibfield  {author} {\bibinfo {author} {\bibfnamefont {V.}~\bibnamefont {Kl\"usener}}, \bibinfo {author} {\bibfnamefont {S.}~\bibnamefont {Pucher}}, \bibinfo {author} {\bibfnamefont {D.}~\bibnamefont {Yankelev}}, \bibinfo {author} {\bibfnamefont {J.}~\bibnamefont {Trautmann}}, \bibinfo {author} {\bibfnamefont {F.}~\bibnamefont {Spriestersbach}}, \bibinfo {author} {\bibfnamefont {D.}~\bibnamefont {Filin}}, \bibinfo {author} {\bibfnamefont {S.~G.}\ \bibnamefont {Porsev}}, \bibinfo {author} {\bibfnamefont {M.~S.}\ \bibnamefont {Safronova}}, \bibinfo {author} {\bibfnamefont {I.}~\bibnamefont {Bloch}},\ and\ \bibinfo {author} {\bibfnamefont {S.}~\bibnamefont {Blatt}},\ }\bibfield  {title} {\emph {\bibinfo {title} {{Long-Lived Coherence on a $\micro\mathrm{Hz}$ Scale Optical Magnetic Quadrupole Transition}}},\ }\href {https://doi.org/10.1103/PhysRevLett.132.253201} {\bibfield  {journal} {\bibinfo  {journal} {Phys. Rev. Lett.}\ }\textbf {\bibinfo {volume} {132}},\ \bibinfo {pages} {253201} (\bibinfo {year}
  {2024})}\BibitemShut {NoStop}%
\bibitem [{\citenamefont {He}\ \emph {et~al.}(2025)\citenamefont {He}, \citenamefont {Pasquiou}, \citenamefont {Escudero}, \citenamefont {Zhou}, \citenamefont {Borkowski},\ and\ \citenamefont {Schreck}}]{He2024}%
  \BibitemOpen
  \bibfield  {author} {\bibinfo {author} {\bibfnamefont {J.}~\bibnamefont {He}}, \bibinfo {author} {\bibfnamefont {B.}~\bibnamefont {Pasquiou}}, \bibinfo {author} {\bibfnamefont {R.~G.}\ \bibnamefont {Escudero}}, \bibinfo {author} {\bibfnamefont {S.}~\bibnamefont {Zhou}}, \bibinfo {author} {\bibfnamefont {M.}~\bibnamefont {Borkowski}},\ and\ \bibinfo {author} {\bibfnamefont {F.}~\bibnamefont {Schreck}},\ }\bibfield  {title} {\emph {\bibinfo {title} {Coherent three-photon excitation of the strontium clock transition}},\ }\href {https://doi.org/10.1103/PhysRevResearch.7.L012050} {\bibfield  {journal} {\bibinfo  {journal} {Phys. Rev. Res.}\ }\textbf {\bibinfo {volume} {7}},\ \bibinfo {pages} {L012050} (\bibinfo {year} {2025})}\BibitemShut {NoStop}%
\bibitem [{\citenamefont {Ferrari}\ \emph {et~al.}(2003)\citenamefont {Ferrari}, \citenamefont {Cancio}, \citenamefont {Drullinger}, \citenamefont {Giusfredi}, \citenamefont {Poli}, \citenamefont {Prevedelli}, \citenamefont {Toninelli},\ and\ \citenamefont {Tino}}]{Ferrari2003}%
  \BibitemOpen
  \bibfield  {author} {\bibinfo {author} {\bibfnamefont {G.}~\bibnamefont {Ferrari}}, \bibinfo {author} {\bibfnamefont {P.}~\bibnamefont {Cancio}}, \bibinfo {author} {\bibfnamefont {R.}~\bibnamefont {Drullinger}}, \bibinfo {author} {\bibfnamefont {G.}~\bibnamefont {Giusfredi}}, \bibinfo {author} {\bibfnamefont {N.}~\bibnamefont {Poli}}, \bibinfo {author} {\bibfnamefont {M.}~\bibnamefont {Prevedelli}}, \bibinfo {author} {\bibfnamefont {C.}~\bibnamefont {Toninelli}},\ and\ \bibinfo {author} {\bibfnamefont {G.~M.}\ \bibnamefont {Tino}},\ }\bibfield  {title} {\emph {\bibinfo {title} {{Precision Frequency Measurement of Visible Intercombination Lines of Strontium}}},\ }\href {https://doi.org/10.1103/PhysRevLett.91.243002} {\bibfield  {journal} {\bibinfo  {journal} {Phys. Rev. Lett.}\ }\textbf {\bibinfo {volume} {91}},\ \bibinfo {pages} {243002} (\bibinfo {year} {2003})}\BibitemShut {NoStop}%
\bibitem [{\citenamefont {Nicholson}\ \emph {et~al.}(2015)\citenamefont {Nicholson}, \citenamefont {Campbell}, \citenamefont {Hutson}, \citenamefont {Marti}, \citenamefont {Bloom}, \citenamefont {McNally}, \citenamefont {Zhang}, \citenamefont {Barrett}, \citenamefont {Safronova}, \citenamefont {Strouse}, \citenamefont {Tew},\ and\ \citenamefont {Ye}}]{Nicholson2015}%
  \BibitemOpen
  \bibfield  {author} {\bibinfo {author} {\bibfnamefont {T.~L.}\ \bibnamefont {Nicholson}}, \bibinfo {author} {\bibfnamefont {S.~L.}\ \bibnamefont {Campbell}}, \bibinfo {author} {\bibfnamefont {R.~B.}\ \bibnamefont {Hutson}}, \bibinfo {author} {\bibfnamefont {G.~E.}\ \bibnamefont {Marti}}, \bibinfo {author} {\bibfnamefont {B.~J.}\ \bibnamefont {Bloom}}, \bibinfo {author} {\bibfnamefont {R.~L.}\ \bibnamefont {McNally}}, \bibinfo {author} {\bibfnamefont {W.}~\bibnamefont {Zhang}}, \bibinfo {author} {\bibfnamefont {M.~D.}\ \bibnamefont {Barrett}}, \bibinfo {author} {\bibfnamefont {M.~S.}\ \bibnamefont {Safronova}}, \bibinfo {author} {\bibfnamefont {G.~F.}\ \bibnamefont {Strouse}}, \bibinfo {author} {\bibfnamefont {W.~L.}\ \bibnamefont {Tew}},\ and\ \bibinfo {author} {\bibfnamefont {J.}~\bibnamefont {Ye}},\ }\bibfield  {title} {\emph {\bibinfo {title} {{Systematic evaluation of an atomic clock at $2 \times 10^{-18}$ total uncertainty}}},\ }\href {https://doi.org/10.1038/ncomms7896} {\bibfield  {journal} {\bibinfo
   {journal} {Nat. Commun.}\ }\textbf {\bibinfo {volume} {6}},\ \bibinfo {pages} {6896} (\bibinfo {year} {2015})}\BibitemShut {NoStop}%
\bibitem [{\citenamefont {Courtillot}\ \emph {et~al.}(2005)\citenamefont {Courtillot}, \citenamefont {Quessada-Vial}, \citenamefont {Brusch}, \citenamefont {Kolker}, \citenamefont {Rovera},\ and\ \citenamefont {Lemonde}}]{Courtillot2005}%
  \BibitemOpen
  \bibfield  {author} {\bibinfo {author} {\bibfnamefont {I.}~\bibnamefont {Courtillot}}, \bibinfo {author} {\bibfnamefont {A.}~\bibnamefont {Quessada-Vial}}, \bibinfo {author} {\bibfnamefont {A.}~\bibnamefont {Brusch}}, \bibinfo {author} {\bibfnamefont {D.}~\bibnamefont {Kolker}}, \bibinfo {author} {\bibfnamefont {G.}~\bibnamefont {Rovera}},\ and\ \bibinfo {author} {\bibfnamefont {P.}~\bibnamefont {Lemonde}},\ }\bibfield  {title} {\emph {\bibinfo {title} {{Accurate spectroscopy of $\mathrm{Sr}$ atoms}}},\ }\href {https://doi.org/10.1140/epjd/e2005-00058-0} {\bibfield  {journal} {\bibinfo  {journal} {Eur. Phys. J. D.}\ }\textbf {\bibinfo {volume} {33}},\ \bibinfo {pages} {161--171} (\bibinfo {year} {2005})}\BibitemShut {NoStop}%
\bibitem [{\citenamefont {J{\"o}nsson}\ \emph {et~al.}(1984)\citenamefont {J{\"o}nsson}, \citenamefont {Levinson}, \citenamefont {Persson},\ and\ \citenamefont {Wahlstr{\"o}m}}]{Jonsson1984}%
  \BibitemOpen
  \bibfield  {author} {\bibinfo {author} {\bibfnamefont {G.}~\bibnamefont {J{\"o}nsson}}, \bibinfo {author} {\bibfnamefont {C.}~\bibnamefont {Levinson}}, \bibinfo {author} {\bibfnamefont {A.}~\bibnamefont {Persson}},\ and\ \bibinfo {author} {\bibfnamefont {C.-G.}\ \bibnamefont {Wahlstr{\"o}m}},\ }\bibfield  {title} {\emph {\bibinfo {title} {{Natural radiative lifetimes in the $^1\mathrm{P}_1$ and $^1\mathrm{F}_3$ sequences of $\mathrm{Sr~I}$}}},\ }\href {https://doi.org/10.1007/BF01439897} {\bibfield  {journal} {\bibinfo  {journal} {Z. Phys. A}\ }\textbf {\bibinfo {volume} {316}},\ \bibinfo {pages} {255--258} (\bibinfo {year} {1984})}\BibitemShut {NoStop}%
\bibitem [{\citenamefont {Metcalf}\ and\ \citenamefont {van~der Straten}(1999)}]{Metcalf1999}%
  \BibitemOpen
  \bibfield  {author} {\bibinfo {author} {\bibfnamefont {H.}~\bibnamefont {Metcalf}}\ and\ \bibinfo {author} {\bibfnamefont {P.}~\bibnamefont {van~der Straten}},\ }\href {https://doi.org/10.1007/978-1-4612-1470-0} {\emph {\bibinfo {title} {{Laser Cooling and Trapping}}}}\ (\bibinfo  {publisher} {Springer New York, NY},\ \bibinfo {year} {1999})\ pp.\ \bibinfo {pages} {50--56}\BibitemShut {NoStop}%
\bibitem [{\citenamefont {Morzy\'nski}\ \emph {et~al.}(2015)\citenamefont {Morzy\'nski}, \citenamefont {Bober}, \citenamefont {Bartoszek-Bober}, \citenamefont {Nawrocki}, \citenamefont {Krehlik}, \citenamefont {\'Sliwczy\'nski}, \citenamefont {Lipinski}, \citenamefont {Mas\l{}owski}, \citenamefont {Cygan}, \citenamefont {Dunst}, \citenamefont {Garus}, \citenamefont {Lisak}, \citenamefont {Zachorowski}, \citenamefont {Wojciech}, \citenamefont {Radzewicz}, \citenamefont {Ciury\l{}o},\ and\ \citenamefont {Zawada}}]{Morzynski2015}%
  \BibitemOpen
  \bibfield  {author} {\bibinfo {author} {\bibfnamefont {P.}~\bibnamefont {Morzy\'nski}}, \bibinfo {author} {\bibfnamefont {M.}~\bibnamefont {Bober}}, \bibinfo {author} {\bibfnamefont {D.}~\bibnamefont {Bartoszek-Bober}}, \bibinfo {author} {\bibfnamefont {J.}~\bibnamefont {Nawrocki}}, \bibinfo {author} {\bibfnamefont {P.}~\bibnamefont {Krehlik}}, \bibinfo {author} {\bibfnamefont {L.}~\bibnamefont {\'Sliwczy\'nski}}, \bibinfo {author} {\bibfnamefont {M.}~\bibnamefont {Lipinski}}, \bibinfo {author} {\bibfnamefont {P.}~\bibnamefont {Mas\l{}owski}}, \bibinfo {author} {\bibfnamefont {A.}~\bibnamefont {Cygan}}, \bibinfo {author} {\bibfnamefont {P.}~\bibnamefont {Dunst}}, \bibinfo {author} {\bibfnamefont {M.}~\bibnamefont {Garus}}, \bibinfo {author} {\bibfnamefont {D.}~\bibnamefont {Lisak}}, \bibinfo {author} {\bibfnamefont {J.}~\bibnamefont {Zachorowski}}, \bibinfo {author} {\bibfnamefont {G.}~\bibnamefont {Wojciech}}, \bibinfo {author} {\bibfnamefont {C.}~\bibnamefont {Radzewicz}}, \bibinfo {author} {\bibfnamefont
  {R.}~\bibnamefont {Ciury\l{}o}},\ and\ \bibinfo {author} {\bibfnamefont {M.}~\bibnamefont {Zawada}},\ }\bibfield  {title} {\emph {\bibinfo {title} {{Absolute measurement of the $^1\mathrm{S}_0 - ^3\mathrm{P}_0$ clock transition in neutral $^{88}\mathrm{Sr}$ over the 330 km-long stabilized fibre optic link}}},\ }\href {https://doi.org/10.1038/srep17495} {\bibfield  {journal} {\bibinfo  {journal} {Sci. Rep.}\ }\textbf {\bibinfo {volume} {5}},\ \bibinfo {pages} {17495} (\bibinfo {year} {2015})}\BibitemShut {NoStop}%
\end{thebibliography}%

\end{document}